%% file: main.tex
\documentclass[journal, article, onecolumn, 12pt]{IEEEtran}

\usepackage{mystyle}
\pdfoutput=1

\begin{document}

\thispagestyle{empty}


\title{Media-Based Modulation for Next-Generation Wireless: A Survey and Some New Developments}

\begin{comment}
\author{Ehsan Seifi$^\ddager,\dagger$, Amir K. Khandani$^\ddagger$, and Mehran Atamanesh $^{*}$ \\
\small {$^\ddagger$E\&CE Dept. University of Waterloo, Waterloo, ON, Canada \\
$^\dagger$ University of Waterloo, Waterloo, ON, Canada, Apple Inc. Cupertino, CA, USA \\
$^\dagger$ University of Waterloo, Waterloo, ON, Canada, Skyworks Solutions Inc. Ottawa, ON, Canada}}  
\end{comment}

\author[1, 2]{Ehsan Seifi}
\author[1]{Amir K. Khandani}
\author[1,3]{Mehran Atamanesh}
\affil[1]{\small E\&CE Dept. University of Waterloo, Waterloo, ON, Canada}
\affil[2]{Apple Inc., Cupertino, CA, USA}
\affil[3]{Skyworks Solutions Inc., Ottawa, ON, Canada}

\begin{comment}
\author{\IEEEauthorblockN{Ehsan Seifi, Amir K. Khandani and Mehran Atamanesh} \\
\IEEEauthorblockA{E\&CE Department, University of Waterloo,\\
Waterloo, ON, Canada}
}
\end{comment}

\maketitle

%\doublespacing
\thispagestyle{plain}
\pagestyle{plain}

\input{abstract.tex}
\input{article.tex}

\input{appendix.tex}

%\bibliographystyle{IEEEtran}
%\bibliography{IEEEabrv, ref}


\end{document}

%% file: abstract.tex
\begin{abstract}

The idea of media-based modulation (MBM) is to embed information in the channel states via intentional perturbations of the transmission media~\cite{c0}\cite{isit2014}. This article covers a broad range of topics regarding MBM, expanding on its benefits and reviewing relevant challenges, alluding to potential future research directions. The article starts by arguing how MBM differs from a source-based modulation; we highlight the key shortcomings in a legacy multiple-input multiple-output (MIMO) system that MBM sets out to address, including the issue of deep fades and MIMO diversity-multiplexing trade-off. The article further explains how MBM works in harmony with other index modulations and improves upon them by providing similar advantages with a more compact transmitter. Numerical results (simulation and analytical) are provided to support the claims on the discussed benefits. The highlights of numerical results include: 1) outage comparison with legacy MIMO systems; 2) comparisons with other state-of-the-art modulation schemes such as generalized spatial modulation; and 3) performance example of sending 32 bits of information in a single transmission with an excellent symbol error rate of $\mathsf{SER} \simeq 10^{-5}$ at ``energy per bit to noise power spectral density ratio'' of $\mathsf{E_b/N_0} \simeq -3.5$ dB.   
The article continues with methods to address the issues of receiver training and decoding for large constellation sets. A number of other research questions, such as pulse shaping to limit spectral growth due to the time-varying nature of MBM and the effect of forward error correcting codes on MBM diversity order are discussed. Finally, an RF transceiver structure is presented to generate independent propagation paths for embedding information. Fabrication and testing of the transceiver structure show close agreement between simulation and measurement.

\end{abstract}

%% file: article.tex
\section{Introduction}
 
\subsection{Motivation}
Shannon capacity results indicate that the transmission rate can increase as a linear function of the available spectrum, multiplied by a logarithmic function of the transmit energy. Wireless communication relies on two key attributes, traditionally considered its inherent bottlenecks. First, the spectrum is shared, causing interference among wireless links operating over the same spectrum. Second, the transmission channel includes a multitude of propagation paths, resulting in multi-path fading. Multi-path fading creates deep fades when signals received through different transmission paths add destructively. In many scenarios of practical interest, the transmission paths change slowly with time (slow fading), potentially resulting in a long-lasting degradation of the received signal-to-noise ratio (SNR), referred to as deep fades. 

Multiple-input multiple-output (MIMO) antenna systems embrace the above attributes towards improving the spectrum/power efficiency~\cite{NN1, NN2, NN3}, as well as dealing with deep fades~\cite{NN4}. The first attribute means signals sent from different transmit antennas add up at each receive antenna. Consequently, the input-output relationship is captured in the form of matrix multiplication. Due to the second attribute, such a channel matrix is, with a high probability, non-singular, resulting in a linear scaling of rate with $\min(A_T, A_R)$, where $A_T$ is the number of transmit antennas and $A_R$ is the number of receive antennas. The term $\min(A_T,A_R)$ is the improvement in spectrum efficiency due to MIMO and is referred to as the multiplexing gain (MG). MG appears as a scale factor times the $\log(\mathsf{SNR})$, in the expression of the rate as a function of SNR for $\mathsf{SNR}\rightarrow\infty$. MIMO also helps to combat deep fades by introducing redundancy among data transmitted/received through separate propagation paths. It is well-known that tackling slow-fading by creating diversity in MIMO systems comes at the cost of a reduction in degrees of freedom (i.e., MG) \cite{NN5}. 

\begin{figure*}[t]
\centering
\includegraphics[scale =0.58]{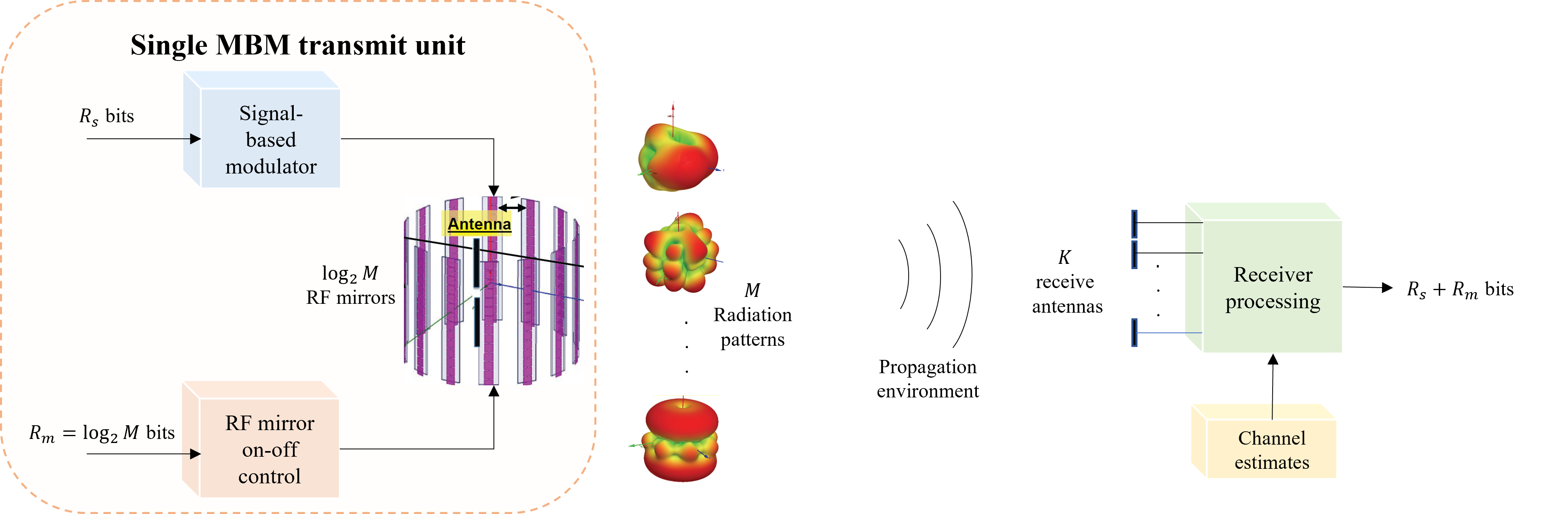}
%\vspace{-2cm}
\caption{SIMO-MBM transceiver diagram.}
\label{SIMO-MBM}
\end{figure*}

Although MIMO systems provide an elegant way to tailor wireless communications to embrace the two fundamental attributes mentioned earlier, three issues limit their achievable rate vs. energy. First, the problem of deep fades can be only (partially) alleviated at the cost of a reduction in the achievable rate (i.e., MG)~\cite{NN5}. Second, MG increases only with the smaller of the number of transmit and receive antennas. Third, the MIMO channel matrix is typically non-orthogonal, reducing the achievable rate compared to an orthogonal channel matrix of the same dimension.

Media-based modulation (MBM) addresses these three issues. The idea is to randomize the wireless channel by perturbing the propagation environment in the vicinity of the transmit antenna(s), which will change the overall transmission path. Perturbing the medium creates a multitude of channel states, each with a different set of transmission paths. The transmitter uses the incoming data (to be transmitted) as an index to select a particular channel state for each transmission. By contrast, in a traditional source-based wireless system, data is embedded in the
variations (e.g., amplitude, phase or frequency) of the radio-frequency (RF) source prior to the transmit antenna, and the wave propagates via random but unaltered paths (media) to the destination.

%In multi-path Rayleigh fading, the received signal points projected over the receive dimensions are modeled as independent, identically distributed (i.i.d.) Gaussian vectors.

%This can be realized if each channel state corresponds to a unique and selectable set of transmission paths. 
%Assuming the set of transmission paths corresponding to different channel states are independent of each other, for a static multi-path channel,  the  states correspond to independent realizations  of the underlying channel gains, which are known to follow a complex Gaussian distribution.  The transmitter will then send a signal through the selected channel state in order to convey the particular selection of the channel state to the receiver. 

Fig. \ref{SIMO-MBM} shows an MBM transceiver diagram (reproduced based on \cite{8758978}). A single transmit antenna is placed within a closure surrounded by walls. Each wall can be switched to operate in one of two states: a transparent state, and a reflecting state. The transmit antenna emits a single pulse-shaped tone. A transparent wall would pass the incident wave to the outside, and a reflecting wall would send it back to the interior of the RF closure. As a result, the transmitted wave bounces back and forth within the RF closure and, in the process, propagates outside from transparent walls. An RF closure with $\log_2M$ switchable walls creates a set of $M$ states for the end-to-end channel. MBM transmitter selects one of these states in each transmission according to incoming data and thereby embeds $R_m=\log_2M$ information bits in selecting the channel state. Additional $R_s$ information bits can be transmitted by modulating the RF signal using a traditional source-based modulator (Fig. \ref{SIMO-MBM}).

In the absence of noise, the received signals (or equivalently the complex fading gains) act as unique signatures for each selected channel state. Let $\mathcal{H} := \{{\rv{h}}_0, {\rv{h}}_1, ..., {\rv{h}}_{M-1}\}$ denote the set of fading gains corresponding to different channel states. Careful design and placement of reflecting walls warrants totally different transmit RF patterns in different states, which, upon propagation in the multi-path (rich scattering) environment, result in the random vectors in the set $\mathcal{H}$ to be mutually independent. 

%Assuming transmission paths for different channel states are independent of each other and considering a rich scattering environment, the channel gains are known to follow independent, identically distributed (i.i.d.) Gaussian distribution. (Independent assumption is reasonable due to random interaction of carefully designed RF mirrors.) 

% Let $\mathcal{H} := \{{\rv{h}}_0, {\rv{h}}_1, ..., {\rv{h}}_{M-1}\}$ denote the set of all channel gains due to different channel states.

A fundamental property of MBM, which makes it distinct from MIMO systems, is that in a system with $K$ receive antennas, MBM signals formed over the receive spatial dimensions span the entire $K$-dimensional vector space. That is to say, with high probability, $span(\mathcal{H}) = K$. In particular, this is true even using a single transmit RF chain and a single transmit antenna. This property allows MBM to achieve full multiplexing gain, i.e., $\mathsf{MG} = K$, with a single transmit antenna (c.f. appendix \ref{appendix:uncoded_diversity}). In a conventional MIMO system, however, the effective number of dimensions is governed by the {\em minimum} of the number of transmit and receive antennas. In particular, source-based single-input multiple-output (SIMO) spans a single complex dimension. From an information theoretic standpoint, this property of MBM is analogous to \say{additivity of information over multiple receive antennas}. Reference \cite{isit2014} (also see \cite{9834835}) shows that a $1 \times K$ MBM over a static Rayleigh fading channel asymptotically achieves the capacity of $K$ parallel complex AWGN channels, where for each unit of transmit energy, the effective energy for each of the $K$ AWGN channels is the statistical average of channel fading. Alternative proof for this feature is presented in \cite{9834835}.

The second key property of MBM is that multiple channel states collectively contribute to enlarging the mutual distance among constellation points. In other words, the constellation signal set is constructed by both good channel states (high channel gains) as well as bad channel states (low channel gains). As a result, the deep-fade bottleneck in the case of legacy source-based modulation (SBM) is avoided. In other words, deep fades result in constellation points closer to the origin. Points closer to the origin are as useful as points further away from the origin (higher fading gains) in filling the constellation signal space uniformly. In a Rayleigh fading channel, constellation points along each spatial complex coordinate follow a Gaussian distribution, which agrees with Shannon's random code-book construction. 
%In the language of information theory, the outage probability (defined as the probability that the mutual information becomes smaller than a target rate) becomes much smaller than signal-based modulation (c.f. Section \ref{sec:outage_comparison} for numerical comparisons). 
Article \cite{9834835} shows that the outage probability due to deep fades is (asymptotically) alleviated as $M$ (number of switchable RF reflectors) increases. Specifically, in a static Rayleigh fading channel with AWGN at receive antennas, the mutual information $I$ in a $1\times K$ MBM is normally concentrated around $K\log(1+\mathsf{SNR})$, that is, 
\begin{IEEEeqnarray}{c}
I \sim \mathcal{N}(K\log(1+\mathsf{SNR}), \sigma^2),
\end{IEEEeqnarray}
where $\mathsf{SNR}$ denotes the signal-to-noise ratio at each receive antenna. As $M$ increases, the variance, and consequently the outage probability, goes to zero according to
\begin{IEEEeqnarray}{c}
\sigma^2 = \frac{K}{M} \left(\frac{\mathsf{SNR}}{1+\mathsf{SNR}}\right)^2.
\end{IEEEeqnarray}
A similar result is derived in \cite{isit2014}, using a different approach in which the loss in rate (vs. a set of parallel AWGN channels) is captured in terms of an additive noise with a vanishing variance. Note that, unlike traditional MIMO systems, diversity in MBM does not come at the cost of a reduction in rate (MG).

Another property of an MBM system with multiple receive antennas is the \say{$K$ times energy harvesting}. This means, assuming a single transmit and $K$ receive antennas and fading with an average statistical gain of one, the average received signal energy will be $K$ times the transmit energy. Legacy $K \times K$ MIMO enjoys a similar property. However, when the channel matrix is non-orthogonal, MIMO can not fully realize the capacity of a set of $K$ parallel channels with independent noise components. Receiver processing techniques, such as channel inversion, can create a set of parallel channels. However, the non-orthogonality of the channel matrix results in statistical dependencies among resulting noise components. Transmitter processing techniques such as eigen beam-forming also diagonalize the channel matrix. However, the underlying issue will surface in another form; it results in different channel gains along different eigen-dimensions. Consequently, in contrast to MBM, which achieves the capacity of a set of parallel channels with equal gains and independent noise components (of equal power), MIMO does not achieve such an upper limit on the achievable rate if the channel matrix is non-orthogonal. Next, we provide a numerical example to elaborate on the above MBM properties.

\begin{table}[]
\begin{center}
\caption{Required transmit energy to achieve $\mathsf{SNR} = 20$ dB at the receiver for the given outage probability with 256QAM over a Rayleigh fading channel.}
\begin{tabularx}{1\columnwidth} { 
  | >{\centering\arraybackslash}X
  | >{\centering\arraybackslash}X 
  | >{\centering\arraybackslash}X 
  | >{\centering\arraybackslash}X 
  | >{\centering\arraybackslash}X 
  | >{\centering\arraybackslash}X
  | >{\centering\arraybackslash}X  | }
  \hline
  Outage probability & Required transmit energy \\
  \hline
   0.1 & 30 dB \\
  \hline
   0.01& 40 dB \\
  \hline
  0.001 & 50 dB\\
  \hline
\end{tabularx}
\label{tab_outage}
\end{center}
\end{table}

\begin{figure}[t]
\centering
%\hspace{-2mm}
%\includegraphics[width = 14.4cm, height = 10.4cm]{256.pdf}
\includegraphics[width = \figsizea]{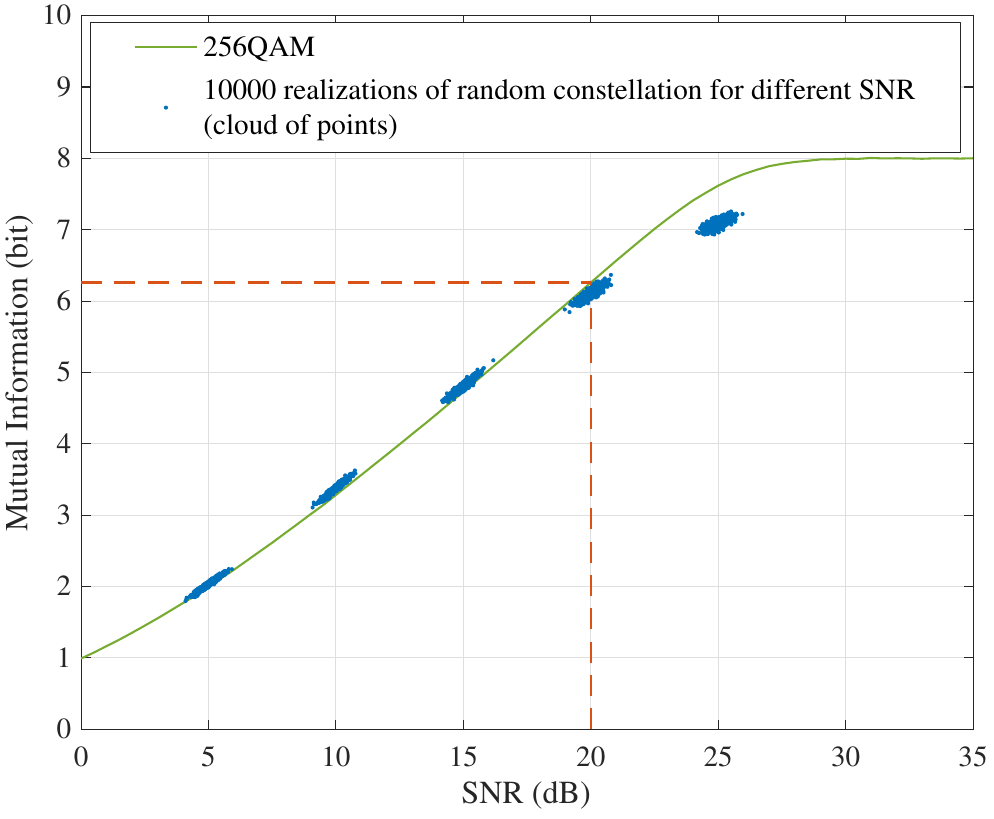}
\caption{Achievable rate (mutual information) of 256QAM vs. random constellation set with 256 points (2-dimensional points). Components of random constellation points follow Gaussian distribution. At each SNR, the cloud of points show  values of mutual information for 10000 realizations of the random constellation.}
\label{FigNew1}
\end{figure}

\begin{figure}[t]
\centering
%\hspace{-2mm}
\includegraphics[width = \figsizea]{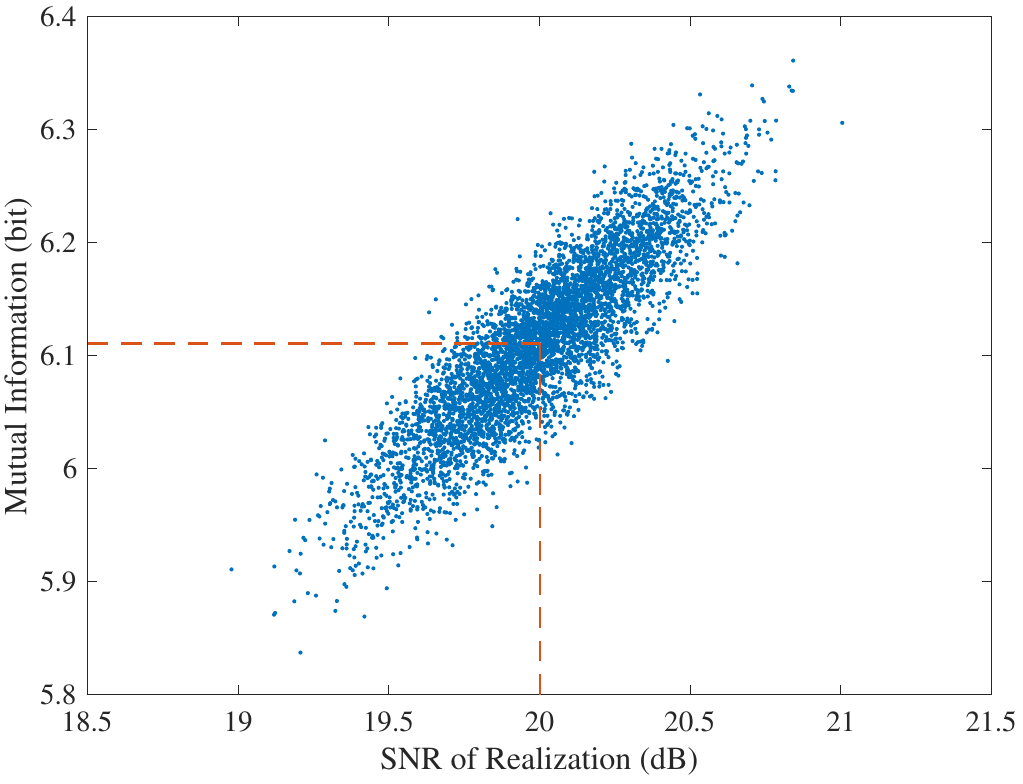}
\caption{Achievable rate (mutual information) for 10000 realizations of random 2-dimensional constellation with 256 points. Transmit energy is equal to 20 dB.}
\label{fig_scatter}
\end{figure}

\vspace{0.3cm}
{\bf Example:} Consider a wireless system with a single transmit and a single receive antenna operating over a static multi-path (static Rayleigh fading) channel with AWGN. The channel has 256 states, each resulting in a constellation point with independent, identically distributed (i.i.d.) Gaussian components over the single complex receive dimension. Each realization of such 256 points achieves a rate equal to the mutual information across the AWGN channel. Fig. \ref{FigNew1} shows achievable rate values for 10000 realizations of a random constellation and the achievable rate of a 256QAM (quadrature amplitude modulation) constellation. We assume AWGN and uniform probability for constellation points in all cases. The distribution of achievable rate values deviates only slightly from the 256QAM rate. Hence, the outage probability for the random constellation becomes negligible at the cost of a small SNR margin. For example, the required margin is $\sim$1dB to guarantee (with a negligible outage probability in the rang of $10^{-4}$) an SNR of 20 dB at the receiver (see Fig.~\ref{fig_scatter} for realized SNR and rates of random constellation). This significantly outperforms the outage behavior of a 256QAM over a static Raleigh fading channel. For instance, Table \ref{tab_outage} shows that 256QAM in a Rayleigh fading channel requires a transmit energy margin of 10, 20, and 30 dB to obtain an SNR of 20 dB at the receiver, with outage probabilities equal to 0.1, 0.01, and 0.001, respectively. Furthermore, as the constellation set size increases, the random constellation realizes the shaping gain (due to the Gaussian distribution of points), which contrasts a QAM constellation with points occurring with uniform probabilities.  $\blacksquare$

\subsection{Literature Survey} 

The idea of embedding information in the state of a communications channel is not new. Mach–Zehnder modulators, widely used for signaling over fiber, modify the light beam after leaving the laser. However, due to the lack of multi-path transmission in single-mode fibers, the advantages realized in the context of wireless do not apply.

Reference ~\cite{c0} coined the term media-based modulation for embedding data in the intentional variations of the transmission media (end-to-end channel) and showed that MBM offers considerable improvement vs. traditional single-input single-output (SISO), single-input multiple-output (SIMO), and multiple-input multiple-output (MIMO) wireless systems. In distinction to MBM, traditional modulation schemes, where data is embedded in the variations of an RF source (for example, in amplitude, phase, or frequency) and propagates via a conventional wireless channel to the destination, are called source-based modulation.  

Following~\cite{c0}, reference \cite{isit2014} (also see~\cite{9834835})
proves that a $1 \times K$ MBM over a
static multi-path channel asymptotically achieves the capacity
of $K$ (complex) AWGN channels, where for each unit of energy
over the single transmit antenna, the effective energy for each
of the $K$ AWGN channels is the statistical average of the channel
fading. It is shown that significant gains can be realized even in a SISO-MBM setup. An example of the practical realization of the system using RF mirrors, accompanied by realistic RF and ray tracing simulations, is presented \cite{lmimo_mbm}\cite{7511273}\cite{isit2014}.

References \cite{new-ak1}\cite{new-ak2} present a method for establishing an unconditionally secure encryption key between two wireless nodes relying on a media-based antenna structure. Different states of the media-based antenna are exploited to measure a set of reciprocal phase values, called masking phase values, between the two nodes. Since each masking phase is uniformly distributed in $[0,2\pi]$, it allows hiding $b$ bits of information within a $2^b$-PSK (phase shift-keying) constellation (by the addition of phase values modulo $2\pi$). A sequence of bits, upon applying forward error correction (FEC), are mapped to a sequence of such rotated PSK constellations (each constellation is masked by the addition of an independent, uniformly distributed phase). Then, the sequence of hidden bits are securely and reliably transmitted over the channel. These hidden bits can be the message itself, or form a key to be subsequently used in conjunction with a conventional encryption algorithm, e.g., advanced encryption standard (AES). 

%There have been some recent works on embedding data in antenna beam-patterns \cite{c1, c2, c3}. Note that, unlike MBM, these works do not fully realize potential advantages due to embedding information in the channel state. Most notably, these advantages, reported for the first time in~\cite{c0, isit2014}, include \say {additivity of information over multiple receive antennas} and \say{inherent diversity without sacrificing transmission rate}.

Authors in \cite{c1, c2} study embedding data in antenna beam patterns, where data is embedded in two orthogonal beam patterns to transmit a binary signal set. Although the use of an orthogonal basis is common in various formulations involving communications systems, it usually does not bring any explicit performance benefits; it merely simplifies the problem formulation, particularly constellation design and detection by keeping the noise projections uncorrelated (independent in signaling over AWGN). The motivation discussed in \cite{c1, c2} has been to reduce the number of transmit chains.

The use of tunable parasitic elements external to the antenna(s) for RF beam-forming is also well established. However, the objective in traditional RF beam-forming is \say{to focus/steer} the energy beam, which does not realize the advantages of MBM (where data is modulated by modifying the RF characteristics of the external parasitic elements). Bains~\cite{c3} discusses using parasitic elements for data modulation and shows energy saving due to the effect of classical RF beam-forming. 

MBM falls within the more general category of \say{index modulation} \cite{Index}, which includes a number of different approaches \cite{N4, N5, N6, N7, N8, N9} for modulating the RF carrier in the spatial domain. In the following, a subset of the earlier works that have played a pioneering role in the area of index modulation is discussed. Due to space limitations and the sheer volume of publications on the topic, readers are referred to \cite{Index, c1, c2, c3, c4-new, N3, N4, N5, N6, N7, N8, N9, N10, N6n, N6nn, N6nnn}, and references therein, for more details. 

%In particular, spatial modulation and its generalizations rely on multiple on/off transmit antennas, wherein part of the data selects one of the transmit antennas (to be on), and the rest of the data modulates the carrier to be transmitted through the selected  antenna. As a result, the number of data bits embedded in the channel state will be equal to” $\log_2$ of the number of on/off antennas. 
 
Spatial modulation (SM)~\cite{N4, N5, N6} uses multiple transmit antennas with a single RF chain, where a single transmit antenna is selected according to the input data (the rest of the data modulates the signal transmitted through the selected antenna). SM is, in essence, a diagonal space-time code, where the trade-off between diversity and multiplexing gain has favored the latter ($\mathsf{MG} = 1$, which is the minimum possible value). As a result, by relying on a single transmit chain, the hardware complexity of SM reduces, but its rate due to the spatial portion increases with $\log_2$ of the number of antennas. A primary difference between MBM and SM is that, in MBM, the rate is scaled linearly with the number of on/off RF mirrors (parasitic RF elements with two states, acting as reflector or as transparent, respectively) vs. logarithmic scaling in the case of SM. The reason is, while MBM relies on a single RF chain too, due to the interaction among RF mirrors, each on/off configuration results in a different transmit antenna pattern. Due to this exponential growth, the number of data bits embedded in the channel state equals the number of on/off mirrors.

In continuation to SM~\cite{N4, N5, N6}, space-shift
keying (SSK)~\cite{N7, N8, N9} and generalized SSK (GSSK) \cite{N10} have been studied for low-complexity implementation of MIMO systems. Again, the key motivation behind the application SM/SSK/GSSK in~\cite{N4, N5, N6, N7, N8, N9, N10} is the use of a single RF chain, and accordingly, one antenna remains active during data transmission. Other advantages include avoiding inter-antenna synchronization and removing inter-channel interference \cite{N9}. In addition to complexity considerations, it is shown that these modulation schemes offer better error
performance as compared to conventional MIMO techniques~\cite{N7}. 

Quadrature spatial modulation (QSM) is proposed in \cite{qsm} to enhance the throughput of SM by creating a new spatial dimension. The input data select two indices corresponding to two transmit antennas. One of the selected antennas transmits the in-phase part of the modulated RF carrier, and the quadrature component of the RF carrier is transmitted from the other selected antenna. QSM allows the transmission of an additional base two logarithms of the number of transmit antennas with respect to ordinary SM. This comes at the expense of synchronizing the transmit antennas.

%Reference \cite{stcm} has introduced space-time channel modulation (STCM) by extending the classical space-time block codes to the channel state dimension using the principle of MBM. STCM increases spectral efficiency and, in addition to that, obtains transmitting diversity by applying a block code along the time and the MBM spatial dimension. 

More recently, a promising research direction, based on using an  \say{intelligent reflecting surface (IRS)} to aid in transmitting a wireless signal, has been introduced in~\cite{new-ak0} (also see \cite{new-ak3} and \cite{new-ak4}). IRS, similar to media-based modulation and spatial modulation, is based on modifying characteristics of a radio frequency signal after it leaves its respective transmit antenna.

%Due to differences in the targeted ranges of rate per channel use, a comparison between MBM and all the schemes mentioned above has not been possible.  Some numerical comparisons between MBM vs. SM (and Generalized Spatial Modulation, GSM) in terms of energy/rate efficiency are provided in later parts of this article (see Tables \ref{tab1} and \ref{tab2}). More detailed comparisons are provided in \cite{N11}. In general, MBM is particularly suitable for transmitting very high data rates per channel use, while most of the works reported in the context of SM/GSM have focused on lower rates. Due to this reason, the scope of comparisons has remained limited by including a subset of articles on SM/GSM that have presented results for medium values of rate per channel use. This limitation has worked to the disadvantage of MBM in these comparisons. In other words, the relative advantages of MBM vs. SM/GSM would be  more pronounced (vs. the comparisons provided in Tables \ref{tab1} and \ref{tab2}, or presented in \cite{N11}), if the rate values were higher. 

%Section \ref{sec: Detection Algorithm} presents a low-complexity successive cancellation list decoder for symbol recovery in LMBM setup. Using this algorithm, we can simulate the performance of MBM for rates as high as 32 bits per complex channel use.
\subsection{Article Arrangement}
The rest of the article is organized as follows. First, the system model for SIMO-MBM is described in Section \ref{sec : System Model}. In Section \ref{sec:lmbm}, we discuss the practical issues raised using MBM with a single transmit unit, i.e., single-input multiple-output MBM (SIMO-MBM). Subsequently, we put forward layered MBM (LMBM) architecture and provide the details on how this configuration addresses the complexities in SIMO-MBM setup with minimal performance degradation. Section \ref{sec : numerical_performance} provides several numerical results, including coded performance and comparisons with legacy and other index modulation techniques. A desirable property of MBM is that one can obtain an increase in the \emph{diversity order} by applying FEC; Section \ref{sec : diversity_gain} shows by applying a FEC with error correction capability $t$, the slope of the diversity order increases by a factor of $t+1$. Section \ref{sec : Pulse Shaping} looks into the bandwidth expansion issue in MBM due to its time-varying nature and discusses a time-limited pulse design that mitigates bandwidth expansion. Finally, Section \ref{RF-sec} studies the practical design for the RF structure of MBM. A new design is presented where the transmit antenna efficiency is improved by adding top and bottom
caps to the antenna cylindrical closure. Furthermore, the RF structure is equipped  with multiple receive antennas.

\section{Single-Input Multiple-Output MBM (SIMO-MBM) System Model \label{sec : System Model}}

Fig. ~\ref{SIMO-MBM} shows the diagram of a $ 1\times K $ SIMO-MBM system with $\log_2 M$  RF mirrors. The transmitter selects a different on/off pattern for the RF mirrors according to message index $m \in \{0, ..., M-1\}$. As a result, a unique fading gain $\rv{h}_m$ will be realized at the receiver. Coordinates of complex vector $\rv{h}_m$ are the fading projections over individual receive antennas. %$\mathbb{E} {\mid\,h_{m, k}\,\mid }^2 = 1$, where $\mathbb{E}$ denotes statistical averaging.
The set of fading gains $\mathcal{H} = \{{\rv{h}}_0, {\rv{h}}_1, ..., {\rv{h}}_{M-1}\}$ due to all possible on/off combinations of RF mirrors constitute \say{MBM constellation points}.
In a Rayleigh fading channel, fading gains $\rv{h}_m$ are modeled as i.i.d. complex Gaussian random vectors. 
%AWGN $\rv{z}$ has independent, identically distributed (i.i.d.) components $z_k, \ k = 1, ..., K$, $ \mathbb{E}{\mid z_k \mid} ^2 = N_0$. 
In the absence of FEC codes, the transmission rate due to MBM is equal to $R_m = \log_{2}{M}$ bits per channel use.

The transmitter does not know the set of fading values (i.e., no channel state information at the transmitter is assumed). The receiver, however, knows fading gains $\mathcal{H}$ and also the one-to-one correspondence between the set of message indices and the fading gains. Receiver training is achieved by selecting each possible RF mirror on/off pattern at the transmitter and sending a signal (unmodulated carrier with possible pulse shaping) for the receiver to measure the associated MBM constellation point. %The order of scanning through channel states in the training phase is prearranged such that the assignment of channel states to data labels is known to both sides. 
For example, the channel state indexed by data $m$ is selected during the $m$th training period to covey $\rv{h}_m$ to the receiver. 

Additional information corresponding to the SBM message may be transmitted by directly modulating the RF signal. Using a linear modulation (for example, amplitude shift keying), the SBM message is communicated as a complex number $s$ multiplying the RF carrier before the transmit antenna. The projected signal at the receiver (ignoring AWGN) will be equal to $s\rv{h}_m$. Here, both complex number $s$ and index $m$ carry information. For brevity, in what follows, we do not consider transmitting additional SBM messages.

%Some complexity issues arise when only a single RF transmit unit is used to realize the advantages of MBM at high data rates. The next section discusses the underlying practical issues and presents methods to overcome MBM shortcomings (in terms of complexity) by relying on a layered structure. Note that, although layered MBM may appear to be relying on multiple transmit antennas, and accordingly it is called Layered Multiple Input-Multiple Output Media-Based Modulation (LMBM),  it still uses a signal transmit chain. LMBM is indeed an approximation to SIMO-MBM, aiming to reduce its algorithmic/training complexities at the cost of some degradation in performance. 

\begin{figure*}[tbp]
\centering
%\hspace{-2mm}
\includegraphics[scale = 0.65]{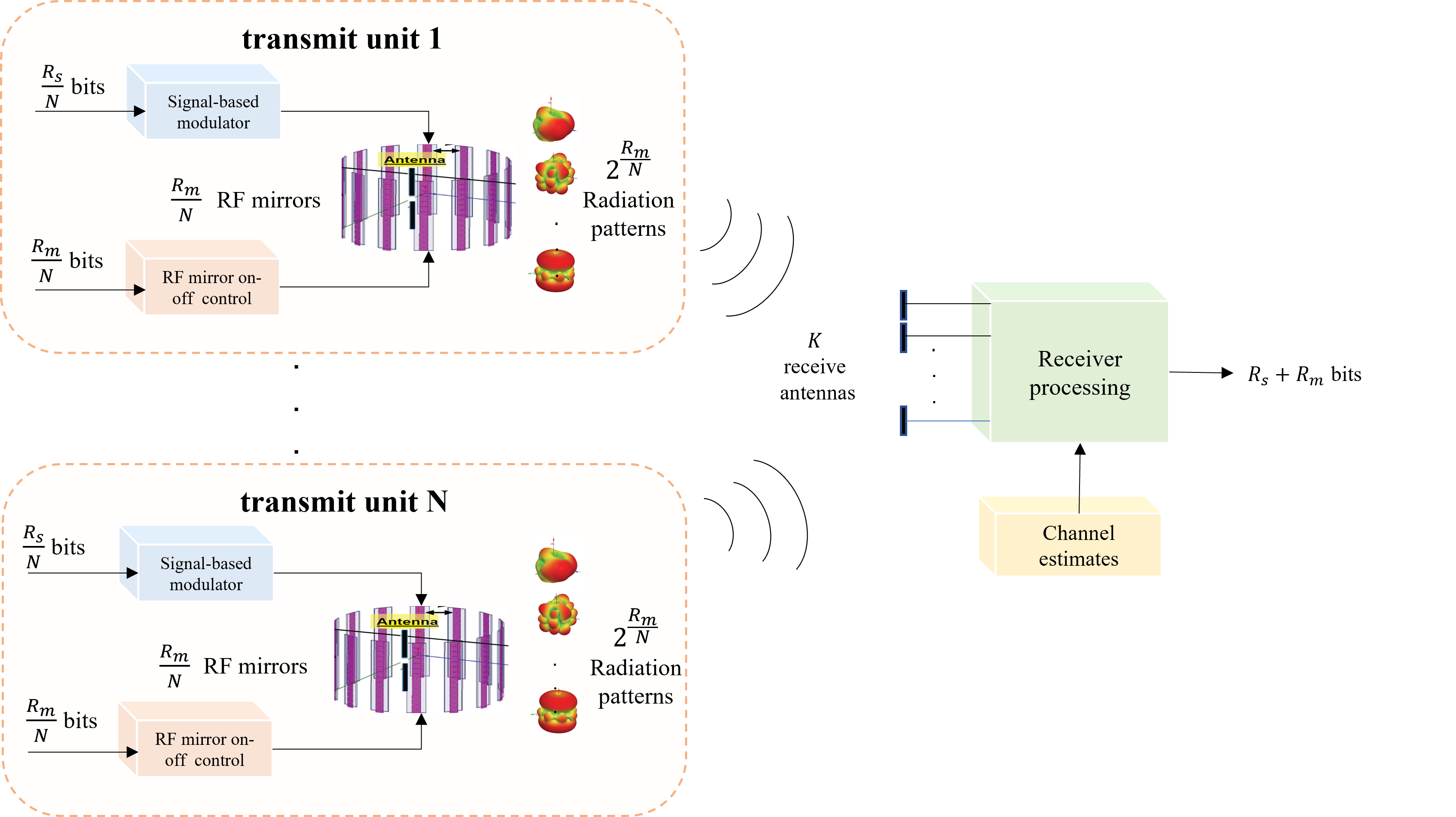}
\caption{LMBM transceiver diagram.}
\label{MIMO-MBM}
\end{figure*}

\section{Layered MBM \label{sec:lmbm}}
%Decoding, training, and physical RF implementation of the MBM system using a single transmit unit become prohibitive when targeting high data rates. This section discusses the issues in detail and puts forward the Layered structure to address them. 
Due to the random nature of constellation points, MBM does not inherit regularity in the constellation's structure (e.g., compared to the conventional 256QAM constellation). Accordingly, some practical complexities, such as receiver training and decoding, arise that are specific to MBM targeting high data rates. In the following, we outline these complexities and subsequently discuss the layered MBM (LMBM) structure which uses multiple transmit units to reduce these complexities significantly.

\begin{enumerate}

\item[A1:] Since MBM constellation points are random, the receiver needs to learn all possible channel states corresponding to different configurations of RF mirrors. As such, receiver training becomes prohibitive for a large set and vulnerable to channel aging.

\item[A2:] To increase the time interval between successive receiver training phases while dealing with channel variations in time, it is desired to track the changes in the position of the constellation points. It is challenging to track $2^{R_m}$ constellation points for large $R_m$.

\item[A3:] Decoding requires computing Euclidean distances to all points in the constellation set, which becomes prohibitive for a large set. 

\item[A4:] It is difficult to efficiently embed many RF mirrors in a single MBM unit.  

\end{enumerate}

%For example, let us assume we are interested in transmitting $32$ bits of information per channel use. The complexity in using a single RF transmit unit to encode all $32$ bits may be excessive. The reasons are:

%\begin{enumerate}
%\item[L1:] It is practically difficult to use $32$ RF mirrors in a single RF transmit unit. 

%\item[L2:]  Training requires transmitting $2 ^ {32}$ test signals, which is resource intensive and is vulnerable to channel time variations (channel aging).

%\item[L3:] Detection requires searching (minimum distance decoding) among $2 ^ {32}$ signal points, resulting in excessive algorithmic and storage complexities.

%\item[L4:]  To deal with channel time variations, it is of interest to track the changes in the position of the constellation points in order to increase the minimum time interval between successive training phases. It is difficult to track $2 ^ {32}$ constellation points.

%\end{enumerate}

%\section{Layered MBM (LMBM) \label{sec:lmbm}}
	
Fig. \ref{MIMO-MBM} shows a $ N\times K $ layered MBM system. The incoming data is demultiplexed into $ N $ streams transmitted over separate media-based transmit units. Each unit independently modifies its respective channel state according to its associated data. The generated signals are intended for a common receiver. The set $\mathcal{H}^0 := \{ \rv{h}_0^0, ..., \rv{h}_{M-1}^0\}$ denotes the fading gains from the first transmit unit to the receiver, likewise $\mathcal{H}^1 := \{ \rv{h}_0^1, ..., \rv{h}_{M-1}^1\}$ are the gains from the second unit to the receiver, and so on. Media-based modulator units are arranged such that there is a negligible coupling among them. As a result, the fading gains from each unit to the receiver (also called \say{constituent vectors}) will be independent of each other. The overall projected signal at the receiver will be the superposition of the constituent vectors due to individual units. More specifically, consider the message sequence $(m[0], ..., m[N-1])$ \footnote{We use square brackets to index elements of an ordered sequence, subscript to index elements of a set, and superscript to index transmit antenna units.}, where $m[n] \in \{0, ..., M-1\}$ corresponds to the random message sent over the $n$th transmit unit, i.e., the $n$th transmit unit selects its own on/off RF mirror configuration according to $m[n]$. Subsequently, a complex $K$ dimensional fading gain $\rv{h}^{n}_{m[n]}$ is realized between the $n$th transmit unit and the receiver. The received constellation point $\rv{c}$ is then formed as 
\begin{IEEEeqnarray}{c}
\label{eq:lmbm_constellaion}
\rv{c} = \sum_{n = 0} ^ {N-1} \rv{h}^{n}_{m[n]}.
\end{IEEEeqnarray}

%To emphasize this superposition property, which forms the basis behind the complexity reduction, the proposed approach is called Layered MIMO-MBM, or LMBM, hereafter. In summary, the overall receive vector will be the sum of the constituent vectors corresponding to different transmit units. 

Since the constituent vectors are random and independent of each other, the cardinality of the set of received constellation set will be equal to the product of the number of constituent vectors corresponding to different units. As a result, using $R_m/N$ RF mirrors at each of the $N$ MBM units creates $2^ {R_m}$ distinct vectors at the receiver, capable of transmitting $R_m$ bits of information per channel use. Like SIMO-MBM, each unit can send additional SBM data by modulating its RF signal. A total of $R_s$ additional bits due to source-based modulation is achieved by transmitting $R_s/N$ bits per unit.

%Here, vector component ${h}_{n, m_n, k}$ is the projected signal over the $k$th receive dimension due to message index $m_n$, transmitted by the $n$th modulator unit. Again, $\mathbb{E} {\mid {h}_{n, m_n, k} \mid}^2 = 1$. 

The following summarizes how the layered structure addresses the shortcomings of SIMO-MBM when transmitting high data rates.

\begin{enumerate}
\item[B1:] For the same transmission rate, the number of RF mirrors used at individual media-based modulator units reduces by a factor of $N$.

\item[B2:] Symbol recovery can be performed using a low-complexity successive cancellation decoder. At each step, the decoder searches for the constituent vector contributed by a single modulator unit and cancels its effect before proceeding to the next modulator unit. A list decoder can further improve the successive cancellation decoder. 

\item[B3:] Training is simplified as it is composed of $N$ separate training tasks, each over a smaller set of an alphabet size  $2^{R_m/N}$, as compared to training over a set with $2 ^ {R_m}$ alphabets.

\item[B4:]  Tracking is simplified as it is composed of $N$ separate tracking tasks, each over a smaller set of an alphabet size $2^{R_m/N}$, as compared to tracking the entire set of $2 ^ {R_m}$ alphabets. 
\end{enumerate}
For example, to send $32$ bits of data per channel, one can use $4$ media-based modulator units, each modulating $8$ bits, meaning only $8$ on/off RF mirrors are required in each unit.  
Training/tracking comprises $4$ separate tasks, each involving a much smaller alphabet size of $2^8 = 256$ elements.

\emph{Remark}: Unlike SIMO-MBM, LMBM no longer fulfills the independence requirement of Gaussian random coding over AWGN channels. In particular, in a Rayleigh fading channel, the constituent vectors $\rv{h}^{n}_{m[n]}$ follow an i.i.d. complex Gaussian distribution. However, the received constellation points $\rv{c}$ are no longer statistically independent. As such, the performance of LMBM may be inferior to SIMO-MBM. However, numerical results show that the degradation in {SNR} performance is modest. For example, Fig. \ref{fig:SER_8Nr_16R}  shows an example of the gap in performance of MBM vs. LMBM, which is less than 0.5 dB when transmitting 16 bits per complex channel use.

\subsection {Low-complexity Decoder for Layered MBM Scheme}
\label{sec: Detection Algorithm}

%In the presence of AWGN $\rv{z}$, for a received signal $\rv{y} = \sum_{n = 0} ^ {N-1} \rv{h}^n_{m[n]} + \rv{z}, $ a maximum likelihood decoder finds a message sequence $(\widehat{m}[0], ...,\widehat{m}[N-1])$ such that the point $\widehat{\rv{ c }} = \sum_{n = 0} ^ {N-1} \rv{h}^n_{\widehat{m}[n]}$ is at minimum Euclidean distance to $\rv{y}$. The complexity of the maximum likelihood decoder grows with the constellation size and becomes prohibitive for large sets. 

The layered structure of LMBM enables implementing a low-complexity successive cancellation list decoder that recovers individual constituent vectors $(\rv{h}^0_{m[0]}, ..., \rv{h}^{N-1}_{m[N-1]})$, leading to the elements of the message $(m[0], ..., m[N-1])$. 
At each step, the successive cancellation decoder makes a decision about the transmitted symbol corresponding to a single modulator unit, say $i$. The estimate of the constituent  vector $\widehat{\rv{h}}^{i}$ is then used to subtract the contribution of message ${m}[i]$ from the received signal and use the remainder to recover the rest of  messages. This process is continued until an estimate for all the elements of the message sequence $({m}[0], ..., m[N-1])$ is recovered.

The successive decoding algorithm details are explained in appendix \ref{appendix:scld}. Using this decoder and Monte-Carlo simulation, we can numerically measure the symbol error probability of LMBM for rates as high as 32 bits per channel use (see Fig. \ref{fig:SER_16Nr_32R}). 

%The decoder starts by computing an estimate of $\rv{h}^0_{m[0]}$, denoted by $\widehat{\rv{h}}^0 \in \mathcal{H}^0$, and subtracts this estimate from the received signal $\rv{y}$. Subsequently, an estimate of $\widehat{\rv{h}}^1$ is computed from $\rv{y} - \widehat{\rv{h}}^0$, and so on.
\section {Numerical Performance \label{sec : numerical_performance}
}
For numerical performance, we assume a static Rayleigh fading channel where the constituent vectors (fading gains) from each modulator unit to each receive antenna are generated as complex i.i.d. Gaussian random vectors. The performance is averaged over many independent runs of fading gains and AWGN noise. 
{\em Energy per bit}, $E_b$, is defined as the sum of total signal energies of all transmit units divided by the total number of bits per channel use, and $N_0$ denotes the AWGN spectral density at individual receive antennas. 
MBM performance reported in the following does not consider transmitting any additional source-based data, i.e., $R_s= 0$ (the total data rate is $R = R_m$). 

\subsection{Uncoded Performance}
\begin{figure}[tbp]
%\hspace{2mm}
\centerline{\includegraphics[width = \figsizea]{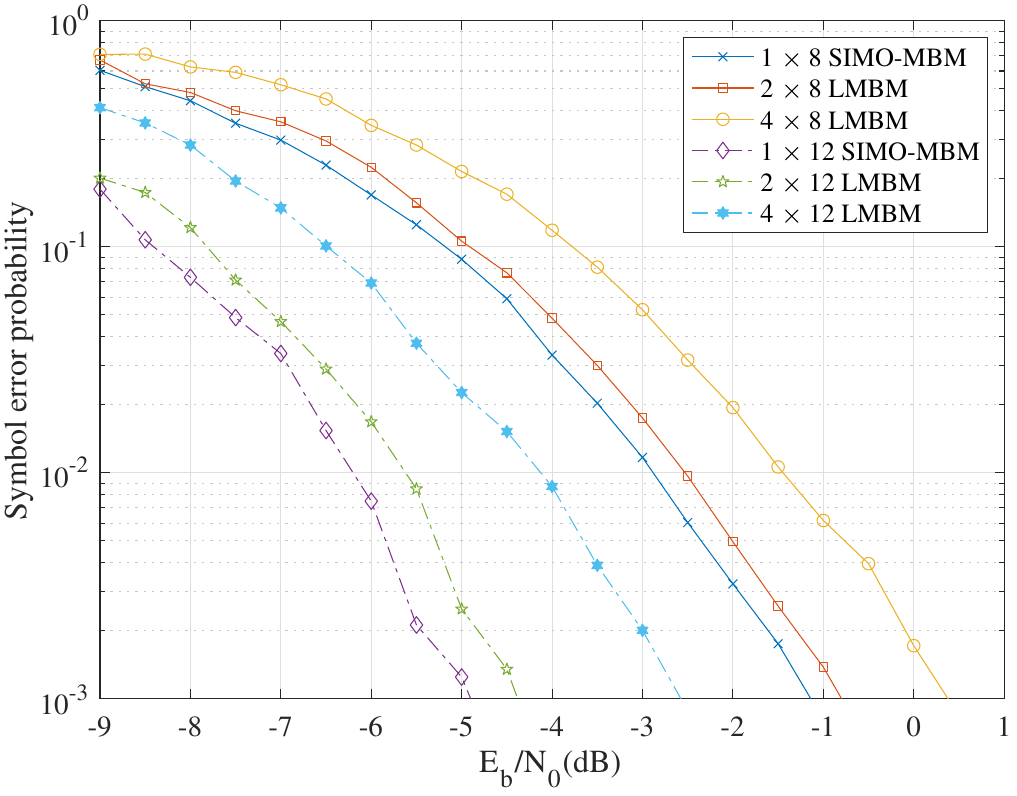}}
\caption{2$\times$8 and 2$\times$12 LMBM schemes rely on 8 RF mirrors in each of the 2 modulator units. 4$\times$8 and 4$\times$12 LMBM schemes rely on 4 RF mirrors in each of the 4 modulator units. SIMO-MBM uses 16 RF mirrors in a single MBM modulator unit. All schemes transmit 16 bits per channel use in a single transmission without any FEC.}
\label{fig:SER_8Nr_16R}
\end{figure}

\begin{figure}[tbp]
%\hspace{2mm}
\centerline{\includegraphics[width = \figsizea]{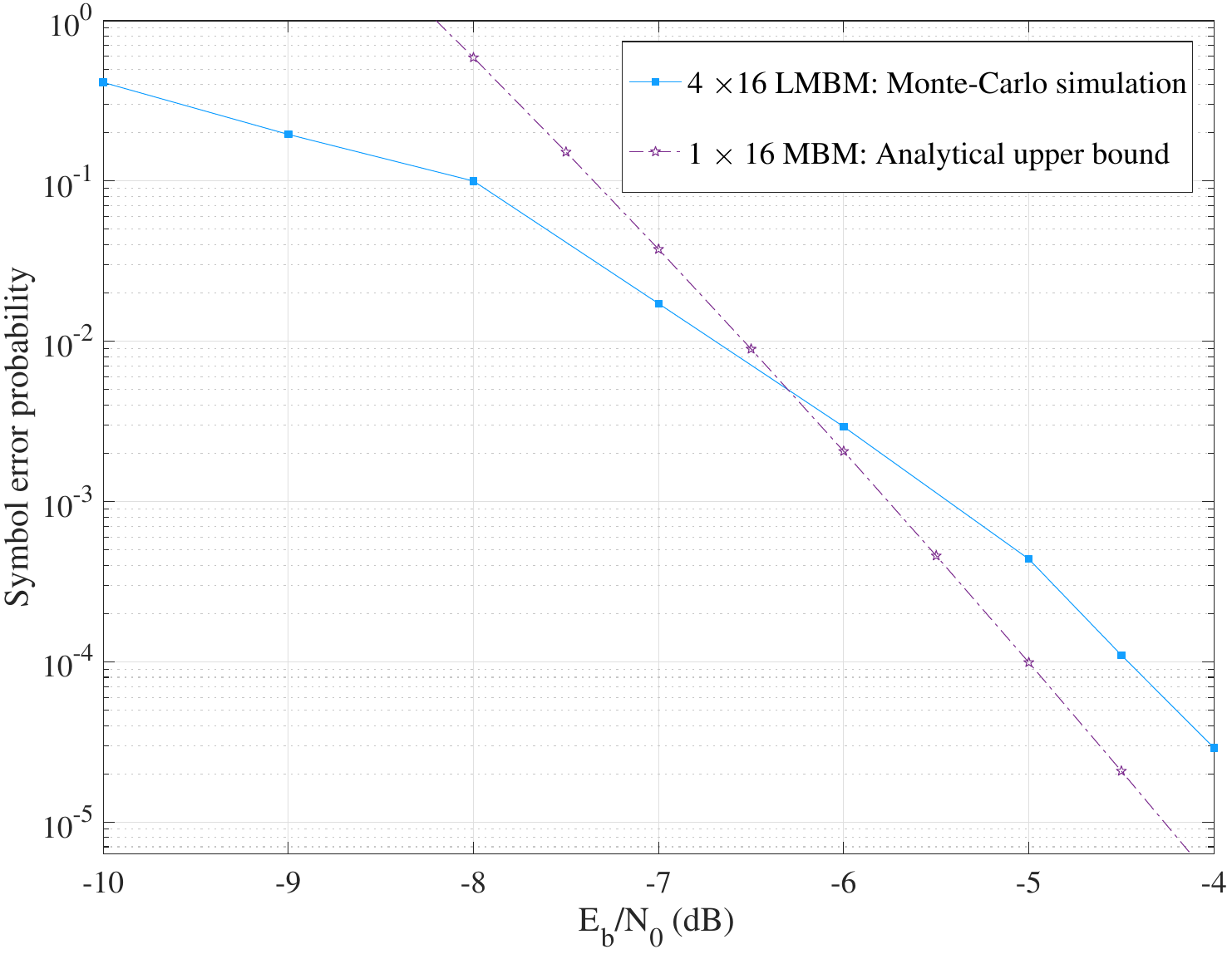}}
\caption{Analytical upper-bound on the symbol error probability for SIMO-MBM (see appendix \ref{appendix:uncoded_diversity}) with 32 RF mirrors in a single transmit unit vs. simulated LMBM performance (8 RF mirrors in each of the 4 transmit units). Both schemes transmit 32 bits in a single complex channel use (without any FEC) using 16 receive antennas. SIMO-MBM upper-bound is tight at high SNR regime, suggesting that the LMBM performance penalty vs. SIMO-MBM is negligible.}
\label{fig:SER_16Nr_32R}
\end{figure}

Fig. \ref{fig:SER_8Nr_16R} demonstrates the performances of SIMO-MBM and LMBM, transmitting $16$ bits per channel use. Fig. \ref{fig:SER_16Nr_32R} shows the performance of 4$\times$16 LMBM system, transmitting $32$ bits per channel use. Since maximum likelihood decoding becomes prohibitive for $2^{32}$ points in the constellation, the performance is obtained using the successive cancellation list decoder explained in appendix \ref{appendix:scld}. These performances are achieved via a single transmission without FEC.
% The decoding is carried out on $24$ independent branches each corresponding to a different ordering of the $4$ modulator units.  Each branch employs 2 iterations of the same successive decoding method on $128$ candidates. The candidate which has the closest projected constellation point to the received signal $\rv{r}$ in Euclidean space is chosen as the final decoded message.

\subsection{Performance Including FEC}
\begin{figure}[tbp]
\centering
{\includegraphics[width = \figsizea]{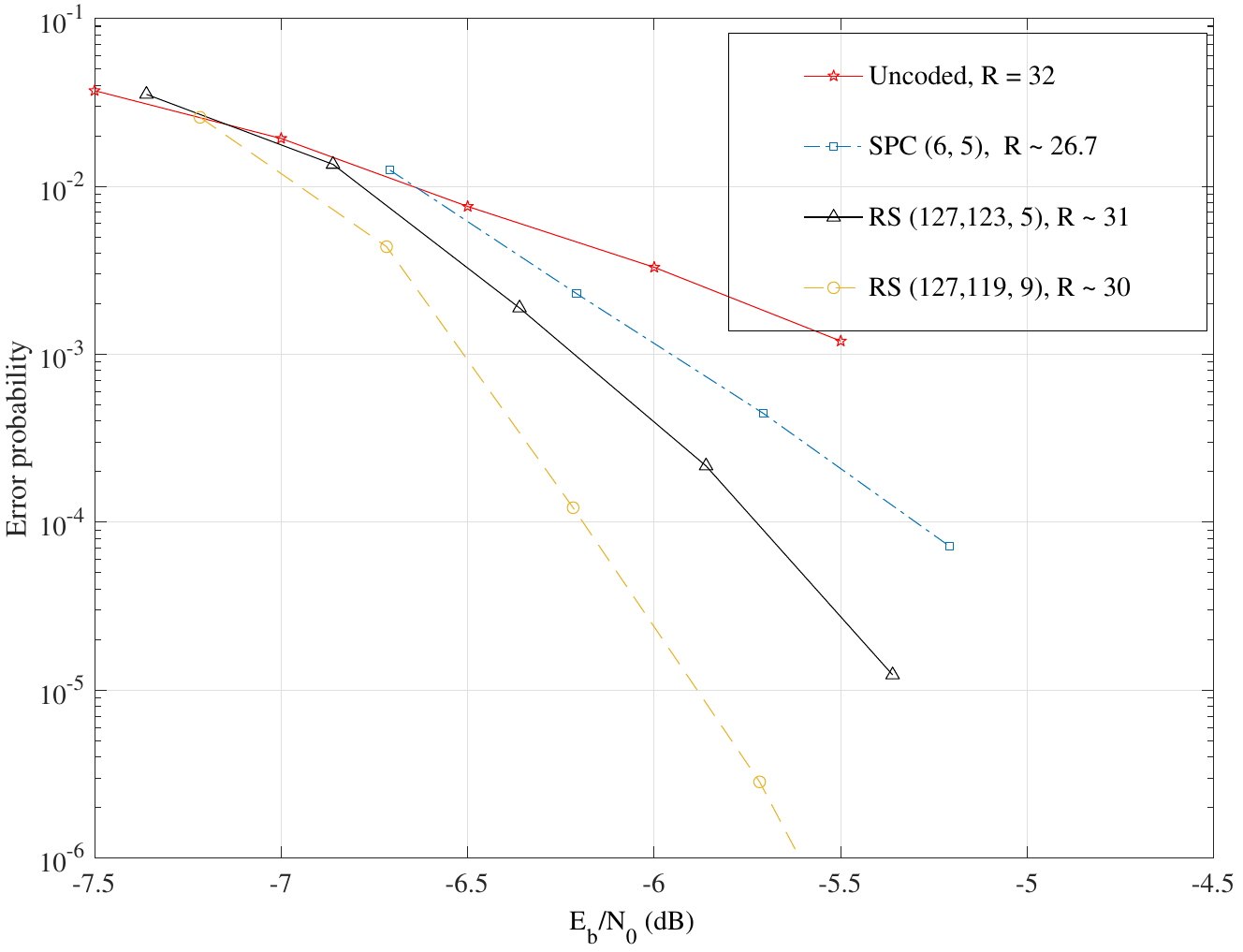}}
%{\includegraphics[trim = {3cm 6cm 0 7cm}, scale = 0.55]{mbm_coded1.pdf}}
\caption{Performance of 4$\times$16  LMBM, uncoded vs. coded using a single parity check \emph {SPC}$(N_c, K_c)$ code, and Reed-Solomon code \emph{RS}$(N_c, K_c, D)$. These channel coding schemes act upon MBM symbols. $R$ indicates the effective transmission rate in bits per channel use. Error probability corresponds to block error probability and symbol error probability for the coded and uncoded schemes, respectively.}
\label{reed_solomon}
\end{figure}

\iffalse
\begin{figure}[bt]
\centering
%\hspace*{-4cm}
{\includegraphics[width = \figsizea]{SEP_coded.pdf}}
\caption{Symbol error probabilities for 4$\times$16 LMBM uncoded and coded with Reed-Solomon codes acting on MBM symbols. $R$ indicates the effective transmission rate in bits per channel use.}
\label{fig:SEP_coded}
\end{figure}
\fi 

\begin{figure}[tbp]
\centering
%\hspace*{-4cm}
{\includegraphics[width = \figsizea]{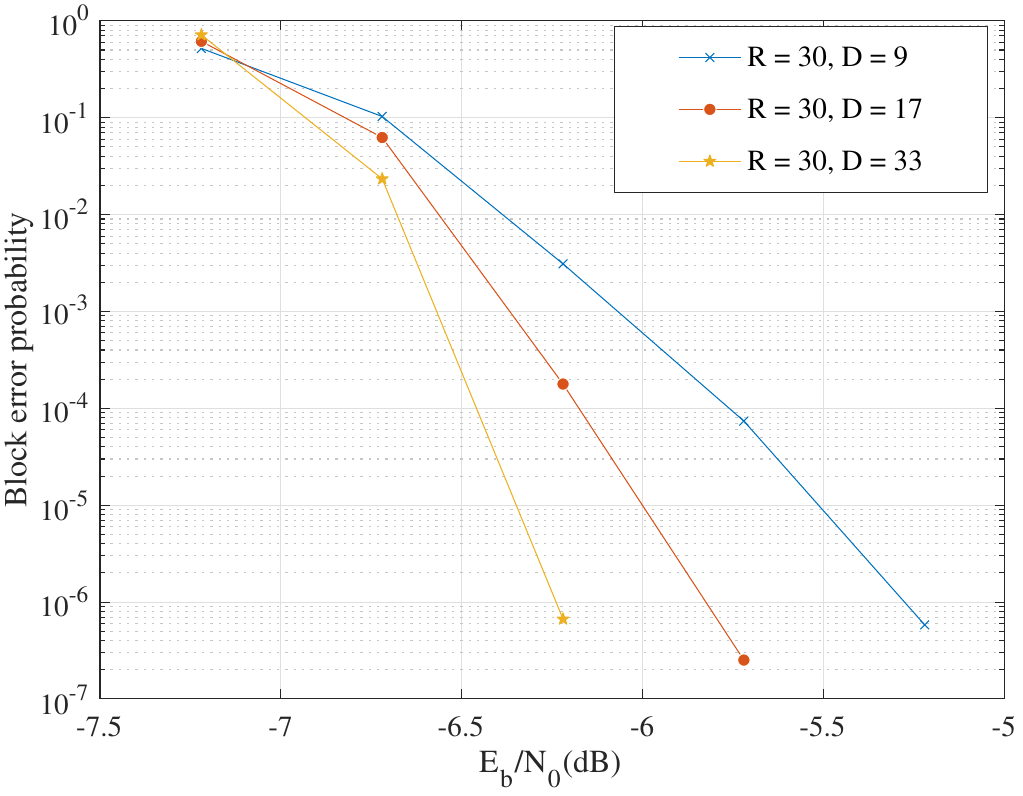}}
\caption{Block error probabilities for 4$\times$16 LMBM coded with Reed-Solomon codes acting on LMBM symbols. $R$ indicates the effective transmission rate in bits per channel use.}
\label{fig:BEP_coded}
\end{figure}
The application of FEC to MBM can be materialized using simple code structures operating on symbols (rather than bits). For MBM with $\log_2 M$ RF mirrors, the class of Group codes with alphabet size $M$ would be a natural choice. Reference~\cite{forney1} proves that in searching for good group codes, one can limit the search to those formed over elementary Abelian groups. Maximum distance separable (MDS) codes, including RS codes, are a subclass of such group codes for which the minimum distance $D$ has the maximum possible value satisfying the Singleton bound. Reed-Solomon codes with alphabet size $M$ and block length $M-1$ are well-known codes fulfilling MDS property.

%Reed-Solomon codes used in this article are obtained by puncturing such a larger code. 
An RS code of block size $N_c$ and dimension $K_c$ with minimum distance $D$, using a hard-decision decoder, can correct up to $t = \left \lfloor {(D-1)/{2}} \right \rfloor$ symbols in errors. Fig. \ref{reed_solomon} shows the simulated performance of the hard-decision decoder where each symbol of the RS code (i.e., Galois field elements) are mapped to LMBM constituent vectors. % Ehsan, this should be constellation point, not constituent vector, no? constituent vector is one of the pieces forming the constellation point
Fig. \ref{fig:BEP_coded} shows the block error probability performance of coded LMBM. Note the increase in the slope of error probability as the minimum distance $D$ (or equivalently error correction capability $t$) is increased. 

\subsection{Selection Gain}

\begin{figure}[tbp]
\centering
%\hspace{2cm}
\vspace{0cm}
%{\includegraphics[scale = 0.60]{selection_gain.pdf}}
{\includegraphics[width = \figsizea]{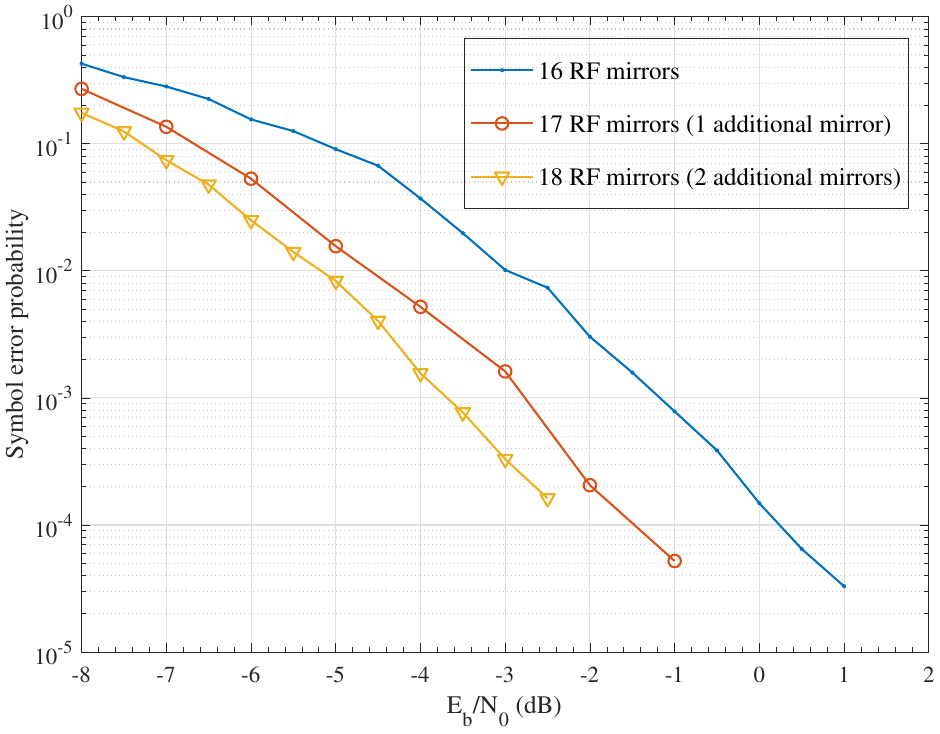}}
\caption{Improvement due to selection of a subset of constellation points with the highest energy. Symbol error probabilities are for SIMO-MBM achieving 16 bits per complex channel use with 8 receive antennas.}
\label{selection_gain}
\end{figure}

It is possible to obtain a pre-coding gain by selecting the subset of points in the MBM constellation set, maximizing a relevant figure of merit such as mutual information. We simply select a subset of constellation points with the highest fading gains to limit complexity. Subset selection requires providing the index of the on/off RF mirror configurations with the higher fading gain to the transmitter (i.e., a minimal channel state information should be provided to the transmitter). Removing the subset of points also means we need additional RF mirrors to maintain the same transmission rate. Fig. \ref{selection_gain} shows the pre-coding gain improvement due to subset selection when using one and two additional RF mirrors.

\subsection{Comparison with Spatial Modulation and its Variants}
\begin{figure}[tbp]
%\hspace{2mm}
\centerline{\includegraphics[width = \figsizea]{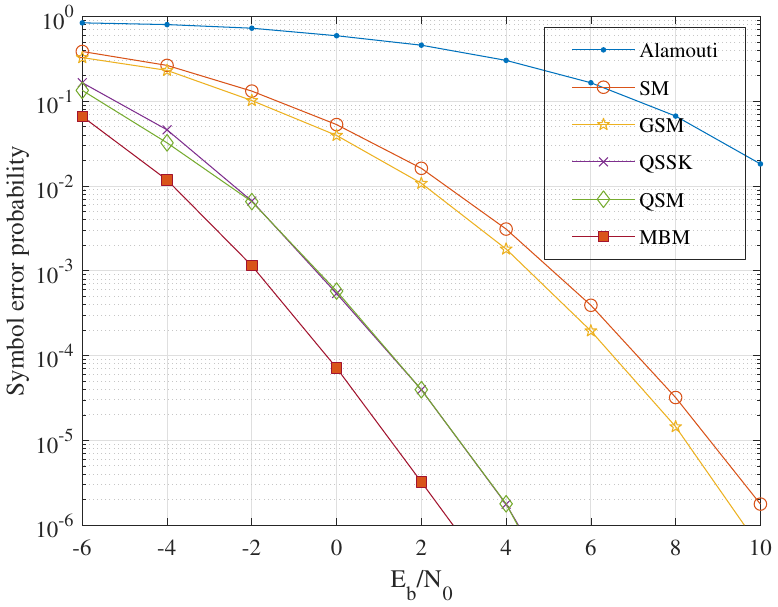}}
\caption{Symbol error probability comparison for schemes achieving transmission rate equal to 8 bits per complex channel use using 8 receive antennas.}
\label{fig:SER_8Nr_8R}
\end{figure}

\begin{figure}[tbp]
%\hspace{2mm}
\centerline{\includegraphics[width = \figsizea]{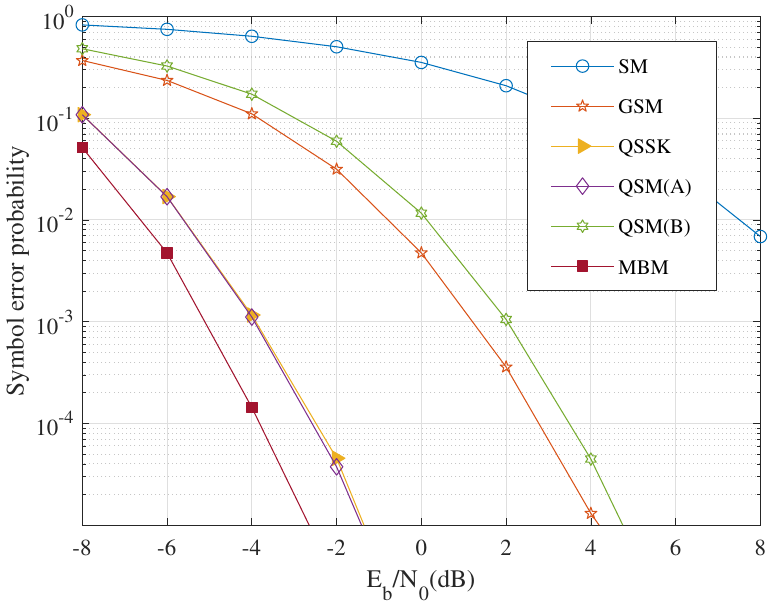}}
\caption{Symbol error probability comparison for schemes achieving transmission rate equal to 12 bits per complex channel use using 12 receive antennas.}
\label{fig:SER_12Nr_12R}
\end{figure}

Fig. \ref{fig:SER_8Nr_8R} and \ref{fig:SER_12Nr_12R} provide a comparison between MBM and other emerging modulation techniques (see also \cite{9817680}). Particularly, the symbol error probability of MBM 
averaged over independent realizations of  a static Rayleigh fading channel is compared to spatial modulation (SM), generalized spatial modulation (GSM), quadrature space shift keying (QSSK), and quadrature space modulation (QSM). Comparison is provided for rates of 8 bits per complex channel use and 12 bits per complex channel use.
The number of transmit and receive antennas, and quadrature amplitude modulation order for each scheme in Fig. \ref{fig:SER_8Nr_8R} and \ref{fig:SER_12Nr_12R} are provided in Table \ref{tab:tab1} and \ref{tab:tab2}, respectively. While other techniques rely on multiple antennas/RF chains at the transmitter as well as high modulation orders to achieve specified transmission rates, MBM performance is for a single transmit unit and in the absence of RF source modulation. 

%To further demonstrate the suitability of MBM for wireless transmission in high data rates, Fig. \ref{fig:SER_8Nr_16R} includes the performance of MBM for rate equal to 16 bits per complex channel use. Comparing to Fig. \ref{fig:SER_8Nr_8R}, we observe that the increase at the rate of MBM from 8 bits per complex channel use to 16 bits per complex channel use is achieved at the cost of a small (about 1dB) increase in $E_b/N_0$. Fig. \ref{fig:SER_16Nr_32R} demonstrate the MBM symbol error probability for rate 32 bits per complex channel use with 16 receive antennas. Layered MIMO-MBM (LMBM) introduced in \cite{lmimo_mbm} provides system architecture for practical (reduced-complexity) implementation of MBM, including reduced decoding complexity and receiver training overhead. Simulated error performance of LMBM in both Fig. \ref{fig:SER_8Nr_16R} and Fig. \ref{fig:SER_16Nr_32R} assert that such lower complexity architecture enjoys similar error performance and diversity order as SIMO-MBM.  

%For interested readers reference \cite{doi:https://doi.org/10.1002/9781119375692.app1} provides codes for simulation of various spatial modulation schemes. 

\begin{table}[tbp]
\renewcommand{\arraystretch}{1} % Default value: 1
\caption{System specifications for Fig. \ref{fig:SER_8Nr_8R}.}
\begin{center}
\begin{tabular}{c c c c c}
\hline \hline
Method & TX  & RX & QAM & Rate in\\
& antennas & antennas & modulation order & bits/s/Hz\\
\hline 
Alamouti & 2 & 8 & 64 & 8 \\ 
\hline
SM & 8 &  8 & 32 & 8 \\ 
\hline
GSM & 8 &  8 & 8 & 8 \\ 
\hline
QSSK & 16 & 8 & 0 & 8 \\ 
\hline
QSM & 8 &  8 & 4 & 8 \\ 
\hline
MBM$^{\mathrm{*}}$ & 1 &  8 & 0 & 8 \\ 
\hline \hline
\multicolumn{5}{l}{$^{\mathrm{*}}$ Here, MBM uses 8 RF mirrors to deliver 8 bits/s/Hz.} 
\end{tabular}
\label{tab:tab1}
\end{center}
\end{table}

\begin{table}[tbp]
\renewcommand{\arraystretch}{1} % Default value: 1
\caption{System specifications for Fig. \ref{fig:SER_12Nr_12R}.}
\begin{center}
\begin{tabular}{c c c c c }
\hline \hline
Method & TX  & RX & QAM & Rate in\\
& antennas & antennas & modulation order & bits/s/Hz\\
\hline 
SM & 16 & 12 & 256 & 12 \\ 
\hline
GSM & 16 &  12 & 8 & 12 \\ 
\hline
QSSK & 64 & 12 & 0 & 12 \\ 
\hline
QSM(A) & 32 &  12 & 4 & 12 \\ 
\hline
QSM(B) & 16 &  12 & 16 & 12 \\ 
\hline
MBM & 1 &  12 & 0 & 12 \\ 
\hline \hline
\multicolumn{5}{l}{$^{\mathrm{*}}$ Here, MBM uses 12 RF mirrors to deliver 12 bits/s/Hz.} 
\end{tabular}
\label{tab:tab2}
\end{center}
\end{table}

\iffalse
\begin{figure}[tbp]
%\hspace{2mm}
\centerline{\includegraphics[trim = {0cm 0cm 0 0mm}, scale = 0.55]{/SER_8Nr_16R.pdf}}
\caption{Symbol error probabilities for SIMO-MBM and LMBM achieving transmission rate equal to 16 bits per complex channel use using 8 receive antennas.}
\label{fig:SER_8Nr_16R}
\end{figure}
\fi 

\subsection{Outage Probability Comparison with Legacy SISO/SIMO/MIMO \label{sec:outage_comparison}}

In this section, we compare MBM with traditional MIMO systems from the standpoint of information theory by comparing outage probabilities. First, we review how the outage probability is computed in the MBM setup, and next, we present the numerical results.

MBM constellation points are selected with an equal probability of $1/M$. The empirical probability mass function over the constellation set is
\begin{IEEEeqnarray}{c}
\hat{P}_{{\mathcal{H}}} = \frac{1}{M} \sum_{i = 1}^{M} \delta({\rv{h}}_i),
\end{IEEEeqnarray}
where  $\delta(\rv{h}_i)$ is the Dirac delta measure at point $\rv{h}_i$. Let $\rv{h}$ denote the random variable drawn from distribution $\hat{P}_{{\mathcal{H}}}$.
The outage is defined as the event that the mutual information of a fading realization does not support a target rate $R$, i.e., 
\begin{IEEEeqnarray}{c}
\big\{{\mathcal{H}}: I(\rv{h; \rv{h}+\rv{z}}) < R\big\}.
\end{IEEEeqnarray}
Mutual information $I(\rv{h}; \rv{h}+\rv{z})$ is a random variable whose value depends on the particular realization of MBM constellation set $\mathcal{H}$. In the presence of AWGN, the distribution of output signal at the receive antennas is given by the convolution of the mass function over the constellation set with Gaussian density, indicated by $\hat{P}_{{\mathcal{H}}} * \varphi$. Accordingly, empirical mutual information is
\begin{IEEEeqnarray}{c}
I({\mathcal{H}}) := I(\rv{h}; \rv{h}+\rv{z}) = h(\hat{P}_{\mathcal{H}} * \varphi_{1}) - h(\varphi).
%&=& h\big(\frac{1}{M}\sum_{i=1}^{M}\varphi_1(\rv{z}-\rv{c}_i)\big)  - h(\varphi_{1})
\label{eq:mutual_info}
\end{IEEEeqnarray}
Function $h$ denotes the differential entropy. The output distribution $\hat{P}_{{\mathcal{H}}} * \varphi$ is a Gaussian mixture, where the mixture components are the realized constellation points. It is shown as the cardinality of the constellation set grows, MBM asymptotically achieves the capacity of $K$ parallel AWGN channels \cite{isit2014}\cite{9834835} and is highly concentrated around $K \log(1+\mathsf{SNR})$. 

\begin{figure}[tbp]
\hspace{-0.3cm}
\centerline{\includegraphics[width = \figsizea]{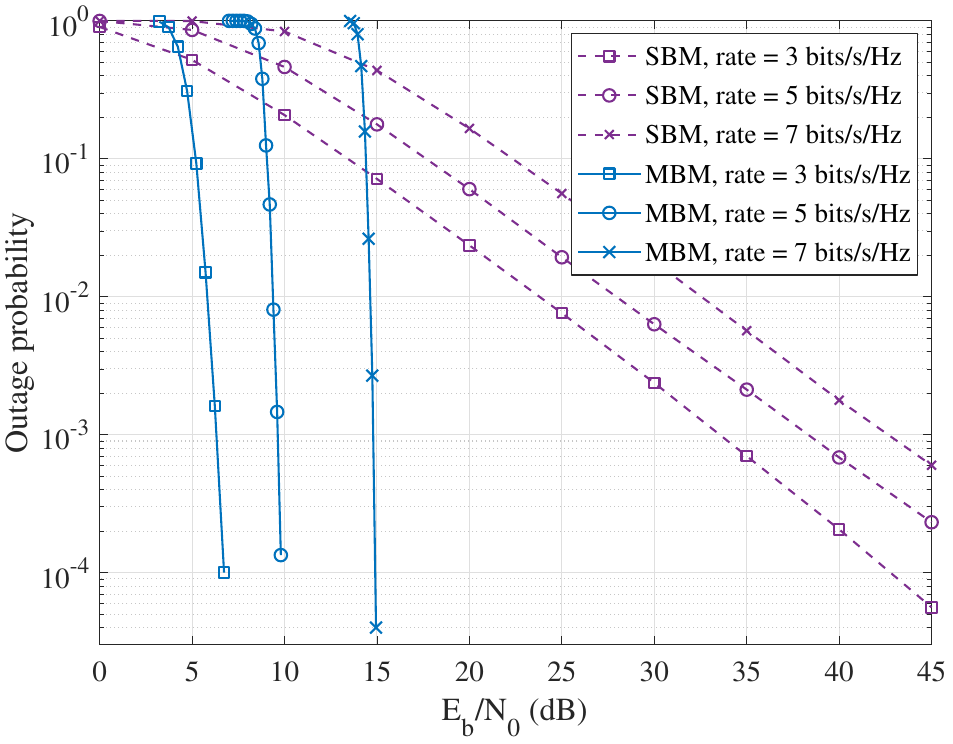}}
\caption{ Outage probability of 1$\times$1 (SISO) MBM (media-based modulation) vs. legacy 1$\times$1 (SISO) SBM (source-based modulation) targeting transmission rates equal to 3, 5, and 7 bits per complex channel use.}
\label{Poutage_SISO}
\end{figure}

Fig. \ref{Poutage_SISO} compares outage capacity curves in SISO setup for rates 3, 5, and 7 bits per complex channel use. The outage probabilities for SISO-MBM are calculated using mutual information corresponding to different realizations of an MBM constellation. The number of points in the constellations used to calculate MBM outage probabilities are 64 (corresponding to 6 RF mirrors), 256 (corresponding to 8 RF mirrors), and 512 (corresponding to 9 RF mirrors) for rates 3, 5, and 7 bits per complex channel use, respectively. This means rates 3, 5, and 7 are achieved by relying on the redundancy of 3, 3, and 2 bits, respectively. It is observed that performance gains are particularly pronounced in SISO setups due to the inherent diversity of MBM. 

\begin{figure}[tbp]
%\hspace{0.5cm}
\centerline{\includegraphics[width = \figsizea]{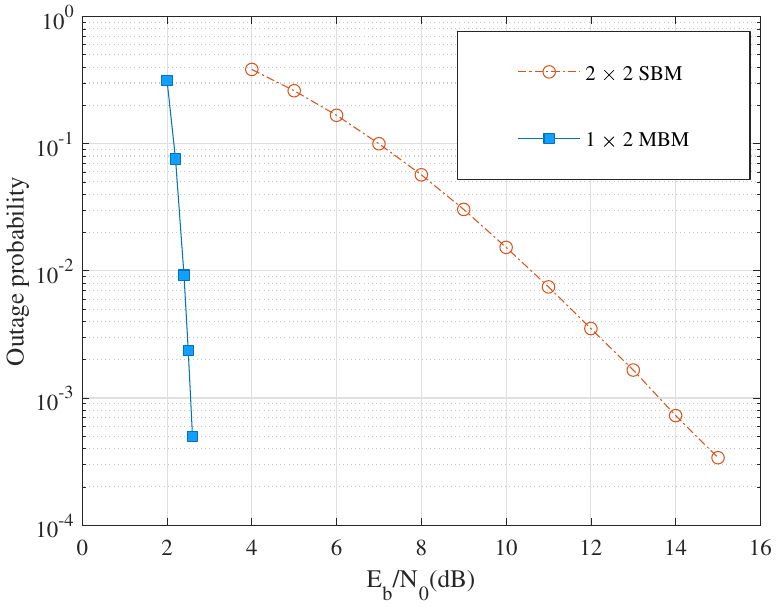}}
\caption{Outage probability of 2$\times$2 MBM compared with 2$\times$2 SBM (legacy) MIMO, targeting a transmission rate equal to 6 bits per complex channel use. The number of RF mirrors used in MBM is 8. } 
\label{Poutage_2Nr}
\end{figure}

\begin{figure}[tbp]
%\vspace{0.4cm}
\centerline{\includegraphics[width = \figsizea]{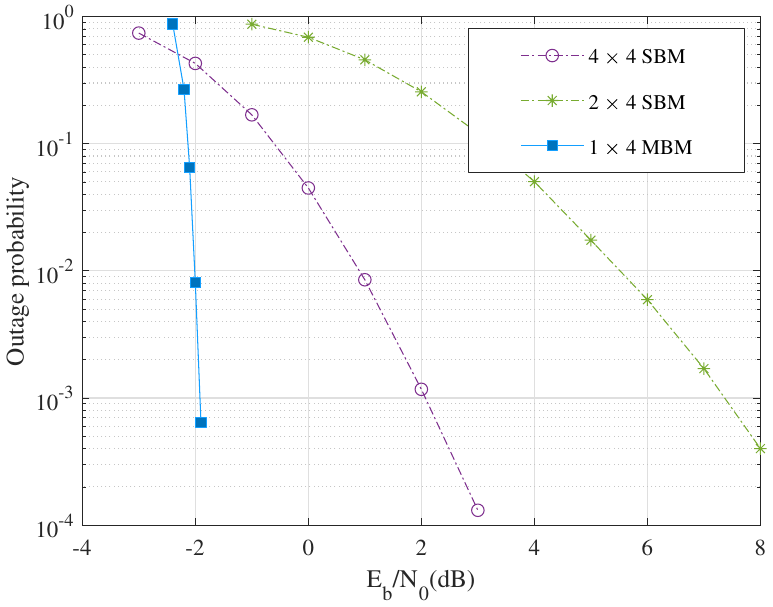}}
\caption{Outage probability of 1$\times$4 MBM compared to 2$\times$4 and 4$\times$4 SBM (legacy) MIMO, targeting a transmission rate equal to 8 bits per complex channel use. The number of RF mirrors used in MBM is 9. } 
\label{Poutage_4Nr}
\end{figure}

Fig. \ref{Poutage_2Nr} and \ref{Poutage_4Nr} compare the outage probabilities in MIMO setup when 2 and 4 receive antennas are used, and target transmission rates are 6 bits per complex channel use and 8 bits per complex channel use, respectively. Note that MBM only uses a single RF chain at the transmit side. Moreover, no additional RF source modulation is used for MBM in any of the scenarios considered. 

Increasing the number of RF mirrors would lower the required energy for a given outage probability. In this article, we confine to outage measurements for up to 4 receive antennas and 9 RF mirrors merely due to the difficulty of computing mixture entropy in higher dimensions. 

%\emph{Remark}: Implementing (relatively complex) FEC structures are typical in legacy wireless systems. MBM achieves great error performance without the need for such FEC structures. In the absence of FEC, symbol detection is performed with the lowest possible delay of a single symbol interval. Due to \say{small decoding delay} and \say{small uncoded symbol error probability}, MBM is great candidate for employing decision feedback methods. Decision feedback can help equalize and track the constellation structure over time. 

%\section {Pulse Shaping, Equalization and Decision Feedback at the Receiver}

\section{Increased Diversity Order Through Application of FEC \label{sec : diversity_gain}}

Fig. \ref{reed_solomon} and \ref{fig:BEP_coded} demonstrate another distinctive property of MBM: the slope of the error probability curve in the MBM scheme increases by a multiplicative factor when applying MDS error correction codes. This property is reminiscent of transmit diversity in the MIMO setup. However, MBM relies on a single transmit unit while achieving the additional diversity gain. In this section, we quantify the realized gain in the diversity order when applying MDS codes to MBM. 

Formally, a scheme is said to achieve diversity gain $d$ and spatial multiplexing order $r$ \cite{NN5}, if it supports the data rate

\begin{IEEEeqnarray}{c}
\lim_{\mathsf{SNR} \to \infty} \frac{R(\mathsf{SNR})}{\log \mathsf{SNR}} = r \> \text{(bit/s/Hz)}
\end{IEEEeqnarray}
with the average error probability 
\begin{IEEEeqnarray}{c}
\lim_{\mathsf{SNR} \to \infty} \frac{\log P_e(\mathsf{SNR})}{\log \mathsf{SNR}} = -d,
\end{IEEEeqnarray}
where $\mathsf{SNR}$ is the average signal-to-noise ratio at each receive antenna. 
Appendix \ref{appendix:uncoded_diversity} shows an uncoded MBM achieves {diversity order} $d = K - r$ without any special processing other than the usual maximum likelihood detection. Furthermore, applying a FEC with minimum distance $D$ increases the {diversity} order by a multiplicative factor of $D$ for a small reduction in the multiplexing gain. More precisely, an MDS code using a maximum likelihood decoder achieves diversity order $d = D \times K-{r}/{\tau}$, where $\tau$ is the code dimensionless rate (c.f. \cite{mbm_dmt}). Moreover, using a simple hard-decision decoder with $t$ error correction capability, MBM achieves a diversity order of $(t+1)\times K$. Appendix \ref{appendix:coded_diversity} provides the analytical analysis for the increase in the diversity order using a  hard-decision decoder.
%Larger code lengths attain such {diversity} gain with a smaller reduction in {multiplexing}. Next, we review the problem formulation and the outline of the results given in the paper.

\emph{Remark}: Unlike legacy MIMO, in MBM, the increase in diversity order is realized using a single transmit unit, i.e., a single RF chain and a single transmit antenna surrounded by RF walls. Therefore, a higher diversity order, resulting in a better error performance, is possible  merely through algorithmic complexity. Furthermore, unlike random-like codes such as turbo codes and low-density parity check (LDPC) codes which typically suffer from error floor, the slope of the error curve in coded MBM will not change as SNR increases.

%Appendix \ref{FEC_appendix} shows that applying an MDS code over a finite field together with hard decision decoding, asymptotically increases the slope of the error rate vs. SNR by a factor of $t+1$.
%Consequently, application of FEC will realize an effect similar to ``diversity over time" with an order determined by $t+1$. 

%\section{Comparison with Spatial Modulation (SM), Generalized Spatial Modulation (GSM) and legacy SISO/SIMO/MIMO.}

\section {Mitigating Bandwidth Expansion via Time-limited Pulse Design} 
\label{sec : Pulse Shaping}

\begin{figure}[tb]
\textbf{}
	\vspace*{0cm}
    \centerline	{\includegraphics[width = \figsizea]{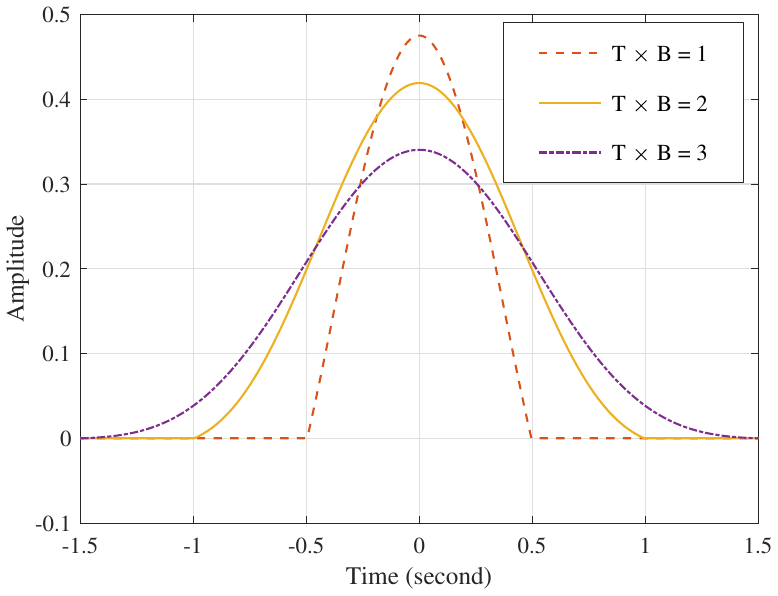}}
	\caption{Optimum pulse shapes for various time-bandwidth products.}
	\label{Fig:pulse_shape}
\end{figure}

\begin{figure}[tb]
\centering
    \centerline{\includegraphics[width = \figsizea]{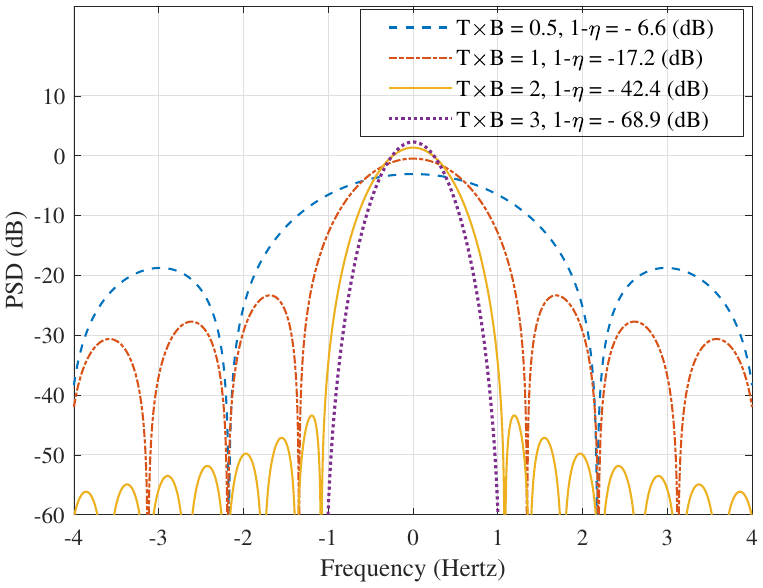}}
	\caption{Power spectral density of optimum pulses for various time-bandwidth products. All optimum pulses correspond to $B = 1$ Hertz.}
\label{Fig:spectral_density}
\end{figure}
\begin{figure}[tb]
    \centerline{\includegraphics[width = 0.5\textwidth]{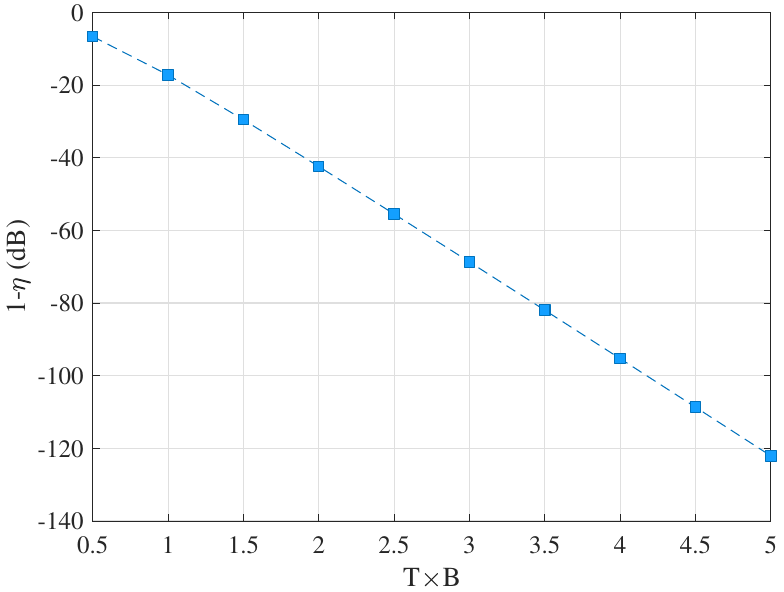}}
	\caption{Portion of the total power leaked outside of the allocated bandwidth as a function of the time-bandwidth product.}
	\label{Fig:efficiency}
\end{figure}
    
MBM and other techniques in the context of ``index modulation", including spatial modulation and its variations, inherently correspond to linear time-variant (LTV) systems. Unlike linear time-invariant (LTI) systems that maintain the spectrum occupancy of the signal, an LTV system  typically increases the occupied bandwidth beyond the spectrum of the pulse used for transmission. The issue of an increase in the bandwidth can be tackled by careful design of the pulse shaping filter at the transmitter. 
%using the combination of the two methods explained next. The first approach is to apply pulse shaping with pulses of a limited length in the time domain and optimized to maintain most of their energy in the allocated frequency band. The second method is to insert periods of silence between successive MBM symbols.  

%\subsection{Pulse Shaping: Time-limited Pulse Design} 
The transmit pulse is designed such that it is contained within successive reconfigurations of RF mirrors. Since each pulse reaches zero value by the time RF mirrors change the configuration, bandwidth expansion beyond the spectrum of the shaped pulse will be avoided. More specifically, the MBM  signal in the time domain is obtained by convolving a sequence of impulses, where the magnitude of impulses is modulated by i.i.d. Gaussian random values (fading realizations), with the time-domain representation of the pulse shaping filter. Since: 1) the sequence of impulses has a flat power spectrum; 2) convolution translates to multiplication in the frequency (power spectral) domain; and 3) the pulse value will be zero at the time of starting the subsequent transmission, the power spectrum of MBM will be proportional to the power spectrum of the pulse shaping filter. 

Let $p(t)$ denote the MBM pulse limited to the interval $[-T/2, T/2] $: 
\begin{IEEEeqnarray}{c}
p(t) = 0 \quad \text{for all} \quad  |t| > \frac{T}{2}.
\end{IEEEeqnarray}
The spectrum of the pulse is
\begin{IEEEeqnarray}{c}
P(f) = \int_{-\frac{T}{2}}^{\frac{T}{2}} p(t) e^{-2\pi i f t} \mathrm{d}t.
\end{IEEEeqnarray}
The optimum time-limited pulse minimizes the energy outside of the allocated frequency band $[-B, B]$. In other words, an optimum pulse maximizes the ratio
\begin{IEEEeqnarray}{c}
\eta := \frac{\int_{-B}^{B} |P(f)|^2  \mathrm{d}f}{\int_{-\frac{T}{2}}^{\frac{T}{2}} |p(t)|^2  \mathrm{d}t},
\end{IEEEeqnarray}
which is the fraction of total power maintained in the allocated band. Reference ~\cite{chalk} studies the solution for this optimization problem. Appendix \ref{appendix:pulse_design} includes the mathematical formulation for the optimum pulse shape. 

The ratio $\eta$ for the optimum pulse depends only upon $T\times B$, i.e., the product of pulse length and allocated bandwidth. The shape of optimum pulses as well as power spectral densities (PSD) corresponding to different time-bandwidth product, $T\times B$, are depicted in Fig. \ref{Fig:pulse_shape} and \ref{Fig:spectral_density}, respectively. The quantity $1-\eta$ in Fig. \ref{Fig:spectral_density} denotes the fraction of the pulse energy outside the given frequency band. Note that $T\times B=1/2$ corresponds to the traditional Nyquist signaling rate. It is observed that, for a $T\times B$ as low as two, the (total) out-of-band energy is about 42.4 dB lower than the pulse's energy. Fig. \ref{Fig:spectral_density} shows how the total out-of-band leaked power changes for a wide range of $T\times B$ values.

\section{Radio Frequency Implementation} \label{RF-sec}

\begin{figure}
\hspace{0 cm}
%\vspace{-0.2cm}
\centering
\includegraphics[width=0.5\columnwidth]{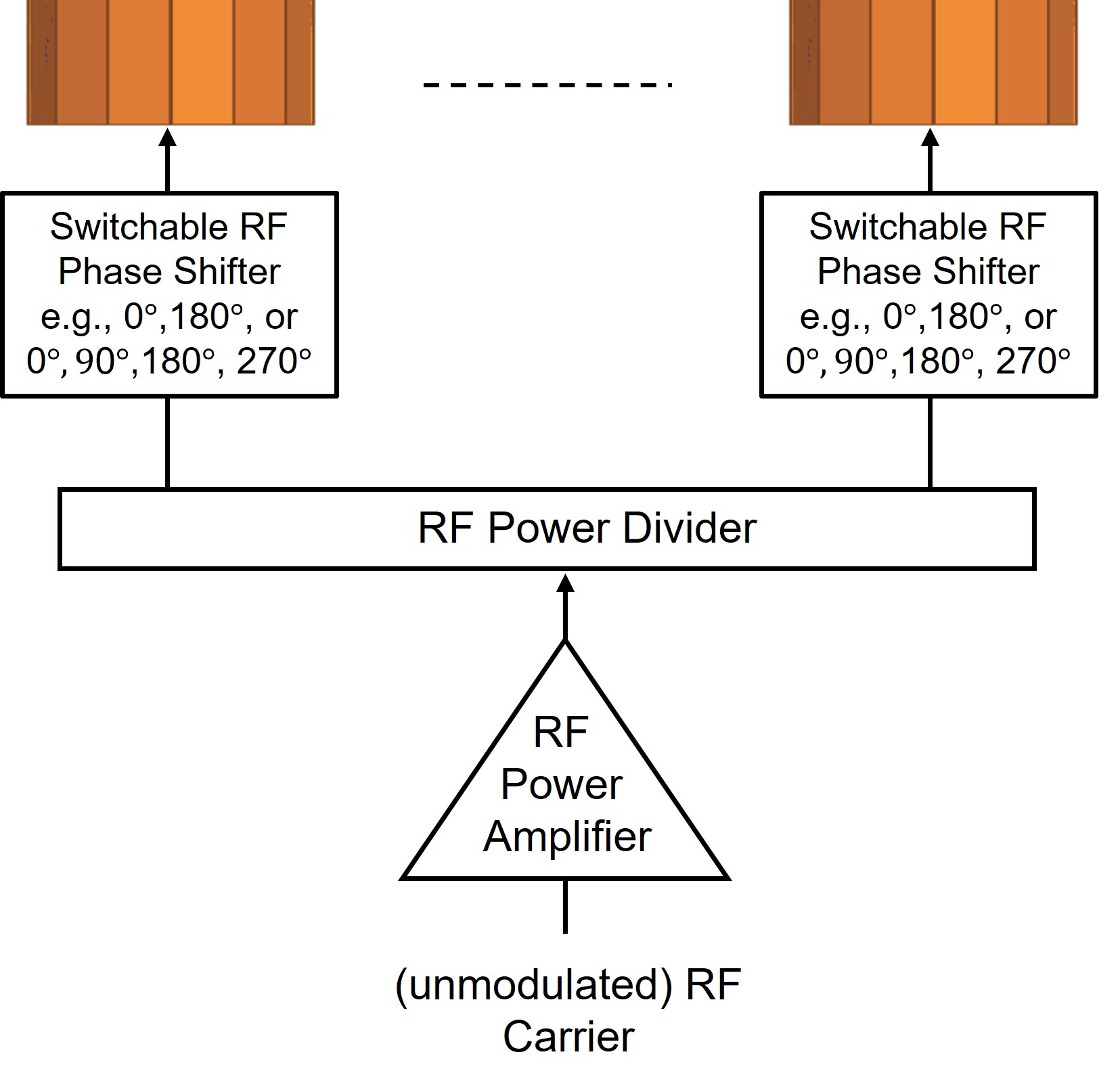}
\caption{Overall structure of a transmitter with multiple transmit antennas.}
\label{single_rf_chain}
\end{figure}

Reference \cite{lmimo_mbm} (also see~\cite{7511273}) reports an MBM antenna unit operating at 5.8 GHz. Fig.  \ref{single_rf_chain} shows an LMBM structure relying on the corresponding antenna. The RF structure in \cite{lmimo_mbm} enjoys low hardware complexity, as it incorporates a single transmit chain, where an RF divider is used to feed multiple transmit antennas. The RF phase shifters in Fig. \ref{single_rf_chain} allow sending additional SBM data. For example, one can select 0\degree, 90\degree, 180\degree, and 270\degree~phase shifts according to two additional bits of SBM data (per transmit unit). Reference \cite{lmimo_mbm} discusses techniques, including the use of RF phase shifters in  Fig.~\ref{single_rf_chain}, to facilitate receiver training. The same techniques apply to the new antenna structure presented in the current article. Readers are referred to~\cite{Tech-rep1}  for examples of antenna patterns and realizations of MBM constellations corresponding to indoor and outdoor propagation environments (also see~\cite{isit2014} for the latter). Simulations are performed using ANSYS high-frequency structural simulator (ANSYS HFSS\footnote{\href{www.ansys.com/products/electronics/ansys-hfss}{www.ansys.com/products/electronics/ansys-hfss}}) to obtain antenna patterns, which are then imported to the ray tracing software (Remcom Wireless Insite\footnote{\href{https://www.remcom.com/wireless-insite-em-propagation-software}{https://www.remcom.com/wireless-insite-em-propagation-software}}) to obtain the constellation points.  
%It is observed that the constellation structure enjoys the desired randomness in all three propagation environments (indoor, outdoor, and line-of-sight).

\begin{figure}[!ht]
\centering
\begin{subfigure}{.5\textwidth}
  \centering
   \includegraphics[width =\columnwidth]{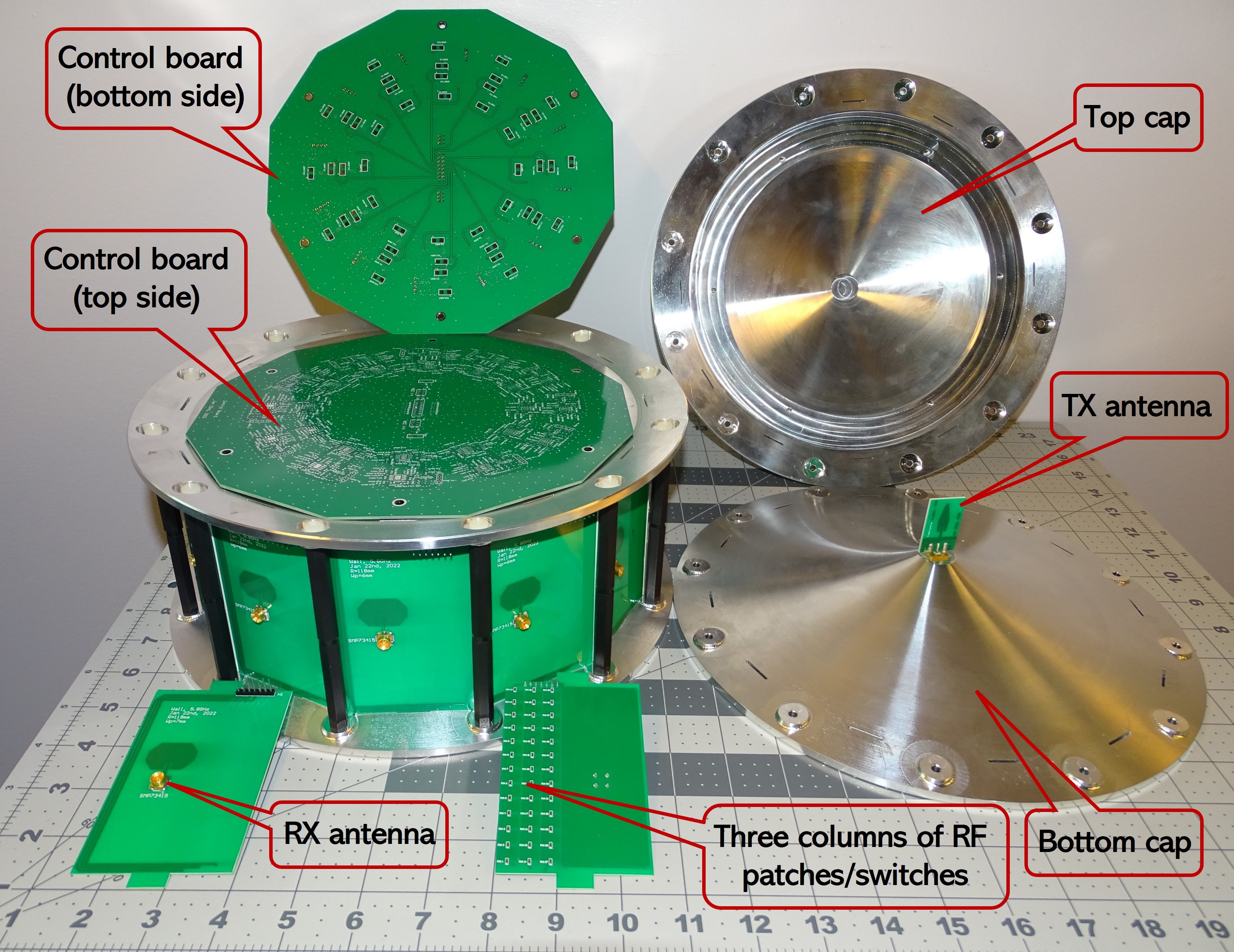}
  \caption{}
\end{subfigure}%
\\
\begin{subfigure}{.5\textwidth}
    \centering
    %$\vspace*{0.1cm}
    \includegraphics[width=1\columnwidth]{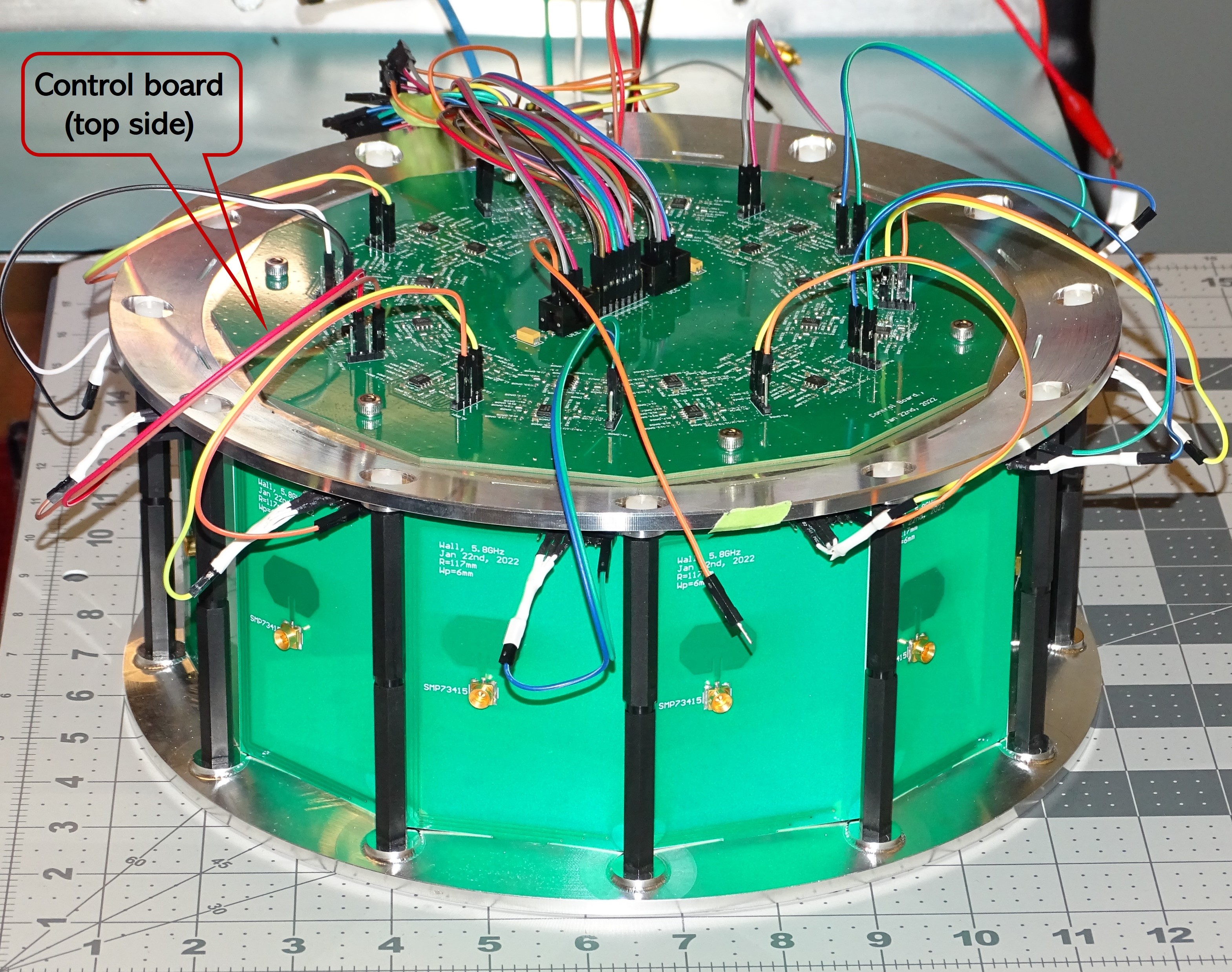}
    \caption{}
\end{subfigure}
\caption{Antenna structure: (a) some components, plus a partially assembled antenna structure, (b) fully assembled antenna structure. There are twelve walls, each with three building block columns  of switchable parasitic elements and one receive antenna in each wall. Building block columns, in total 36, are controlled independently.}
\label{fig:multi-pattern_structure}
\end{figure}

\begin{figure}
\centering
\includegraphics[width=0.45\columnwidth, trim={0cm, 0cm, 0, 1cm}]{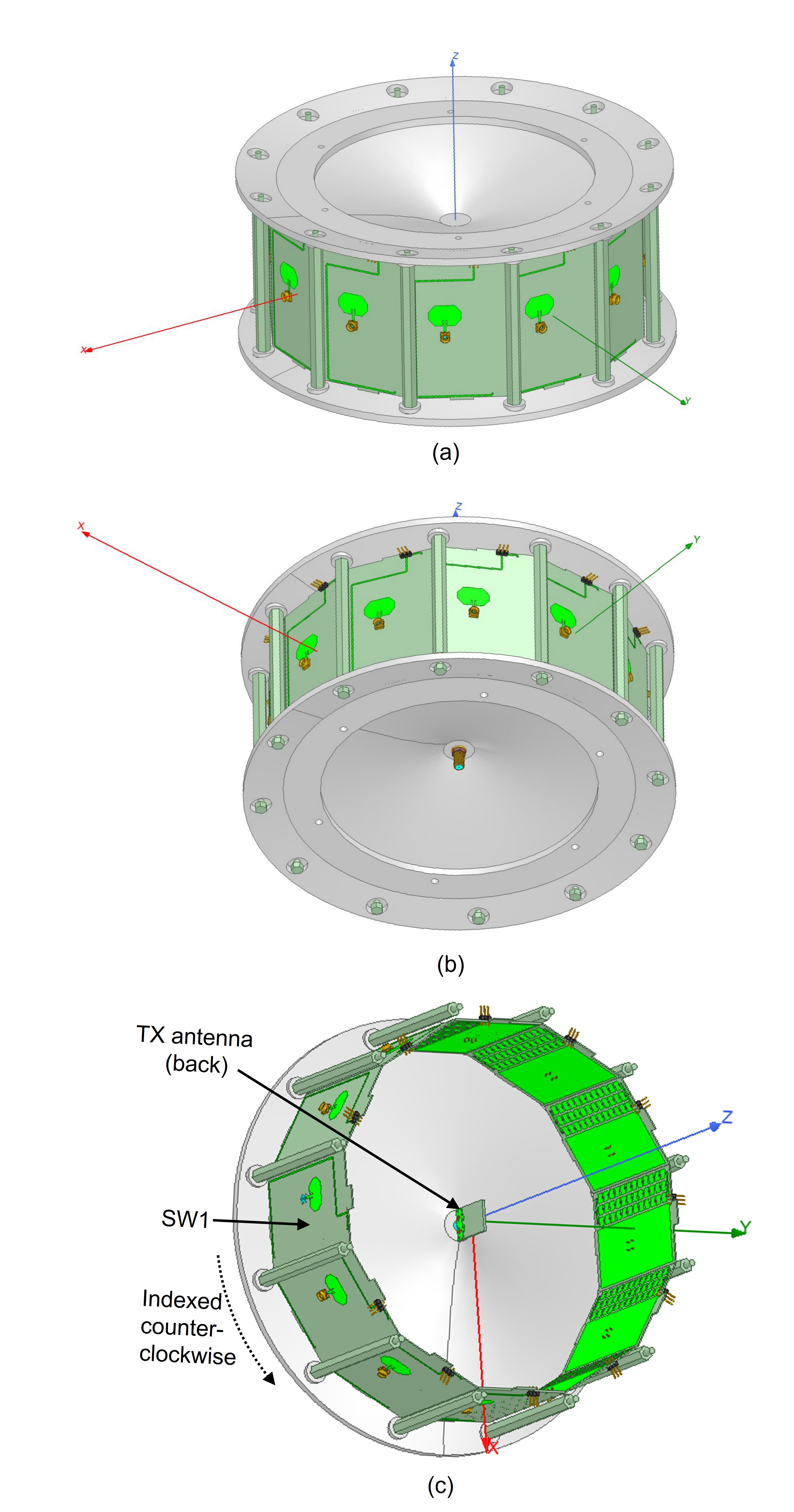}
\caption{Antenna structure: (a) view from the top, (b) view from the bottom, (c) view from the inside - including indexing of walls used in Table~\ref{Tabclw2}.}
\label{fig:multi-pattern_structure2}
\end{figure}

Antenna structure reported in \cite{Tech-rep1}, although tested for functionality (see \cite{new-ak1}, \cite{new-ak2}), suffers from the following shortcomings: 
\begin{enumerate}
    \item The top and bottom of the antenna's cylindrical shape are left open. Consequently, part of the energy of the radio frequency wave leaves the transmitter through the top/bottom openings. This, in turn, reduces the antenna gain in the horizontal plane.  
    \item It lacks any receive (RX) antenna. 
\end{enumerate}
Fig. \ref{fig:multi-pattern_structure} and \ref{fig:multi-pattern_structure2} show the new antenna structure designed to address the above shortcomings. The new structure includes: 1) metallic reflectors closing the top and the bottom; and 2) each wall includes a receive (RX) antenna. 

\begin{figure}[tb]
    \centerline{\includegraphics[scale = 0.7]{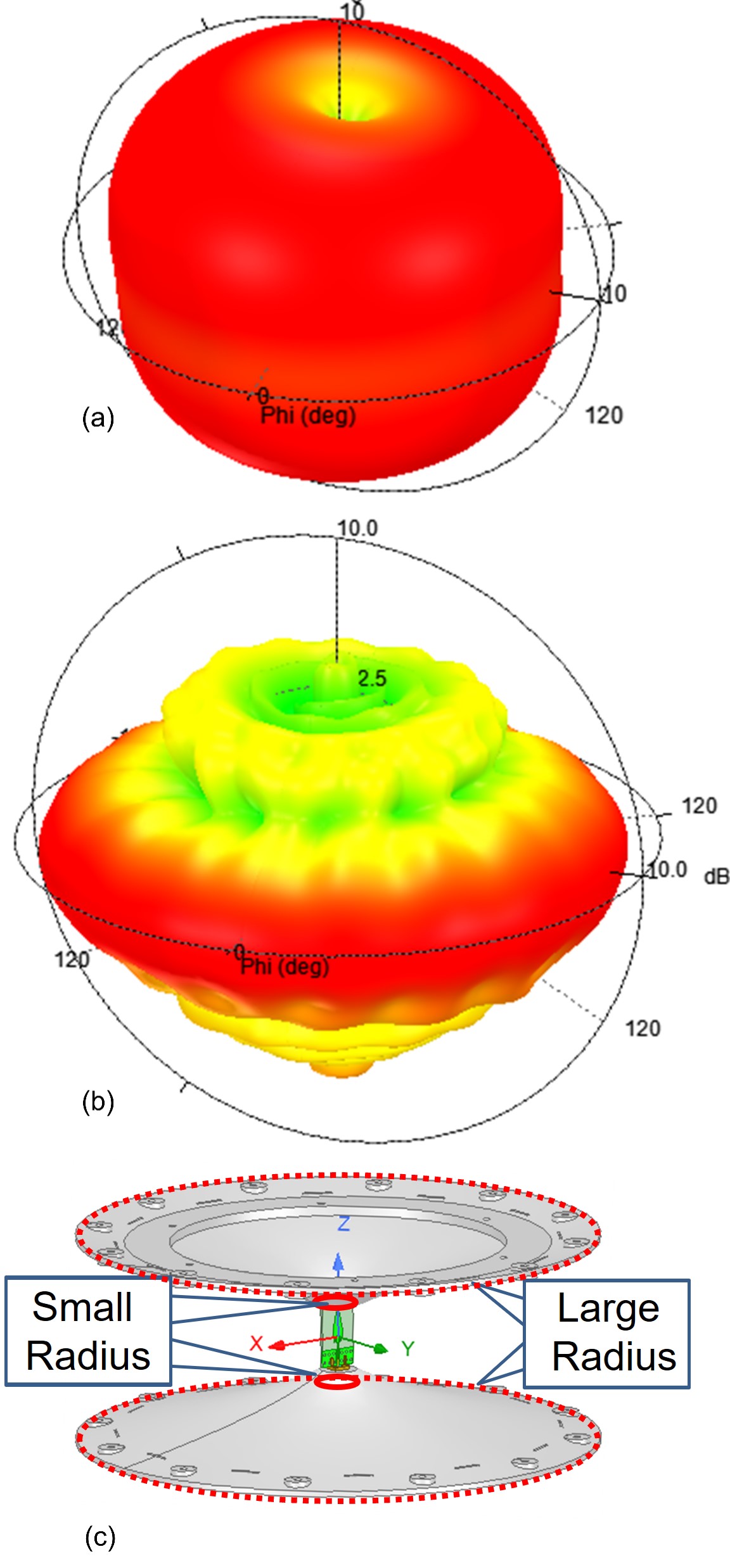}}
	\caption{Pattern of the transmit antenna: (a) in the absence of the top and bottom metallic closures, (b) in the presence of the top and bottom metallic closures shown in (c). }
	\label{TX-all}
\end{figure}

Fig.~\ref{TX-all}(c) shows the transmit (TX) antenna placed at the center of the new structure and fixed to the bottom cap. TX antenna is designed to generate an Omni-directional pattern (see Fig.~\ref{TX-all}(a)) in the absence of the other parts. Next, we explain the reason for choosing Omni-directional pattern. 
Due to the symmetry of the antenna walls along the spherical angle $\phi$ (i.e., in the X-Y plane), it is desirable to have a transmit pattern with a similar spherical symmetry. Such symmetry guarantees that each of the 12 RF walls is exposed to (approximately) the same amount of RF energy. Consequently, when switched on and off, each RF wall has (approximately) the same impact on the outgoing RF signal. Accordingly, the transmitter can provide coverage for any receiver located at an angle of $\phi\in[0,360\degree]$ (with respect to the transmit unit). 
The two metallic reflectors/closures, called metallic caps hereafter, are responsible for constraining the radiation in the horizontal directions - compare the TX pattern in Fig. \ref{TX-all}(a) vs. the one in Fig. \ref{TX-all}(b). 
%The top and bottom ribs in each wall PCB pass through respective slots on top and bottom caps, holding the entire structure together.  
%We have used a 100mm long Polyamide Hex spacer at each corner, and have fixed the entire structure using Nylon screws and nuts. 

In the new design, each wall includes three building-block columns, one large metallic rectangular patch on the interior, and a single RX antenna on the exterior. Each building-block column comprises thirteen small RF patches. The thirteen small patches in each building-block column are connected using RF switches (PIN diodes). RF switches corresponding to each column can be independently controlled (turned on/off). When the switches on a given building-block column are on (low impedance connection), the corresponding patches form a connected metallic strip that reflects the incident wave to the interior of the antenna structure. If the switches are off, the incident wave results in an RF signal propagating outside the antenna structure (i.e., the corresponding building block column acts as if it were transparent to the incident wave). 

\begin{figure}[tb]
    \centerline{\includegraphics[width = \figsizea]{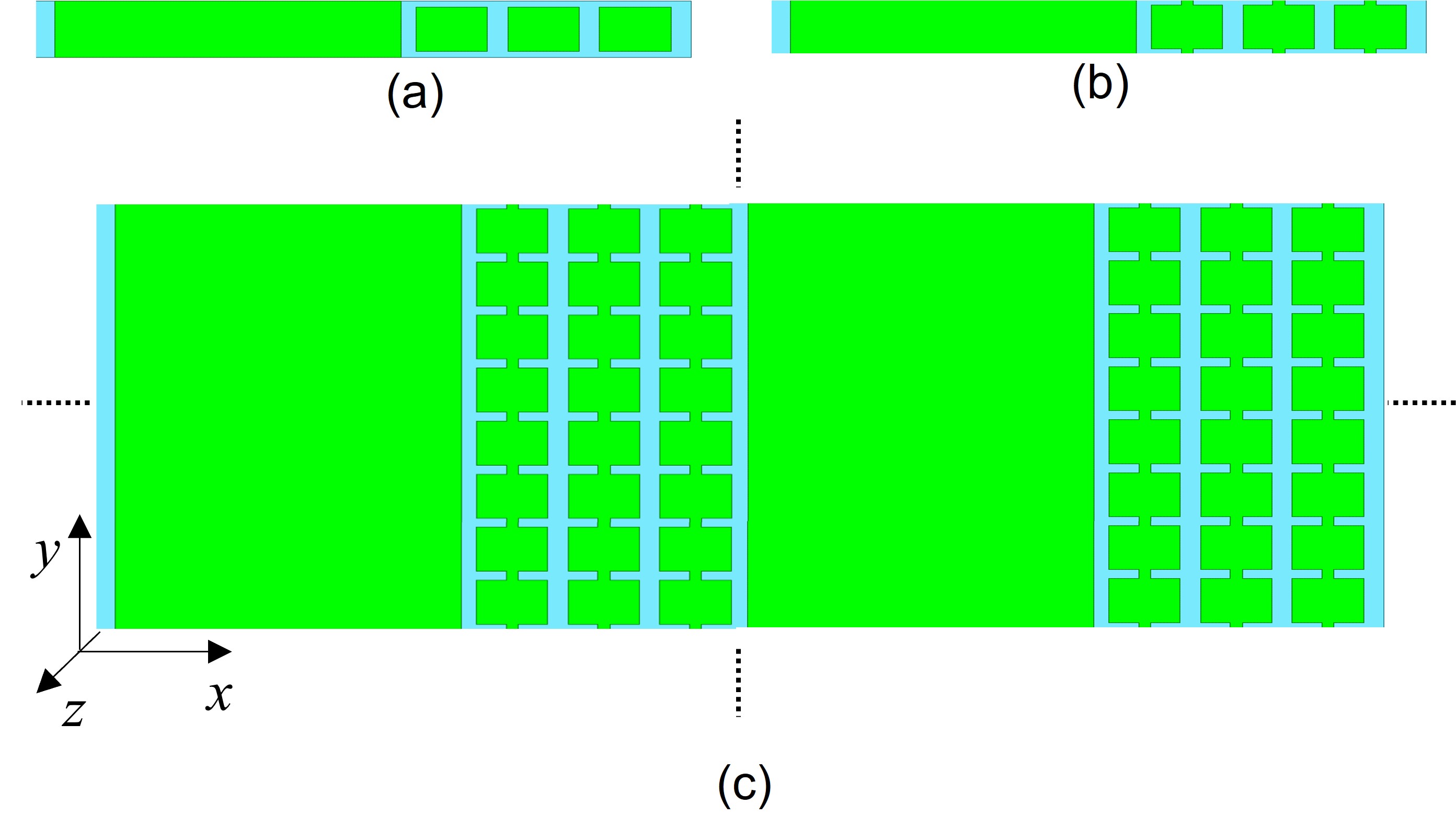}}
	\caption{(a) An RF building block with RF switches being open (approximated as ``open circuit''). (b) An RF building block with RF switches being closed (approximated as ``short circuit''). (c)  Periodic repetition of the RF building block in (b), i.e., with switches being closed.}
	\label{Fig:build}
\end{figure}

\begin{figure}[]
\centering
\begin{subfigure}{\textwidth}
    %\centerline{\hspace*{-9.5cm}\includegraphics[width=0.5\linewidth {./fig_rf/Fig30_b.jpg}}
     \centerline{\includegraphics[scale=0.5] {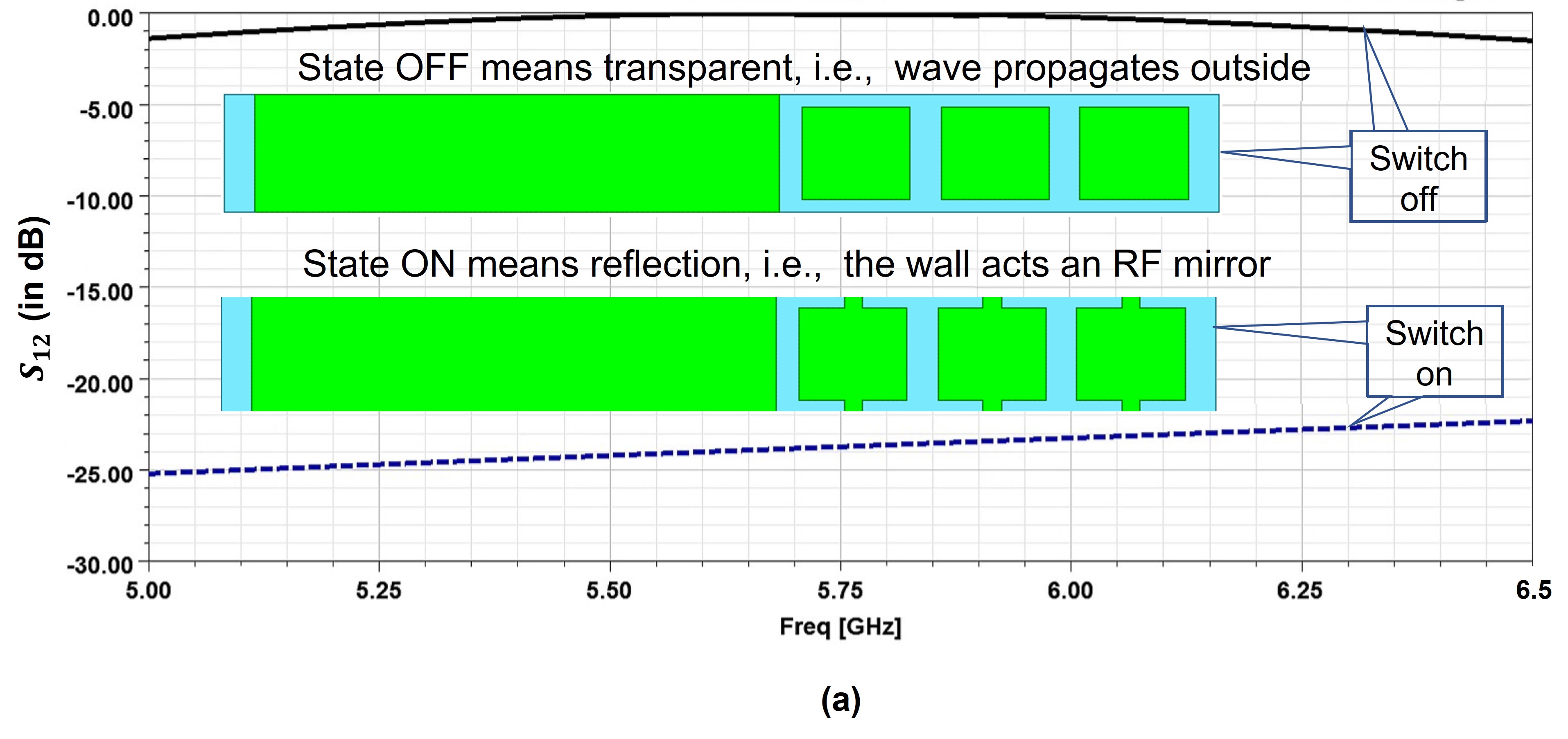}}
    %\centerline{\hspace*{-9.5cm}\includegraphics[width=0.5\linewidth, trim={0, 0cm, 0, 0cm}]{./fig_rf/square_patch_S12.pdf}}
\end{subfigure}
\\
\begin{subfigure}{\textwidth}
\vspace{0cm}
   %\centerline{\hspace*{-9.5cm}\includegraphics[width=0.5\linewidth, trim={0, 0cm, 0, 0cm}]{./fig_rf/Fig30_a.jpg}}
   \centerline{\includegraphics[scale = 0.5]{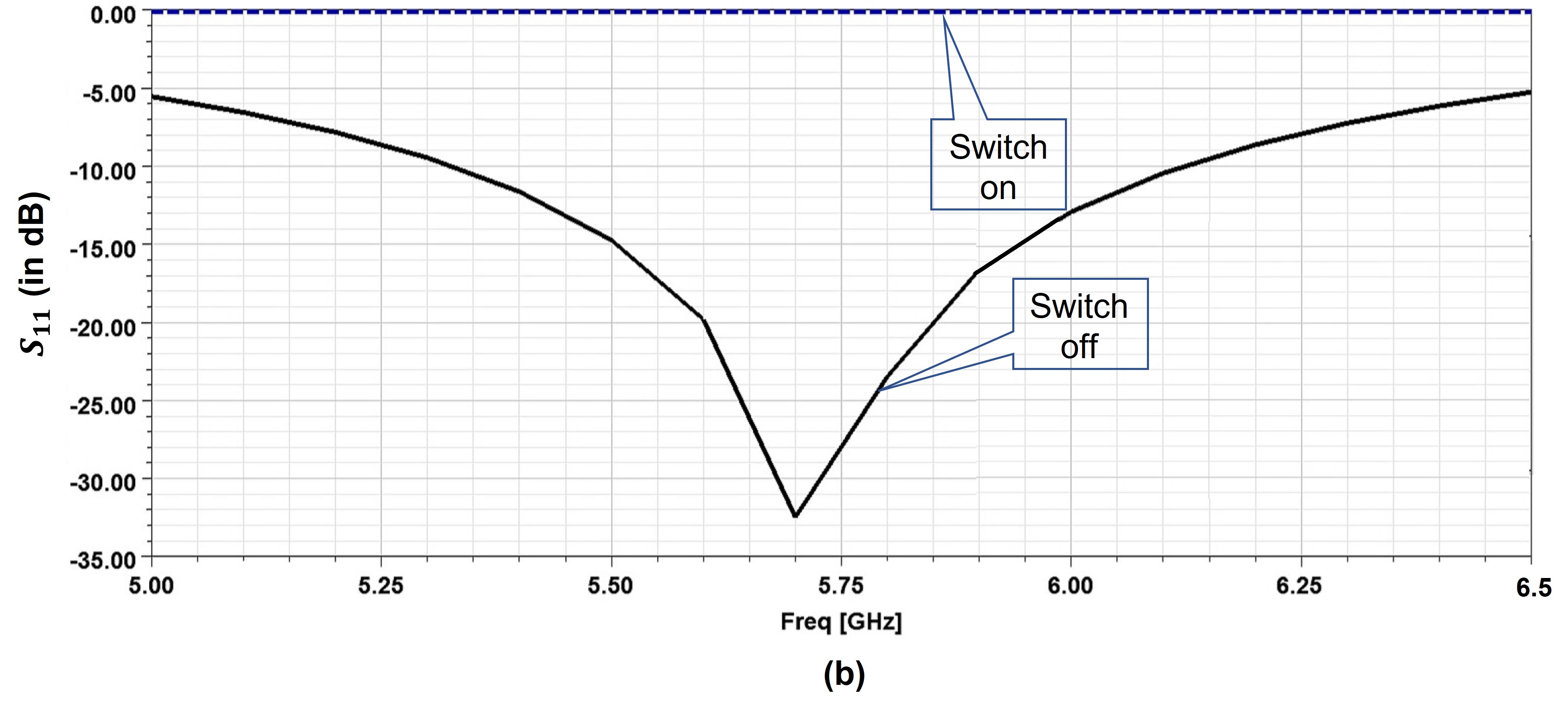}}
   %\centerline{\hspace*{-9cm}\includegraphics[width=0.45\linewidth, trim={0, 0cm, 0, 0cm}]{./fig_rf/square_patch_S11.pdf}}
\end{subfigure}
\caption{Simulation results for the rectangular patch in two states of on and off: (a) transmission and (b) reflection. In these simulations, dimensions of the basic patch (optimized relying on periodic extension) is 4\,mm by 5\,mm.  Upon integrating this initial patch geometry within the rest of the antenna structure, and further tuning, the final patch dimensions are set at 5.6\,mm by 6\,mm. }
\label{simulation_result}
\end{figure}

To optimize the geometry of each wall, one needs to devise a computational method to find an initial approximate solution (with acceptable performance) in a reasonable time. Then, one can include the initial solution obtained in this manner within the entire antenna structure and rely on HFSS to tune (optimize) the underlying dimensions. The basic building block structure is repeated periodically to achieve this goal, forming a plane extending to infinity. Fig. \ref{Fig:build}(a) and \ref{Fig:build}(b) show a single building block column in off and on states, respectively, and Fig.   \ref{Fig:build}(c) shows the periodic extension of a single building block column in the on state. 
The periodic extension in Fig.~\ref{Fig:build} provides a good starting point for subsequent HFSS optimization. This is because, in the actual antenna structure, columns are placed around a circle mimicking an infinite extension along the X axis in Fig.~\ref{Fig:build}(c), while reflections in the top and bottom metallic caps mimic an infinite extension along the Y axis. Relying on HFSS, a plane wave (with a vertical polarization), propagating along the $-z$ axis, is incident on the periodically extended structure in Fig.\ref{Fig:build}(c), then the energy of the wave propagating to the opposite side is measured and used to compute $S_{12}$ and $S_{11}$ (indices 1 and 2 refer to the two sides of the periodically extended surface). In the RF domain, this is achieved by relying on the concept of \emph{Floquet ports} \footnote{https://courses.ansys.com/index.php/courses/ansys-hfss-floquet-port.}. Upon including the resulting initial design within the rest of the antenna structure, the dimensions of the basic patch element are further optimized (using HFSS) to achieve the curves shown in Fig.~\ref{simulation_result}(a) and (b) for $S_{12}$ and $S_{11}$, respectively. The large patch on each wall acts as the RX antenna ground plane.    

%\begin{figure}[]
 %\centerline{\includegraphics[width=1\linewidth, trim={0, 0cm, 0, 0cm}]{./fig_rf/Fig32.jpg}}
 %\centerline{\includegraphics[width=1\linewidth, trim={0, 0cm, 0, %0cm}]{./fig_rf/RX1.jpg}}

 %\caption{RX antenna input matching (a) HFSS simulation).}
%\label{fig:input_matching}
%\end{figure}

\subsection{Top and Bottom Caps: Detailed Derivations}

\begin{figure}[]
\centering
\includegraphics[width=0.5\linewidth, trim={0, 0cm, 0, 3mm}]{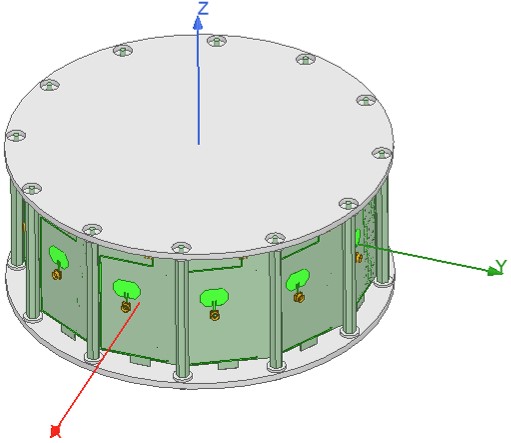}
\caption{Antenna with flat caps.}
\label{flat}
\end{figure}

\begin{figure}[]
\centering
\hspace{-0.5cm}\includegraphics[width=\figsizea, trim={0, 0cm, 0, 0cm}]{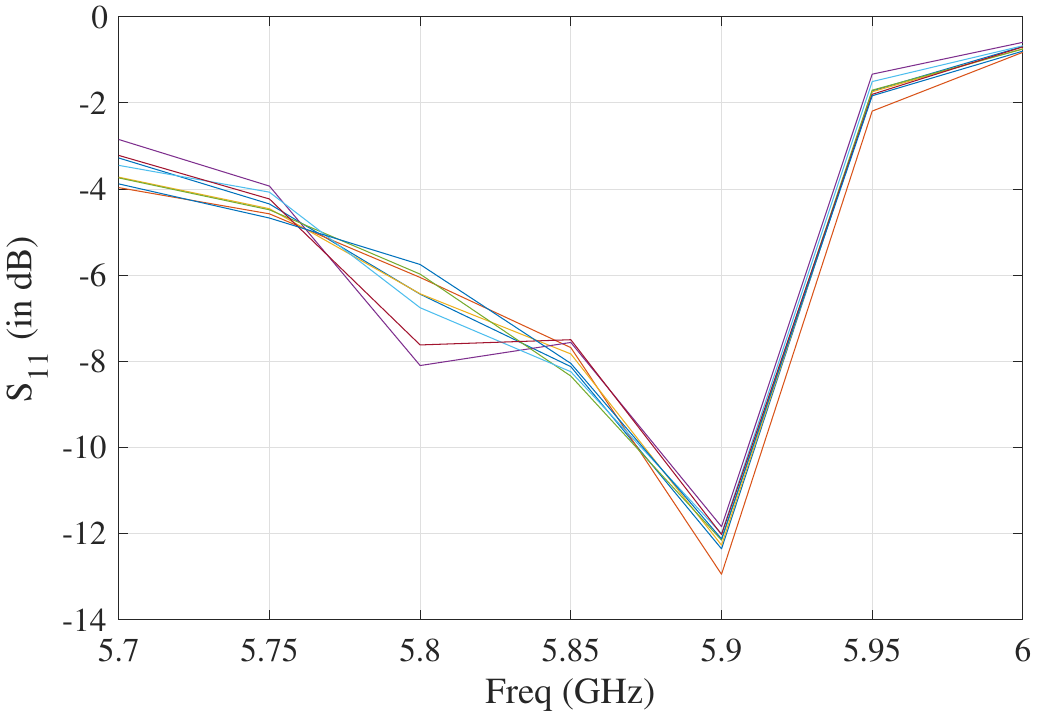}

\caption{$S_{11}$ of the TX antenna with flat caps (HFSS simulation).}
\label{fig:S11_flat}
\end{figure}

\begin{figure*}[t]
\centerline{\includegraphics[width=1.\linewidth]{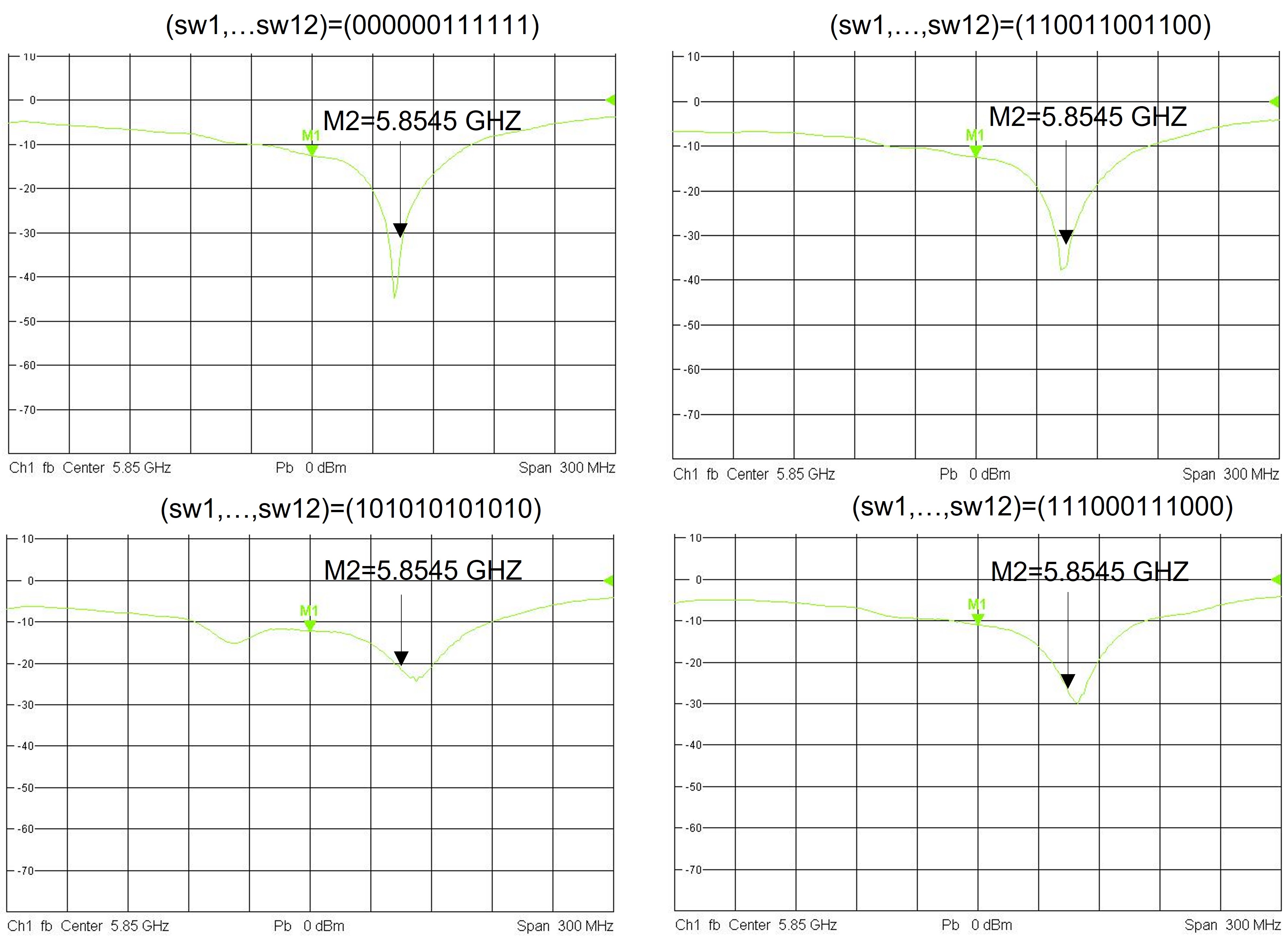}}
\caption{$S_{11}$ of the TX antenna (with optimized caps) for some configurations of on/off switches (measurement results).}
\label{tx_input_matchingm}
\end{figure*}

In a preliminary design, we used two flat metallic caps to close the antenna structure's top and bottom (see Fig.~\ref{flat}). It was observed that the impedance matching of the TX antenna could not be improved beyond -10 dB. Fig.~\ref{fig:S11_flat} shows the corresponding $S_{11}$ graphs.
A new cap structure is designed to improve the impedance matching by guiding the incident RF wave (see Fig.~\ref{fig:multi-pattern_structure}, \ref{fig:multi-pattern_structure2} and ~\ref{TX-all}). Furthermore, the new caps provide other benefits, including higher efficiency (see Table~\ref{TabIII}) and a pattern concentrated around the horizontal plane (see Fig.~\ref{TX-all}). Fig.~\ref{tx_input_matchingm}, obtained through measurement, confirms that using the curved caps, the $S_{11}$ parameter is rather insensitive to the switching pattern, and values for different switching patterns remain at an acceptable level. 

\begin{figure}[tbp]
%\centerline{\includegraphics[width=.9\linewidth]{./fig_rf/Fig2.pdf}}
\centerline{\includegraphics[scale=0.3, trim = {0, 3cm, 0, 0}]{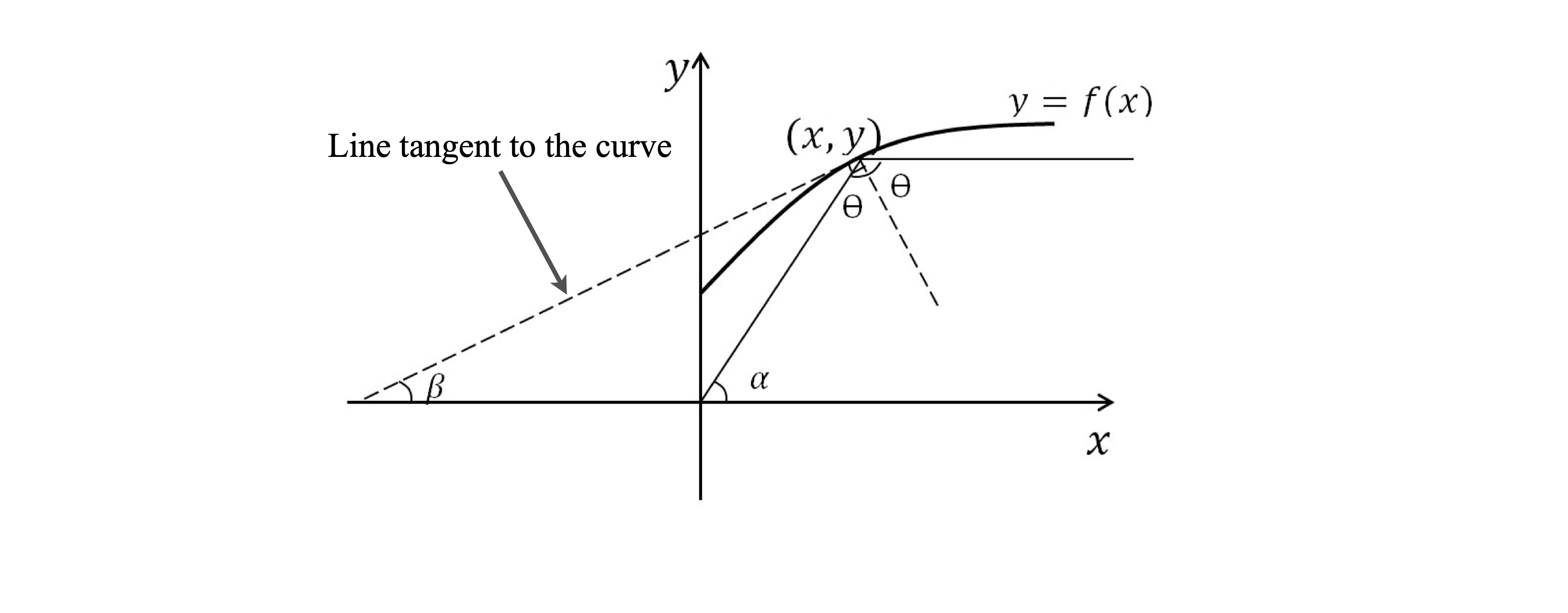}}
\caption{The curvature of the proposed cap.}
\label{fig:curvature}
\end{figure}

%In the initial design, we used two flat metallic caps on top and bottom of the structure, but it resulted that the radiation efficiency fell around 60\% and also make the TX antenna matching very difficult and sensitive to the state of PIN diodes on walls, so we decided to design the top and bottom caps such that they reflect the rays initiating from the origin parallel to the horizontal plane (Fig. \ref{fig:curvature}). The curvature of the interior surface of the proposed cap is shown in Fig. \ref{fig:curvature}.

The top and bottom caps are designed to reflect a ray initiating from the origin such that it propagates parallel to the horizontal axis. Fig. \ref{fig:curvature} shows the curvature of a cut in the interior surface of the proposed cap. Let us assume a radiating source point is located at the origin, i.e.,  $(0, 0)$. The cap surface, which is obtained by rotating the function of $y=f(x)$ around the $y$-axis, should reflect a ray originating from such source point parallel to the $x$-axis. Referring to Fig.~\ref{fig:curvature}, this means 

\begin{IEEEeqnarray}{c}
\alpha + 2 \theta = \pi.
\label{eq:rf1}
\end{IEEEeqnarray}
We also have
\begin{IEEEeqnarray}{c}
\alpha = \beta  + \frac{\pi}{2} - \theta.
\label{eq:rf2}
\end{IEEEeqnarray}
By eliminating $\theta$ in \eqref{eq:rf1} and \eqref{eq:rf2}, we obtain 
\begin{IEEEeqnarray}{c}
\alpha=2\beta.	
\label{eq:rf4}
\end{IEEEeqnarray}
The dashed line in Fig.~\ref{fig:curvature} intersecting with the $x$ axis is tangent to the curve $y=f(x)$. Therefore, 
\begin{IEEEeqnarray}{c}
f^{\prime}(x) = y^{\prime} = \tan(\beta).
\label{eq:rf3}
\end{IEEEeqnarray}
By substituting \eqref{eq:rf4} in \eqref{eq:rf3}, and noting $\tan{\left(\alpha\right)}={y}/{x}$,
%\begin{equation}
%\tan{\left(\alpha\right)}={y}/{x},
%\end{equation} 
it follows that 
\begin{IEEEeqnarray}{c}
{\tan}^{-1}\left(\frac{y}{x}\right) = 2{\tan}^{-1}\left(y^\prime\right). 
\label{eq:tan_inv}
\end{IEEEeqnarray}
Taking $\tan$ of the two sides of \eqref{eq:tan_inv} and using the identity $\tan(2a) = 2\tan(a)/(1-\tan^2(a))$, we obtain
% Ehsan, is it the "above equation"? -> referenced it with a number
\begin{IEEEeqnarray}{LL}
\frac{y}{x} &= \tan {\left(2{\tan}^{-1}\left( y^\prime \right)\right)} \\
 &= \frac{y^{\prime}}{1-{y^{\prime}}^2}.
\label{eq:rf5}
\end{IEEEeqnarray}
Equation \eqref{eq:rf5} is a quadratic equation of the form
\begin{IEEEeqnarray}{c}
{y^{\prime}}^2+2\frac{x}{y}y^\prime-1 = 0
\label{eq:quadratic_diff}
\end{IEEEeqnarray}
in $y^{\prime}$. Referring to Fig.~\ref{fig:curvature}, since $y>0$ and $y^{\prime}>0$, the quadratic equation \eqref{eq:quadratic_diff} has a single valid solution of the form
\begin{IEEEeqnarray}{c}
\frac{y^\prime}{y}=\frac{1}{x+\sqrt{x^2+y^2}}.	
\end{IEEEeqnarray}
To solve the above differential equation, we perform some manipulations as follows:
\begin{IEEEeqnarray}{c}
x\frac{y^\prime}{y}+\frac{y^\prime}{y}\sqrt{x^2+y^2}-1=0.
\end{IEEEeqnarray}
Since $y>0$, we have
\begin{IEEEeqnarray}{c}
\frac{x\frac{y^\prime}{y}+\frac{y^\prime}{y}\sqrt{x^2+y^2}-1}{\sqrt{x^2+y^2}}\times\frac{-x+\sqrt{x^2+y^2}}{-x+\sqrt{x^2+y^2}}=0.
\label{eq:rf9}
\end{IEEEeqnarray}
Let us define $U:=-x+\sqrt{x^2+y^2}$ and ${U^\prime} := \mathrm{d}U/ \mathrm{d}x$. Then, \eqref{eq:rf9} simplifies to
\begin{IEEEeqnarray}{c}
\frac{U^\prime}{U}=0\rightarrow\ln{\left(U\right)}=c,
\label{Eq-ak2}
\end{IEEEeqnarray}
 where $c$ is a constant. After some simplifications, we obtain
\begin{IEEEeqnarray}{LLL}
-x &+& \sqrt{x^2+y^2}=e^c := S, \\
y^2 &=& S^2+2Sx,
\label{ParS}
\end{IEEEeqnarray}
where $S$ is a parameter to be optimized (through HFSS simulations) to attain the best $S_{11}$ for the entire structure. The caps are fabricated using aluminum with a thickness of 5 mm. 

\begin{figure}[]
\centering
\hspace{-0.0cm}\includegraphics[width=0.9\textwidth, trim={0, 0cm, 0, 0cm}]{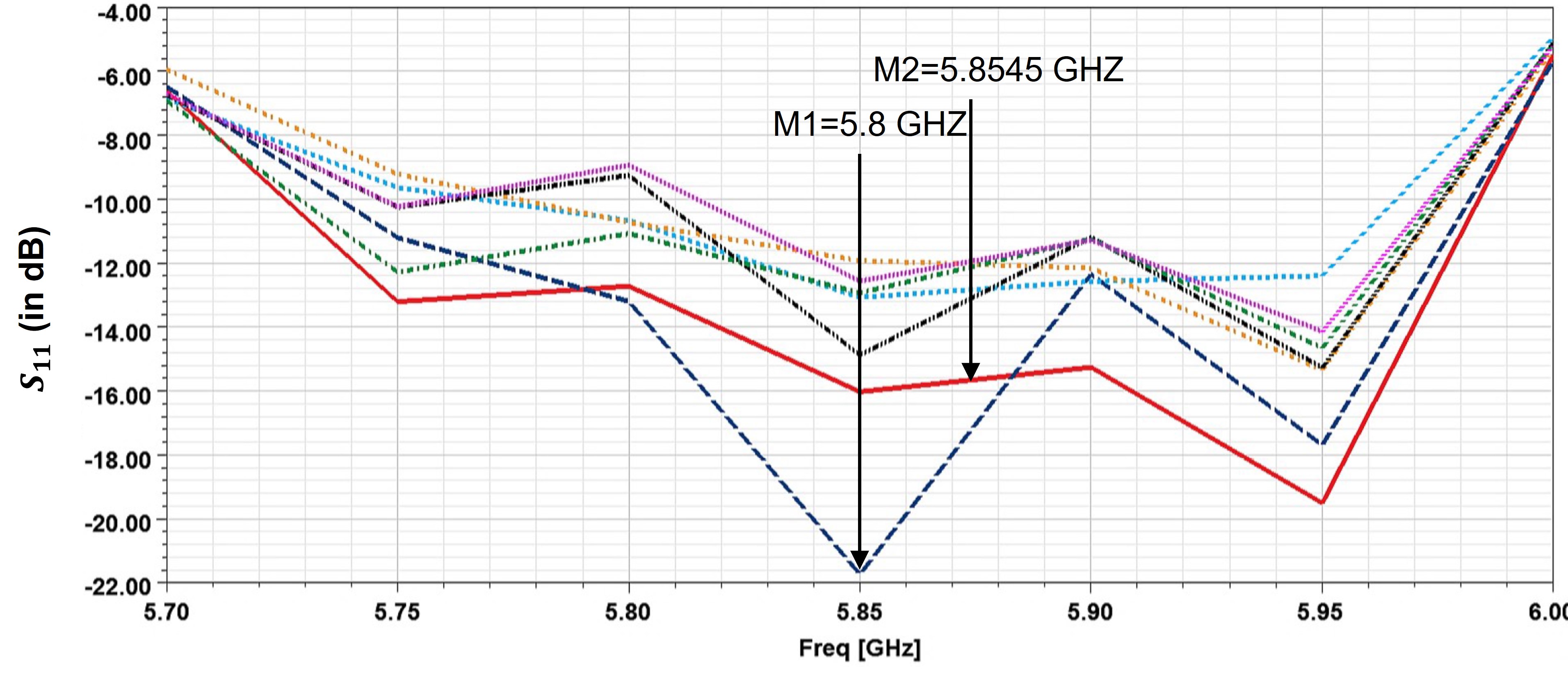}
\caption{$S_{11}$ of the TX antenna, with optimized caps, for 7 different walls' states mentioned in Table \ref{Tabclw2}  (HFSS simulation).}
\label{tx_input_matching}
\end{figure}

We have simulated two different scenarios to study the relative merits of optimized (curved) caps. Both scenarios targeted a center frequency of 5.8545 GHz. The objective was to improve TX antenna matching, i.e., reduce the $S_{11}$ of the TX antenna over the desired bandwidth centered at 5.8545 GHz. In the first scenario, two flat metallic caps were used to close the antenna structure's top and bottom. Tuning was performed on two parameters: 1) structure diameter; and 2) TX antenna length. These parameters were optimized (using HFSS simulations) while keeping all other parameters at their optimized values obtained beforehand. In the second scenario, caps with optimized curvature were used. Details of tuning are explained in Section~\ref{Tune}.
The conclusion was that to have a reasonable TX impedance matching, the diameter of the structure in the first scenario (flat caps) should be at least 7\% larger than the case of the second scenario. Nevertheless, for flat caps, the TX $S_{11}$ was rarely (for a small number of switching patterns) below -10 dB (see Fig.~\ref{fig:S11_flat}). On the other hand, in the second scenario, it reaches -20 dB for almost all switching patterns (see Fig.~\ref{tx_input_matching}, supported by measurement results shown in Fig.~\ref{tx_input_matchingm}). Table~\ref{TabIII} compares the radiation efficiency and maximum gain for the structure with flat caps (case 1) vs. the structure with curved caps (case 2). We observe that the overall efficiency increases from 82\% (for flat caps) to 90\% (for curved caps). More importantly, the use of curved caps has guided the outgoing RF wave to concentrate around the horizontal plane - see improvement in maximum gain for case 2 vs. case 1 in Table~\ref{TabIII}, and the focused TX pattern in Fig.~\ref{TX-all}(c).
Finally, measurement results (see Fig.~\ref{tx_input_matchingm}) confirm that using the curved caps, the $S_{11}$ parameter is acceptable for different switching patterns. 

\begin{comment} 
In the following study, we measured the ability of the antenna to focus the radiated power close to the horizontal plane. For this reason, two partial far-field spheres were defined in HFSS, one with an elevation range of ±20\degree and the other ±2\degree from the horizontal plane. In other words, in the first case, the far-field parameters were calculated on spherical coordinate when $\theta$ varies from 70\degree to 110\degree and $\phi$ varies from 0\degree to 360\degree, while in the second case $\theta$ varies from 88\degree to 92\degree and $\phi$ varies from 0\degree to 360\degree. The figures below show the antenna parameters calculated for the two mentioned cases at 5.8545\,GHz.

For different combinations of on and off walls, we observed that the radiated power is higher in the second scenario when the proposed caps are used. We can see that out of 1W, only about 820mW is radiated from the first structure while more than 900mW is radiated from the second structure. 
As Fig below shows, the maximum amount of radiated power intensity in the first scenario is about 186mW/sr and 145mW/sr for the first and second far-field surface cases, respectively, while we can see that when the proposed curved caps are used, this parameter increases to about 474mW/sr and 325mW/sr, for ±20\degree and ±2\degree beamwidths, respectively. This will prove that the radiation mostly takes place at the directions and beam-width which are quite desirable when we use the proposed caps.
\end{comment}

\begin{table*}
\begin{center} 
\caption{Antenna radiation parameters at 5.8545\,GHz  for flat caps (case 1) and for caps with optimized curvatures (case 2). Maximum gain values are computed over a partial spherical far-field surface with $\theta$ deviating $\pm 20$\degree~from the horizontal plane. Accepted power accounts only for the impact of $S_{11}$. Radiated Power accounts for the impact of $S_{11}$ and the waste of energy in the antenna structure. Radiated Power is computed by integrating the energy flux density, i.e., the Poynting vector, over a sphere (at far-field) surrounding the entire antenna structure. Efficiency is defined as the ratio: Radiated Power/Incident Power.}
\begin{tabularx}{0.9\textwidth} { 
  | >{\centering\arraybackslash}X
  | >{\centering\arraybackslash}X 
  | >{\centering\arraybackslash}X 
  | >{\centering\arraybackslash}X 
  | >{\centering\arraybackslash}X 
  | >{\centering\arraybackslash}X
  | >{\centering\arraybackslash}X 
  | >{\centering\arraybackslash}X
  | >{\centering\arraybackslash}X 
  | >{\centering\arraybackslash}X 
  | >{\centering\arraybackslash}X
  | >{\centering\arraybackslash}X 
  | >{\centering\arraybackslash}X
  | >{\centering\arraybackslash}X
  | >{\centering\arraybackslash}X  | }
  \hline
  Quantity & case 1 & case 2 \\
  \hline
   Incident Power & 1 W & 1 W \\
  \hline
    Accepted Power & 851.6 mW & 957.74 mW \\
  \hline
  Radiated Power & 820.9 mW  & 903.28 mW\\
  \hline
  Efficiency & 82.1\% & 90.3\% \\ \hline
    Peak Realized Gain & 2.35 & 5.96 \\
  \hline
\end{tabularx}
\label{TabIII}
\end{center}
\end{table*}

\begin{table*}
\begin{center}
\caption{Different walls' states selected for simulation of $S_{11}$ in Fig.~\ref{fig:S11_flat} and \ref{tx_input_matching}. Even though each of the three RF building blocks forming a wall can be switched independently, in Figs.~\ref{fig:S11_flat}, \ref{tx_input_matching} and \ref{tx_input_matchingm}, all switches on any of the walls ($1$ to $12$) have the same state (i.e., all three building block columns on a given wall are on for value 1, or all three are off for value 0).}
\begin{tabularx}{0.9\textwidth} { 
  | >{\centering\arraybackslash}X
  | >{\centering\arraybackslash}X 
  | >{\centering\arraybackslash}X 
  | >{\centering\arraybackslash}X 
  | >{\centering\arraybackslash}X
  | >{\centering\arraybackslash}X 
  | >{\centering\arraybackslash}X
  | >{\centering\arraybackslash}X 
  | >{\centering\arraybackslash}X 
  | >{\centering\arraybackslash}X
  | >{\centering\arraybackslash}X 
  | >{\centering\arraybackslash}X
  | >{\centering\arraybackslash}X  | }
  \hline
  State & SW1 & SW2 & SW3 & SW4 & SW5 & SW6 & SW7 & SW8 & SW9 & SW10 & SW11 & SW12\\ 
  \hline
  1 & 0 & 0 & 1 & 1 & 0 & 0 & 1 & 1 & 0 & 1 & 1 & 0\\ 
  \hline
  2	& 1	& 0	& 0	& 1	& 1	& 1	& 0	& 0 & 0 & 1 & 1 & 0 \\
  \hline
  3	& 1	& 1 & 1 & 0 & 0 & 1 & 1 & 0 & 1 & 1 & 1 & 0 \\
  \hline
  4	&1	&1	&1	&0	&0	&0	&1	&1	&1	&0	&0	&0 \\
  \hline
  5	&1	&0	&0	&0	&0	&1	&1	&0	&0	&1	&1	&1 \\
  \hline
  6	&1	&1	&1	&0	&0	&0	&0	&0	&0	&1	&1	&1 \\
  \hline
  7	&1	&0	&1	&0	&1	&0	&1	&0	&1	&1	&0	&1 \\
  \hline
\end{tabularx}
\label{Tabclw2}
\end{center}
\end{table*}

\begin{comment}
\begin{figure}[]
\begin{subfigure}{.5\textwidth}
  \centerline{\includegraphics[width=1\linewidth]{./fig_rf/Fig13_a.pdf}}
  \caption{}
\end{subfigure}%
\\
\begin{subfigure}{.5\textwidth}
   \centerline{\includegraphics[width=1\linewidth]{./fig_rf/Fig13_b.pdf}}
     \caption{}
\end{subfigure}
\caption{Control board PCB, (a) top layer, (b) bottom layer.}
\label{fig:control_board}
\end{figure}
\end{comment}
%
%\begin{figure}[t]
%\centerline{\includegraphics[width=1.\linewidth]{./fig_rf/S11-RX}}
%\caption{$S_{11}$ of the RX antenna (measurement result).}
%\label{fig:S11_RX}
%\end{figure}

\subsection{Details of Surrounding Walls \label{sec:pcb_wall}}
The 12 side walls are made of RO4003C material with a thickness of 32 mil. On the outer face of each wall, the RX antenna and the PIN diodes control command signal traces are located. Part of the inside face of the wall is a large copper patch that acts as the RX patch antenna ground. The rest consists of three building block columns of small patches connected using SMP1320-040LF RF PIN diodes.
%As mentioned earlier, dimensions of patches are optimized such that by turning the PIN diodes on and off, the value of reflection coefficient (or the transmission coefficient) changes significantly. 
%For the initial estimation of the dimensions, we have considered that the wall consists of a building block which can be seen in Fig. \ref{fig:side_wall} with the two on and off states. The PIN diode is modeled as an ideal short or open circuit at this state. Also, we assumed that this building block will be repeated both in the x and y directions.  For this purpose, we defined Master/Slave boundary conditions both for the zx- and yz- planes.
%(Fig. \ref{fig:boundary_condition}). 
To optimize the patch dimensions, we defined two \emph{Floquet ports} at the two sides of a boundary box. Then, the reflection and the transmission coefficients, $S_{11}$ and $S_{21}$, respectively, are computed for different patch dimensions and also for the two states of on and off. Dimensions of the basic patch are optimized such that by turning the PIN diodes on and off, the value of the reflection coefficient (or the transmission coefficient) changes significantly. Relying on this initial solution, we subsequently fine-tune the patch dimensions by simulating the entire antenna structure using HFSS. 

We randomly selected seven different wall states to optimize the overall structure. Table~\ref{Tabclw2} shows these selected states. Fig.~\ref{tx_input_matching} shows the TX antenna impedance matching, i.e., the $S_{11}$ values corresponding to these seven states obtained through simulation.
The final patch size at 5.8545\,GHz is 5.6\,mm by 6\,mm wide. Furthermore, the width of the ground patch is around 37.5\,mm. There is a 1.6\,mm gap along the vertical axis between two adjacent patches in a building block column and a 1.5\,mm gap between two adjacent building block columns.  The radius of the structure where the PCBs are located is around 119\,mm from the center axis (z-axis). 
%In the following figures, Sw represents the building block states where 1 represents the on state, and 0 represents the off state.
%In the HFSS simulation, for each PIN diode, three boundary conditions are defined. 
% HERE 

The equivalent circuit for the PIN diode is composed of an inductance $L_s$ in series with a diode die equivalent circuit, where $L_s$ models the effect of the diode packaging. For SMP1320-040LF, we have\footnote{www.skyworksinc.com/-/media/SkyWorks/Documents/Products/101-200/SMP1320\_Series\_200047S.pdf} $L_s=0.45$\,nH.
The diode die itself is modeled by a parallel RC circuit in which values of the capacitor, $C_d$, and the resistor, $R_d$, depend on the diode state. Since $R_d$ in the on state (forward bias) is very small, i.e., $\simeq 0$, and in the off state (reverse bias) is very large,  i.e., $\simeq \infty$, it turns out that only the value of  $C_d$ in the off state affects the simulation outcome. Referring to MP1320-040LF application note$^{5}$, the value of $C_d$ in the off state (reverse bias capacitance) is set at $0.23$ pF. Exact values for $R_d\simeq 0$ in the on (forward bias) and $R_d\simeq \infty$ in the off (reverse bias) states do not affect the simulation outcomes. However, in the HFSS simulations, we have used $R_d$(forward bias)\,=1\,Ohm and $R_d$(reverse bias)\,=10\,MOhm.

\subsection{Details of the RX Antenna}
\begin{figure}[]
\centering
\includegraphics[width=0.65\linewidth, trim={0, 0cm, 0, 0cm}]{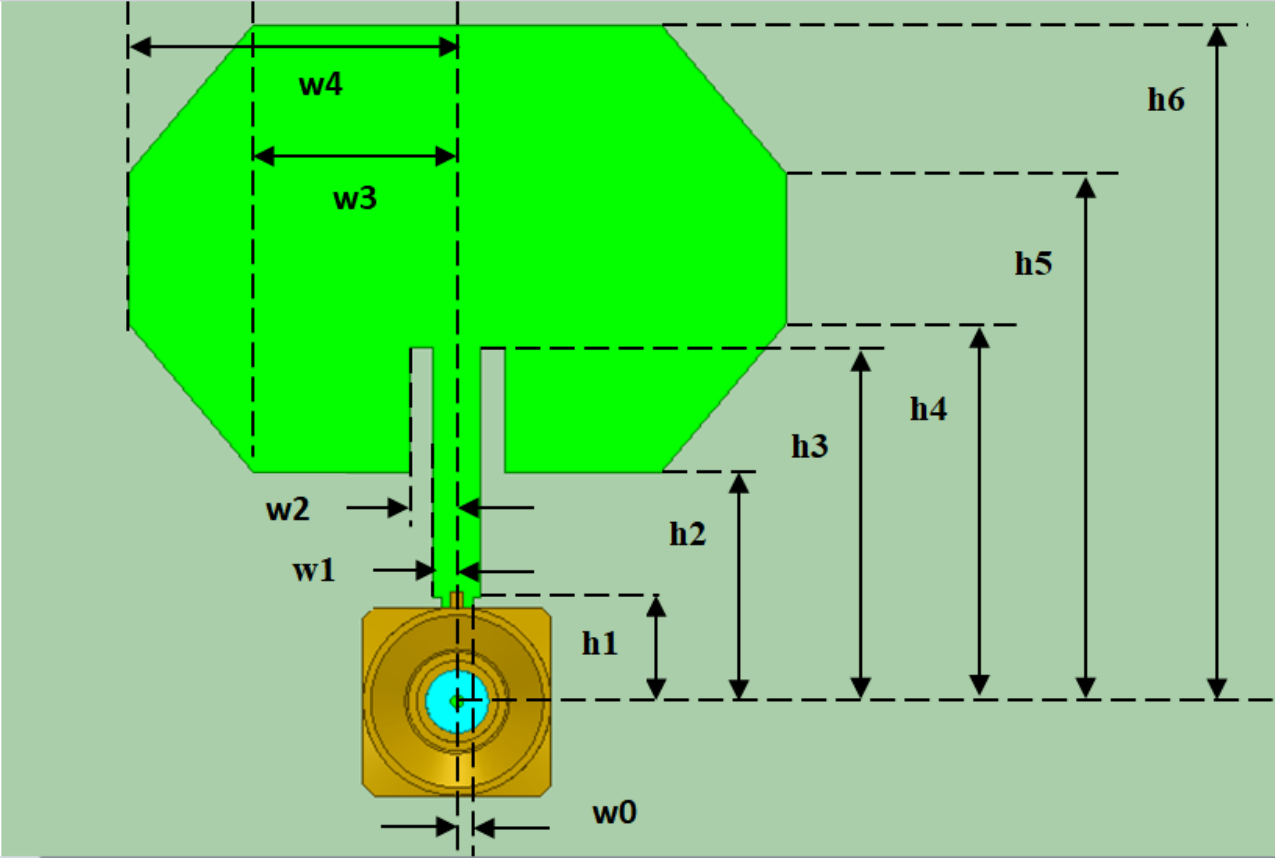}
\caption{RX antenna (refer to Table~\ref{tab:rx_antenna_dimensions} for values of optimized dimensions).}
\label{fig:rx_antenna}
\end{figure}

\begin{figure}[]
 \hspace{-.4cm} 
 \centerline{\includegraphics[width=0.72\columnwidth, trim={0, 0cm, 0, 0cm}]{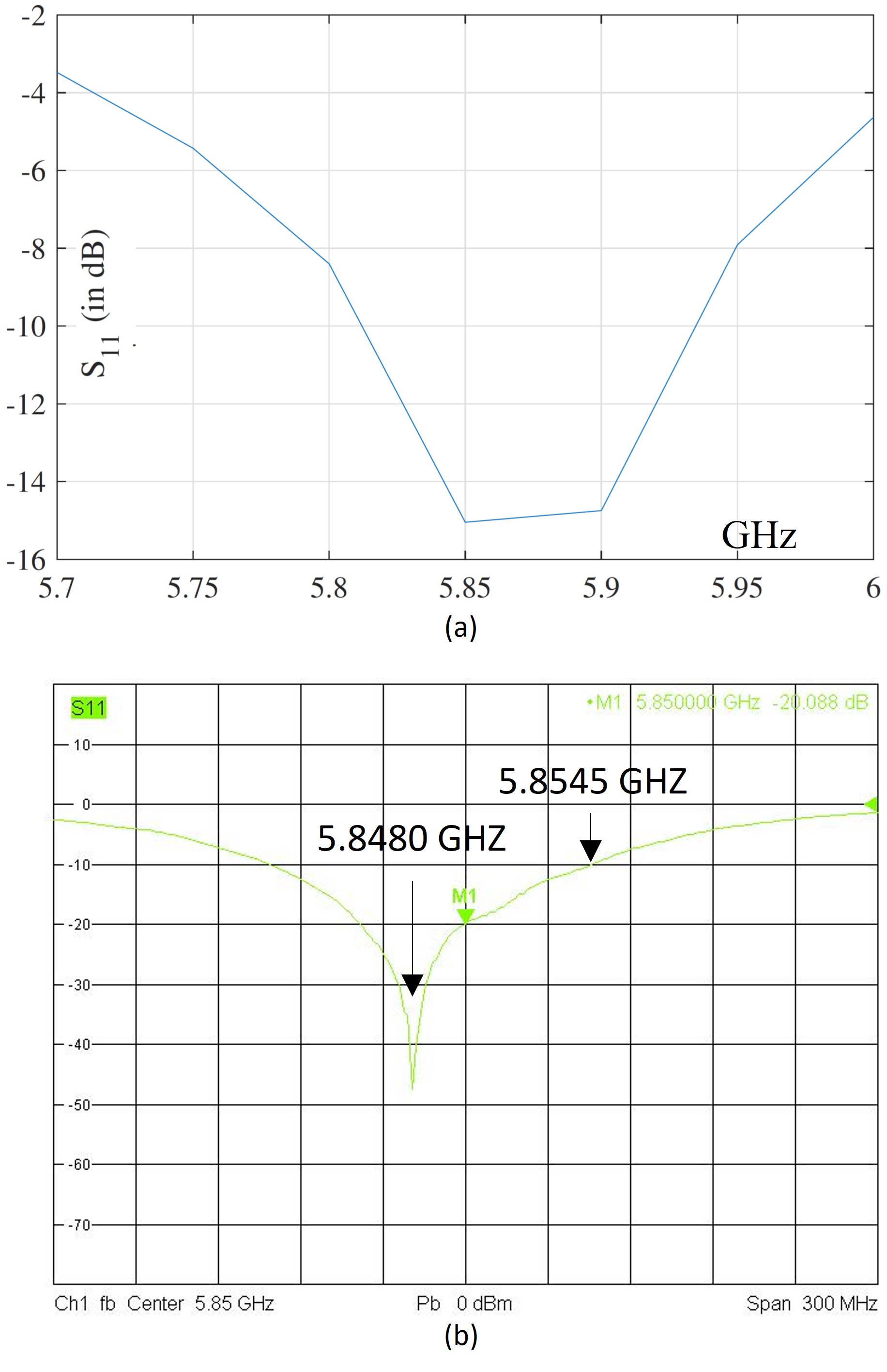}}
\caption{RX antenna input matching: (a) HFSS simulation and (b) measurement.}
\label{fig:input_matching2}
\end{figure}

\begin{figure}[htbp]
    \centerline{\includegraphics[width=0.45\columnwidth]{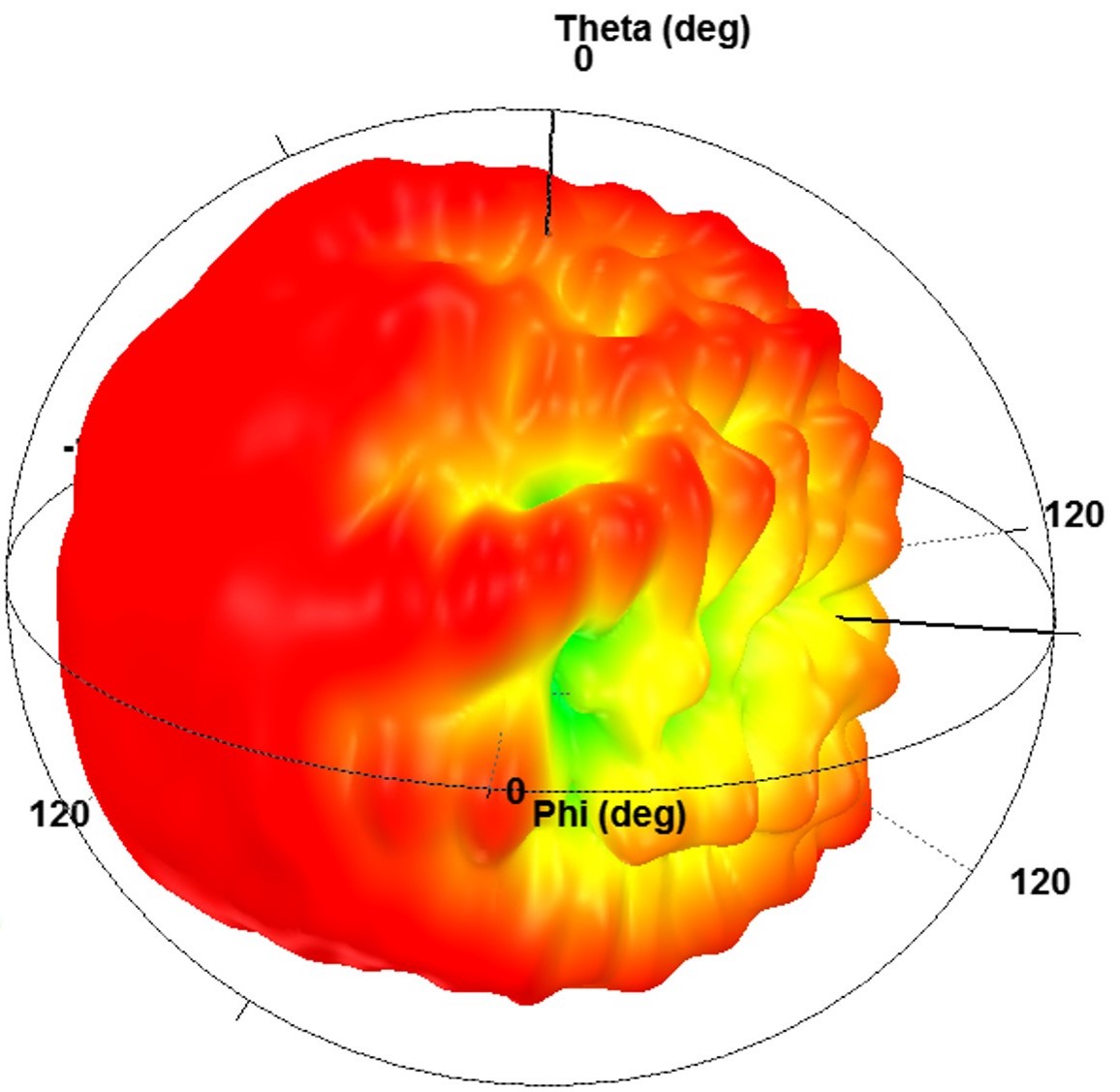}}
	\caption{Pattern of a receive antenna in the presence of the entire antenna structure. }
	\label{RX-pattern}
\end{figure}

The RX antenna is a planar patch element that is fed through a microstrip trace. On each wall, we have one RX antenna and a plane that acts as the patch antenna ground, i.e., the body of the antenna connector will be soldered to it. To have a higher bandwidth, the corners of the patch have been chamfered. Also, the inset feeding technique has been used for matching purposes (similar to the TX antenna).
Several parameters in the RX antenna were tuned at the desired center frequency of 5.8545\,GHz. These parameters include patch width, patch length, chamfer dimensions, inset length, and trace width. Fig.~\ref{fig:rx_antenna} shows the RX antenna, and Table~\ref{tab:rx_antenna_dimensions} shows final dimensions after fine-tuning in the presence of the entire antenna structure.
 Parameter $S_{11}$ for the RX antenna is shown in Fig.~\ref{fig:input_matching2}(a)
(HFSS simulation) and in Fig.~\ref{fig:input_matching2}(b) (measurement). It is observed that the frequency of the point with minimum $S_{11}$ in Fig.~\ref{fig:input_matching2}(a) is slightly different from that of Fig.~\ref{fig:input_matching2}(b). Despite this slight disagreement between simulation and measurement, the measured $S_{11}$ at the target frequency of 5.8545\,GHz is at an acceptable level (about -10 dB). In RF, such a small mismatch between simulation and measurement is quite common and can be corrected.  The mismatch can be due to fabrication error or the difference between the parameters of materials used in fabrication vs. what is modeled in HFSS.  Computational inaccuracies can be corrected by targeting a (proportionally) higher frequency in simulation.  Fig. \ref{RX-pattern} shows the pattern of the receive antenna.
\begin{table*}[h]
\begin{center}
\caption{RX antenna dimensions reported in millimeters.}
\begin{tabularx}{0.9\textwidth} { 
  | >{\centering\arraybackslash}X
  | >{\centering\arraybackslash}X 
  | >{\centering\arraybackslash}X 
  | >{\centering\arraybackslash}X 
  | >{\centering\arraybackslash}X
  | >{\centering\arraybackslash}X 
  | >{\centering\arraybackslash}X
  | >{\centering\arraybackslash}X 
  | >{\centering\arraybackslash}X 
  | >{\centering\arraybackslash}X
  | >{\centering\arraybackslash}X 
  | >{\centering\arraybackslash}X
  | >{\centering\arraybackslash}X  | }
  \hline
  w0 & w1 & w2 & w3 & w4 & h1 & h2 & h3 & h4 & h5 & h6\\ 
  \hline
  0.5 & 0.75 & 1.5 & 6.5 & 10.5 & 3.3 & 7.3 & 11.3 & 12.05 & 16.8 & 21.55\\ 
  \hline
\end{tabularx}
\label{tab:rx_antenna_dimensions}
\end{center}
\end{table*}
%Table~\ref{tab:rx_antenna_dimensions} contains the dimensions of the RX antenna.

\subsection{Details of the TX Antenna}
\begin{figure}[!h]
\centering
\begin{subfigure}{.4\textwidth}
  \centerline{\includegraphics[width=1.1\linewidth]{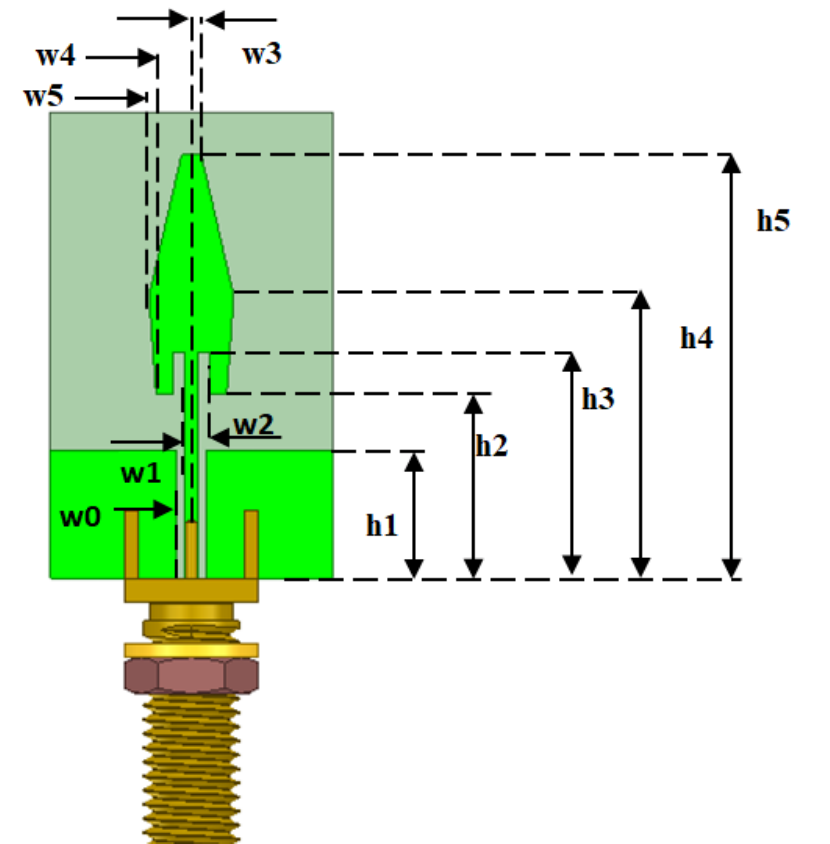}}
  \caption{}
\end{subfigure}%
\\
\begin{subfigure}{.45\textwidth}
  \centerline{\includegraphics[width=0.91\linewidth]{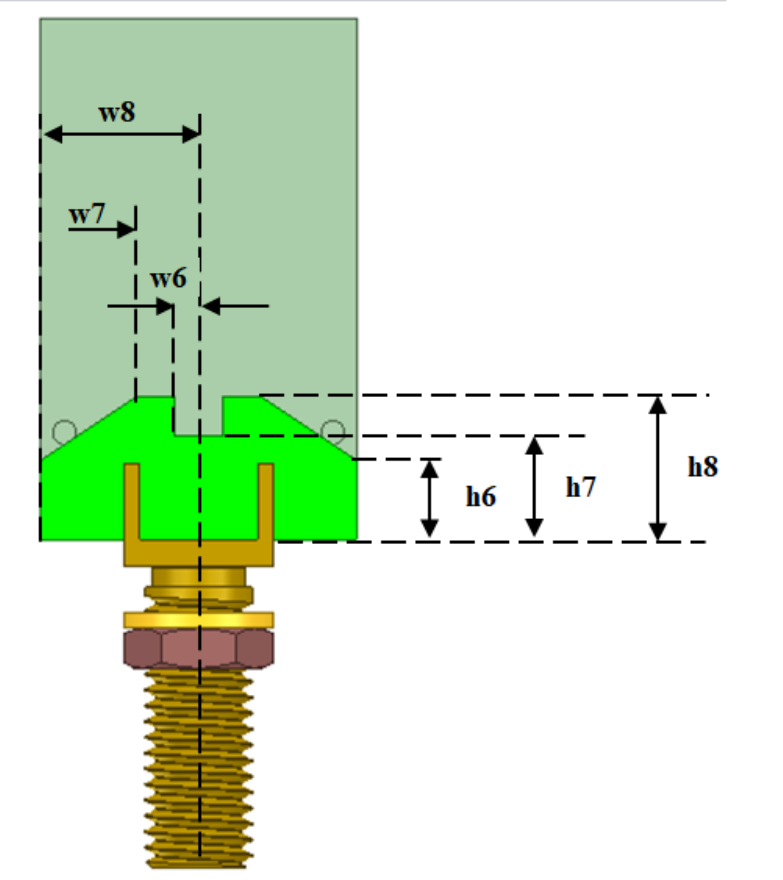}}
     \caption{}
\end{subfigure}
\caption{TX antenna: (a) front view and (b) back view.}
\label{fig:tx_antenna}
\end{figure}

Noting the circular symmetry of the structure, an omni-directional TX antenna is needed to operate at the center of the cylindrical structure. A good candidate for this purpose is a planar monopole antenna. Fig. \ref{fig:tx_antenna} shows the designed antenna. The antenna is printed on a 60-mil RO4003C Rogers substrate. The top and the bottom ground patches are connected using vias and soldered to the connector's body. To improve TX matching, the inset feeding technique has been used. The following parameters are tuned in optimizing TX matching: the length of ground patches, the notch dimensions in the bottom ground patch, the length and width of the narrow trace of the monopole, dimensions of the wider part of the monopole, and finally, the inset feed length. TX antenna tuning has been performed in the presence of caps and PCB walls for a center frequency of 5.8545\,GHz. 
Fig. \ref{fig:tx_antenna} shows the TX antenna and Table \ref{tab:tx_antenna_dimensions} shows final dimensions after fine-tuning in the presence of the rest of the antenna structure (refer to Section~\ref{Tune}). 
\begin{table*}
\begin{center}
\caption{TX antenna dimensions (in mm).}
\begin{tabularx}{0.9\textwidth} { 
  | >{\centering\arraybackslash}X
  | >{\centering\arraybackslash}X 
  | >{\centering\arraybackslash}X 
  | >{\centering\arraybackslash}X 
  | >{\centering\arraybackslash}X
  | >{\centering\arraybackslash}X 
  | >{\centering\arraybackslash}X
  | >{\centering\arraybackslash}X 
  | >{\centering\arraybackslash}X 
  | >{\centering\arraybackslash}X
  | >{\centering\arraybackslash}X 
  | >{\centering\arraybackslash}X
  | >{\centering\arraybackslash}X
  | >{\centering\arraybackslash}X
  | >{\centering\arraybackslash}X
  | >{\centering\arraybackslash}X
  | >{\centering\arraybackslash}X  | }
  \hline
  w0 & w1 & w2 & w3 & w4 & w5 & h1 & h2 & h3 & h4 & h5 & w6 & w7 & w8 & h6 & h7 & h8\\ 
  \hline
  1.008 & 0.5 & 1.25 & 0.7 & 2.5 & 3 & 9 & 13 & 16 & 20 & 30 & 1.5 & 4 & 10 & 5 & 6.5 & 9\\ 
  \hline
\end{tabularx}
\label{tab:tx_antenna_dimensions}
\end{center}
\end{table*}

\subsection{Control Board}
For driving the PIN diodes, we have designed a control board that is attached to the upper cap from outside of the structure (see Fig.~\ref{fig:multi-pattern_structure}). 
%which are connected to the pin header through hole connectors embedded on both the control board and each wall.
We have used an AD828 dual, low-power Op-Amp to drive the PIN diodes. This Op-Amp has 130 MHz 3-dB bandwidth at a gain equal to 2 dB and 450 V/µs slew rate. Since we have thirteen PIN diodes in each building block column, considering the forward voltage of 0.85 V for each PIN diode, one needs around 11 V to turn on all the diodes. The selected Op-Amp is able to perform this task since it operates using a ±15 V power supply.

\subsection{Iterative Optimization of Design Components} \label{Tune}
RX antenna is initially designed as a standalone element and then tuned in the presence of the rest of the structure to optimize the corresponding $S_{11}$ parameter.   Similarly, the TX antenna is initially designed as a standalone element by optimizing the corresponding $S_{11}$ parameter while achieving an omni-directional pattern.  
The patch dimensions are initially designed using Floquet ports, where the reflection coefficient, $S_{11}$, and the transmission coefficient, $S_{12}$, of the building block are monitored to approach a reflecting surface in the on state and a transparent surface in the off state.
Next, the above initial designs are integrated into the larger structure and the underlying four design decisions, namely geometries of: 1) RX antenna; 2) TX antenna; 3) surrounding walls; and 4) caps, including curvature, i.e., the parameter $S$ in \eqref{ParS} and the two radii in Fig. \ref{TX-all}(c)) are tuned iteratively. Objectives have been to realize good $S_{11}$ for antennas, a TX radiation pattern concentrated around the horizontal plane, and realize desirable $S_{11}$ and $S_{12}$ for the walls focusing on the seven different test states mentioned in Table \ref{Tabclw2}.

\subsection{Simulation vs. Measurement Results}
Fig. \ref{tx_input_matchingm} shows some measurement results, which closely match the simulation results in Fig. \ref{tx_input_matching}. Due to space limitations, only four cases of measurement results are shown. Several other cases have been tested, including some that were not considered in the initial design optimization. Measurement results show that, in all cases, the achieved value of $S_{11}$ remains at an acceptable level and is rather insensitive to the switches state. 

\section{Concluding Remarks}
Recently, several new ideas have emerged that show advantages when a radio frequency signal is selectively modified after leaving the transmitter. Most notable examples include \say{index modulation}, with media-based modulation as a special case, and \say{intelligent reflecting surfaces}. Media-based modulation offers unprecedented power/bandwidth efficiency improvements while enjoying a simple transmitter that sends a carrier. Random placement of constellation points over receive dimensions results in handling deep fades and can realize negligible error rates using a simple forward error correcting code. Despite mentioned benefits, media-based modulation suffers from challenges and shortcomings that need further research. These challenges include: 1) spectral growth due to the time-varying nature of the transmit unit requires new techniques for designing filters that could be integrated within the media-based transceiver; 2) training for constellation sets of large sizes is a cumbersome task. Parametric channel models, with a small number of parameters, could be used to capture channel variations over time. Such a channel model could assist in tracking the coordinates of MBM constellation points over time. Novel training signals need to be designed to efficiently and accurately update the underlying model parameters, while relying on a low training overhead.

%\subsection{Shaping gain and selection gain}
%In legacy modulation, using constellation points with equal probability results in losing the shaping gain. However, in a media-based modulation, constellation points will asymptotically have a Gaussian distribution (assuming Raleigh fading), which in turn means the shaping gain will be inherently realized. Motivated by this observation,  constellation points in MBM are used with equal probability.  Under this restriction, selecting a subset of points can increase the mutual information. The optimum solution is too complex. However, it is easy to show that the slope of the mutual information vs. energy at low SNR is determined by the sample second moment of the selected subset (assuming a symmetrical constellation, the sample first moment will be zero). This observation motivates selecting the subset of points with the highest energy.  Readers are referred to \cite{c00} for related mathematical derivations and simulation results. A comparison is provided between {\em optimum selection} and {\em selection with the highest sample second moment}, which performs very close to the optimum~\cite{c00}. 

%\subsection{Simulations using ray tracing}

%% file: appendix.tex
% !TEX root = ../main.tex

\begin{appendices}

\section{LMIMO-MBM Successive Cancellation List Decoder \label{appendix:scld}}
\subsection{Successive Cancellation Decoder}

In the presence of AWGN $\rv{z}$, for a received signal $\rv{y} = \sum_{n = 0} ^ {N-1} \rv{h}^n_{m[n]} + \rv{z}, $ an optimum decoder aims to find a message sequence $(\widehat{m}[0], ...,\widehat{m}[N-1])$ such that the point $\widehat{\rv{ c }} = \sum_{n = 0} ^ {N-1} \rv{h}^n_{\widehat{m}[n]}$ is at minimum Euclidean distance to $\rv{y}$. 
The successive cancellation decoder makes a decision about the transmitted symbol over a single modulator unit,  subtracts this estimate from the received signal, and continues the process recursively until all messages (corresponding to all transmit units) are recovered.
%Let us assume at decoding step $0$ the symbol transmitted over the $n^{(0)}$th unit is being recovered. Once a decision $\widehat{m}_{n^{(0)}}$ is reached, the constituent vector $\rv{h}_{n^{(0)}}(\widehat{m}_{n^{(0)}})$ corresponding to the recovered symbol is subtracted from the received signal $\rv{r}$ to reduce the interference on decoding data streams of other units.
%This process is continued until an estimate for all the elements of the message sequence $({m}[0], ..., m[N-1])$ is recovered.

Denote the code-book comprised of all the constituent vectors for all transmit units by ${\mathcal{U}}  := \cup_{i = 0}^{N-1} \mathcal{H}^i$. Initially, at decoding step $t = 0$, the decoder search through all the fading gains in the set ${\mathcal{U}} $ to find
\begin{IEEEeqnarray}{c}
\rv{g}[0] \gets \argmin_{\rv{h}  \in {\mathcal{U}}} { \lVert \rv{y}  - \rv{h} \rVert }_2. 
\end{IEEEeqnarray}
Let ${n}[0]$ denote the transmit unit corresponding to the estimate $\widehat{\rv{g}}[0]$ recovered at step $t=0$, i.e., ${\rv{g}}[0] \in \mathcal{H}^{{n}[0]}$. Therefore, we set
\begin{IEEEeqnarray}{c}
\widehat{\rv{h}}^{{n}[0]} \gets {\rv{g}}[0].
\end{IEEEeqnarray}
For decoding step $t = 1$, the received signal is updated according to $\rv{y}[1] \gets \rv{y} - {\rv{g}}[0]$, and since the message corresponding to unit ${n}[0]$ is already recovered, the search space for the next constituent vector is limited to the smaller set ${\mathcal{U}} \setminus \mathcal{H}^{{n}[0]}$, i.e, 

\begin{IEEEeqnarray}{c}
\rv{g}[1]\gets \argmin_{\rv{h}  \in {{\mathcal{U}} \setminus \mathcal{H}^{{n}[0]}}} { \lVert \rv{y}[1]  - \rv{h} \rVert }_2, \\
\widehat{\rv{h}}^{{n}[1]} \gets {\rv{g}}[1].
\end{IEEEeqnarray}
The general update rules for step $t$ are 
\begin{IEEEeqnarray}{c}
\rv{y}[t] \gets \rv{y}[t-1] - {\rv{g}}[t-1],
\end{IEEEeqnarray}
%\begin{IEEEeqnarray}{cc}
%\nonumber
%& {\mathcal{H}} ^{(N-t)\times M} (t) :=  \big\{{\rv{h}}_{n, m_n} \vert m \in \{0, ..., M-1\}, \\
%& n \in \{1, ..., N\}\setminus \{\widehat{n}(0), ..., \widehat{n}(t-1)\} \big\}, 
%\IEEEeqnarraynumspace
%\end{IEEEeqnarray}
\begin{IEEEeqnarray}{c}
{\rv{g}}[t] \gets \argmin_{\rv{h}  \in {{\mathcal{U} \setminus \cup_{t^\prime = 0}^{t-1} \mathcal{H}^{{n}[t^\prime]}}}} { \lVert \rv{y}[t]  - \rv{h}\rVert }_2, \\
\widehat{\rv{h}}^{{n}[t]} \gets {\rv{g}}[t].
\end{IEEEeqnarray}
The decoding is concluded at step $t = N-1$, where estimates $(\widehat{\rv{h}}^0, ..., \widehat{\rv{h}}^{N-1})$ for the constituent vectors corresponding to all transmit units, and consequently, $(\widehat{m}[0], ..., \widehat{m}[N-1])$ are recovered.

\subsection{Improved List Decoder}
Successive cancellation decoding may suffer from error propagation due to decision errors in intermediate steps. As a remedy to error propagation, we maintain a list of $L$ unique candidates. A larger list size improves the error performance at the cost of higher complexity.

Consider a list of candidates, where each candidate comprises sequences of the form $(\widehat{m}[0], ..., \widehat{m}[N-1])$ of tentative solutions for symbols transmitted over the $N$ transmit units. 
At each step $t$, exactly one constituent vector for each candidate is recovered. The constituent vector recovered for the $j$th candidate  at step $t = 0$ satisfies
\begin{IEEEeqnarray}{c}
{\rv{g}}[0, j] \gets \argjmin_{\rv{h}  \in {\mathcal{U}} } { \lVert \rv{y}  - \rv{h}\rVert }_2, \\
\widehat{\rv{h}}^{{n}[0, j]} \gets {\rv{g}}[0, j].
\end{IEEEeqnarray}
Here, ${n}[t, j]$ denotes the index of the transmit unit associated with the estimate $\rv{g}[t, j]$, i.e., $\rv{g}[t, j] \in \mathcal{H}^{n[t, j]}$. At a general step, $t \neq 0$, each candidate uses the same successive cancellation decoder independently of other candidates to obtain an estimate for a new constituent vector:

\begin{IEEEeqnarray}{c}
\rv{y}[t, j] \gets \rv{y}[t-1, j] - {\rv{g}}[t-1, j],
\end{IEEEeqnarray}
\begin{IEEEeqnarray}{c}
{\rv{g}}[t, j] \gets \argmin_{ \rv{h}  \in {{\mathcal{U} \setminus \cup_{t^\prime = 0}^{t-1} \mathcal{H}^{{n}[t^\prime, j]}}} } { \lVert \rv{y}[t, j]  - \rv{h}\rVert }_2, \\
\widehat{\rv{h}}^{{n}[t, j]} \gets {\rv{g}}[t, j].
\end{IEEEeqnarray}
Once the list decoder concludes the last step $t = N-1$, the candidate in the list which is closest to the received signal $\rv{y}$ (in terms of the Euclidean distance) is chosen as the final decoded message.

%\subsection{Parallel Branches Corresponding to Different Ordering of Transmission Units}
%Another method to improve the performance of successive decoding is to use different ordering for sequence of modulator units $n^{(i)}$ at each iteration. For $N$ modulator units there are $N!$ different possible ordering of the units for successive decoding, that can potentially result in a different solution. ``Multiple Candidate Solutions'' and `` Different Ordering of Transmission Units'' can be used together to reduce the possible error floor in successive decoding. This can be realized by using $N!$ branches, with multiple candidate solutions in each branch. Candidates in each branch are successively decoded using a different ordering of transmission units. Once all the candidates in all branches have concluded their successive iterations, they will be sorted according to the Euclidean distance of their projected $K$ dimensional constellation point to the original received signal $\rv{r}$. The candidate which is closest to received signal $\rv{r}$ in Euclidean space is chosen as the final decoded message.
\section{Time-limited Pulse Design \label{appendix:pulse_design}}
The solution maximizing the ratio $\eta$ is given by the characteristic function corresponding to the smallest characteristic value $\lambda$ of the following Fredholm integral equation:
\begin{IEEEeqnarray}{c}
p(x) = \frac{\lambda}{\pi} \int_{-\frac{T}{2}}^{\frac{T}{2}} p(t)\frac{\sin( 2\pi B(x-t))}{(x-t)}  \mathrm{d}t.
\end{IEEEeqnarray}
Particularly, $p(t)$ is the first characteristic function corresponding to the smallest characteristic value $\lambda$ for the Sinc kernel. The series expansion of the solution is according to
\begin{IEEEeqnarray}{c}
p(t) = \sum_{n=0}^{\infty}a_{n}\frac{J_{n+\frac{1}{2}}(t)}{\sqrt{t}},
\end{IEEEeqnarray}
where $J_{n+1/2}(t)$ is Bessel function of order $n+1/2$, and coefficients $a_n$ are determined from set of linear equations of the form
\begin{IEEEeqnarray}{c}
\frac{a_n}{2n+1} = \frac{\lambda}{2} \sum_{m} a_m  \int_{-TB}^{TB}J_{n+\frac{1}{2}}(t)J_{m+\frac{1}{2}}(t) \frac{ \mathrm{d}t}{t}.
\end{IEEEeqnarray}
%Coefficients $V_{nm}$ are determined according to,
%\begin{equation}
%V_{nm} = \frac{1}{2} \int_{-Tf_0}^{Tf_0}J_{n+1/2}(t)J_{m+1/2}(t) \frac{ \mathrm{d}t}{t}. 
%\end{equation}
%\newpage

\section{Analytical Bounds on Average Symbol Error Probability\label{appendix:uncoded_diversity}}
This section provides tight analytical bounds on MBM symbol error probability averaged over the sample space of channel gains and AWGN.
\subsection{Upper Bound}
Using maximum likelihood decoding, given that message $m = 0$ is transmitted, (i.e. fading vector $\rv{h}_0$ is projected at the receiver), a decoding error occurs if $\lVert \rv{y} -\rv{h}_i \rVert_2 \leq \lVert \rv{y} -\rv{h}_0 \rVert_2$
%\begin{IEEEeqnarray}{c}
%\left\{\rv{h}_0 \rightarrow \rv{h}_i \> | \> \rv{h}_0 \right\}  = \left\{ \sum_{j=1}^{n} \left(y_j-\sqrt{\mathsf{SNR}} x_{i,j}\right)^2  \leq \sum_{j=1}^{n} \left( y_j-\sqrt{\mathsf{SNR}} x_{0,j} \right)^2 \right \},
%\left\{\rv{h}_0 \rightarrow \rv{h}_i \> | \> \rv{h}_0 \right\}  =  \big \{ \lVert \rv{y} -\rv{h}_i \rVert^2 \leq \lVert \rv{y} -\rv{h}_0 \rVert^2 \big\},
%x\end{IEEEeqnarray}
for some $i \neq 0$.
Let ${\phi}(\rv{h}_0, \rv{h}_i)$ denote the pairwise error probability between $\rv{h}_0$ and $\rv{h}_i$, given $\rv{h}_0$ is sent. 
%\begin{IEEEeqnarray}{L}
%{\phi}(\rv{h}_0, \rv{h}_i) := \mathrm{Pr}\left\{\rv{h}_0 \rightarrow \rv{h}_i \> | \> \rv{h}_0 \right\}. 
%\end{IEEEeqnarray}
Since $\rv{h}_0$ and $\rv{h}_i$ are distributed normally, the squared Euclidean norm of their difference follows a chi-squared distribution with $n = 2K$ degrees of freedom. Therefore, the exact pairwise error probability averaged over the ensemble of MBM realizations is computed as
\begin{IEEEeqnarray}{c}
%\mathbb{E} \big[ \mathrm {Pr}\left\{\rv{h}_0 \rightarrow \rv{h}_i \> | \> \rv{h}_0 \right\} \big] 
\mathbb{E}{ [{\phi}(\rv{h}_0, \rv{h}_i)]} = \int_{z = 0}^{\infty} Q\left(\sqrt{\frac{\mathsf{SNR} z}{2}}\right) f(z; n) \ \mathrm{d}z.
\label{integral_exact}
\end{IEEEeqnarray}
Here, $\mathbb{E}$ indicates averaging over the ensemble of MBM constellation (channel realizations), $f(z; n)$ is the probability density function for chi-squared distribution with $n$ degrees of freedom, and $Q$ is the tail distribution function of the standard normal distribution. Note that $\mathbb{E}{[{\phi}(\rv{h}_0, \rv{h}_i)]}$ is independent of $\rv{h}_0$ and $\rv{h}_i$. Therefore, we use simplified notation $\mathbb{E}{[\phi]} := \mathbb{E}{[{\phi}(\rv{h}_0, \rv{h}_i)]}$.

%In \cite{789668}, a closed-form solution for this integral is provided as
%\begin{IEEEeqnarray}{L}
%\mathbb{E}{[\phi]} = \left[P(c)\right]^{\frac{n}{2}} \sum_{k=0}^{\frac{n}{2}-1} \binom{\frac{n}{2}-1+k}{k} [1-P(c)]^k, \\
%P(c) := \frac{1}{2} \left( 1-\sqrt{\frac{c}{1+c}} \right),
%\end{IEEEeqnarray}
%where $c$ is defined as $c := {\mathsf{SNR}}/2$. 

\begin{figure}[t]
\centering
\hspace{0cm}
\includegraphics[width=\figsizea]{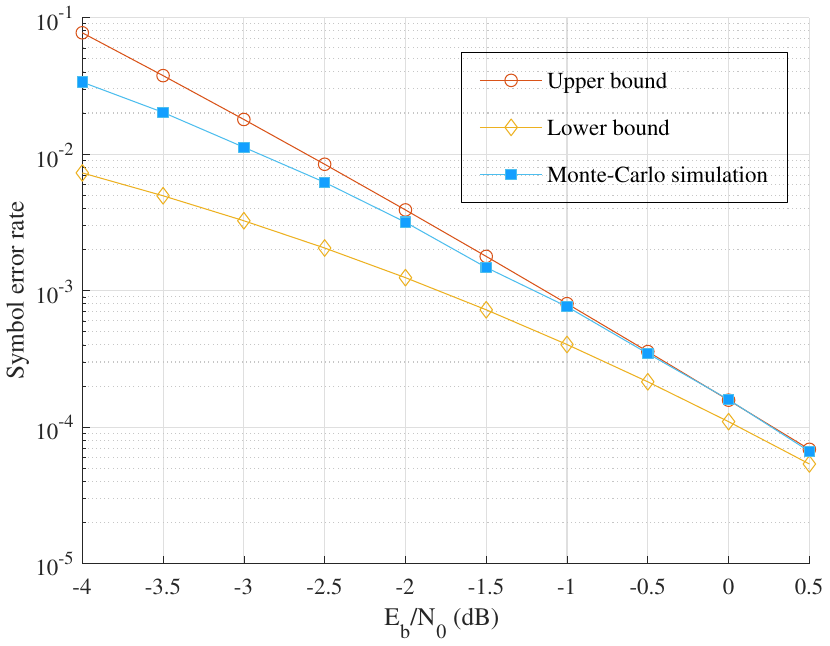}
\caption{Upper and lower bounds on the probability of error for $1\times8$ SIMO-MBM at rate $R=16$ bits per complex channel-use.} \label{fig:AK2}
\end{figure}

Reference \cite{789668} provides a closed form bound on pair-wise error probability. The alternative form for $\mathbb{E}{[\phi]}$ is given by
\begin{IEEEeqnarray}{c}
\frac{\sqrt{c/\pi}\,\Gamma\left(\frac{n+1}{2}\right)}{2(1+c)^{\frac{n+1}{2}}\Gamma\left(\frac{n+2}{2}\right)} \> _{2}F_1\left(1, \frac{n+1}{2}; \frac{n+2}{2}; \frac{1}{1+c}\right),
\IEEEeqnarraynumspace
\label{eq:closed_form}
\end{IEEEeqnarray}
where $c$ is defined as $c := {\mathsf{SNR}}/2$ and $_{2}F_1(.,.;.;)$ indicates  \textit{Gauss hyper-geometric function}. Applying the bound given in  \cite{gauss_ref} for {\em Gauss hyper-geometric function}, we get
\begin{IEEEeqnarray}{c}
 %& \overline{{\phi}(\rv{h}_0, \rv{h}_i)} \\
%& = \int_{x = 0}^{\infty} Q\left(\sqrt{\frac{\mathsf{SNR}x}{2}}\right) f(x; n) \ \mathrm{d}x \\
%&= \frac{\sqrt{c/\pi} \Gamma(\frac{n+1}{2})}{2(1+c)^{\frac{n+1}{2}}\Gamma(\frac{n+2}{2})} \> _{2}F_1(1, \frac{n+1}{2}; \frac{n+2}{2}; \frac{1}{1+c}) \label{eq:closed_form}\\
%&\leq \frac{\Gamma(\frac{n+1}{2})}{2 \sqrt{\pi}\Gamma(\frac{n+2}{2})}  \frac{\sqrt{c}}{(1+c)^{\frac{n+1}{2}}} \left(1+ \frac{n+1}{(n+2)c} \right)  \\
\mathbb{E}{[\phi]} \! \leq \frac{\Gamma\left(\frac{n+1}{2}\right)}{ 2 \sqrt{\pi}\Gamma\left(\frac{n+2}{2}\right)} (1+c)^{-\frac{n}{2}}.
\label{eq:tight_upper_bound}
\end{IEEEeqnarray}
%The inequality gives a tight upper bound on pairwise probability of error based on SNR and number of receive dimensions. 
Using the union-bound, the average probability of error over the ensemble of media-based constellations is bounded by
\begin{IEEEeqnarray}{LL}
P_e &\leq (M-1) \mathbb{E}{[\phi]} \\
&\leq \frac{ (M-1)\Gamma(\frac{n+1}{2})}{ 2 \sqrt{\pi}\,\Gamma(\frac{n+2}{2})} (1+c)^{-\frac{n}{2}}.
\IEEEeqnarraynumspace
\label{eq:union_bound}
\end{IEEEeqnarray}
%Upper bound \eqref{eq:union_bound} provides an achievable diversity-multiplexing trade-off for un-coded media-based modulation. 
Fig. \ref{fig:AK2} verifies that \eqref{eq:union_bound} provides a tight bound on average error probability in MBM compared to Monte-Carlo simulated performance. Plugging $\log M = r \log \mathsf{SNR}$, the average error probability at high SNR is bounded by
\begin{IEEEeqnarray}{c}
\label{eq:diversity_mg}
P_e(\mathsf{SNR}) \leq \frac{2^{\frac{n}{2}}\Gamma(\frac{n+1}{2})}{ \sqrt{\pi}\Gamma(\frac{n+2}{2})} (\mathsf{SNR})^{-(K -r)}.
\end{IEEEeqnarray}
%\begin{tcolorbox}[colback=green!3!white,colframe=black!20!white]
Equation \eqref{eq:diversity_mg} demonstrates that an uncoded MBM achieves diversity gain $d = K - r$, while maintaining multiplexing gain equal to $r$ i.e., date rate $R = \log M = r \log \mathsf{SNR}$.
%Fig. \ref{fig:slope} demonstrates the change in error slope of symbol error probability (i.e. diversity gain) for various rates (i.e. multiplexing gains).
%where $K=2n$ is the number of receive antennas and $ r $ denotes the \textit{spatial multiplexing gain}. 
%This means un-coded media-based, simultaneously supports transmission rate
\iffalse
\begin{IEEEeqnarray}{c}
R(\mathsf{SNR}) = \log M \approx r \log \mathsf{SNR} \>  \text{(bit/s/Hz)}
\end{IEEEeqnarray}
and average error probability 
\begin{IEEEeqnarray}{c}
P_e(\mathsf{SNR}) \approx b \> {\mathsf{SNR}}^{-(K - r)}
\end{IEEEeqnarray}
\fi
%\end{tcolorbox}
\iffalse
\begin{figure}[tbp]
\centerline{\includegraphics[scale  = 0.6 ]{./figs/analytical_bound.pdf}}
\caption{Monte-Carlo vs. bound \eqref{eq:union_bound} with $K = 8$ receive antennas and transmission rate $R = 16$ bits/sec/Hz. ${E_b}/{N_0} := {\mathsf{SNR}}/{R}$.}
\label{fig:analytical_bound}
\end{figure}
\begin{figure}[b]
\label{fig:slope}
\centerline{\includegraphics[scale  = 0.6 ]{./figs/diversity_multiplexing_pe.pdf}}
\caption{Upper bound on symbol error probability for different rates (i.e., spatial multiplexing gains) with $K = 8$ receive antennas.}
\end{figure}
\fi
\subsection{Lower Bound}
Given $\rv{h}_0$ is transmitted, the pairwise error probability between $\rv{h}_0$ and its closest neighbor in the constellation set, $\rv{h}^{\prime}$, provides a lower bound on the symbol error probability. The probability density function for Euclidean distance $|| \rv{h}^{\prime}  - \rv{h}_0||$, of closest neighbour to the transmitted point, is obtained using ordered statistics. Therefore, the lower bound on the expected probability of symbol error averaged over MBM channel realizations is given by
\begin{IEEEeqnarray}{c}
P_e \geq \int_{0}^{\infty}  f(z; n) \big(1-F(z; n)\big)^{M-2} Q \left(\sqrt{\frac{\mathsf{SNR}z}{2}} \right) \mathrm{d}z,
\IEEEeqnarraynumspace
\end{IEEEeqnarray}
%& \times Q(\frac{z}{\sqrt{2} \sigma _n}) \mathrm{d}z.
where $F(z; n)$ denotes the cumulative density function for chi-squared distribution with $n=2K$ degrees of freedom.

\section{Increasing Diversity Order via FEC  \label{appendix:coded_diversity}}
\subsection{MDS Codes and Information Sets}
Let $\Delta$ denote an $(N_c, K_c, D)$ code over alphabet $\mathcal{A}$. Here, $N_c$, $K_c$, and $D$, respectively, denote the code block length, dimension, and minimum distance such that any two codewords of $\Delta$ differ in at least $D$ coordinates. The dimensionless rate of the code is given by $\tau = {K_c}/{N_c}$. For the specific case of MDS codes, we have $D=N_c-K_c+1$.

Reference~\cite{forney1} defines the concept of ``information set''. Let $J(N_c) := \{1, 2, ..., N_c \}$. A subset of coordinates $I \subseteq J(N_c)$ is an information set if, in the projection\footnote{See~\cite{forney1} for definition and further details.} of the codewords over $I$, every possible $\vert I \vert$-tuple of elements (i.e., $\vert I \vert$-fold Cartesian product of alphabet $\mathcal{A}$) occurs exactly once. Reference~\cite{forney1} proves that for any MDS code of dimension $K_c$, any subset of coordinates with size less than or equal $K_c$ is an information set. In other words, an MDS code $\Delta$ has the property that for any $K_c$-tuple of elements of $\mathcal{A}$ on any given $K_c$ coordinates, there is a unique codeword of $\Delta$ which agrees with these $K_c$-tuple on the given $K_c$ coordinates. We refer to this property as the ``maximum information property" (MIP).

A consequence of codes satisfying MIP is that the coded symbols in each coordinate $I$ with $\vert I \vert \leq K_c$ are independent of each other. Each coordinate reveals the same amount of $\log_2 q$ bits of information about the coded message. This in turn means that each of the $q$ constellation points occurs with probability $1/q$. Noting that the assignment of constellation points to $q$ code alphabets is one-to-one and random, we can conclude that the constellation points in the $\Delta$ positions (selected by various codewords in a code satisfying MIP) are independent of each other. This property is used in the decoder analysis section to show that using simple MDS codes in MBM increases the diversity order. An example of  MDS codes satisfying MIP is Reed-Solomon codes. 
%Most known codes are defined over finite fields. In what follows we consider codes defined over Galois field $GF(M)$, where $M$, the field order, is a prime power.
\subsection{Encoder Mapping}
Given an MDS code over Galois field $GF(q)$, the sequence of symbols forming a codeword in $\Delta$ are mapped into sequences of MBM constellation points. 
%We shall assume that the constellation set $\mathcal{H}$ is partitioned into $N_c$ disjoint subsets $\mathcal{S}^1, ..., \mathcal{S}^{N_c}$. We assign the subset $\mathcal{S}^i$ to the $i$th coordinate of the codewords $\rv{u} := (u[1], u[2], ..., u[{N_c}]) \in \Delta$.
%This ensures that even identical symbols of $GF(q)$ in different coordinates of the codeword sequence are mapped to different points of the constellation set $\mathcal{H}$. Hence, we take $M$, the cardinality of MBM constellation set, according to $M = q \times N_c$. This requires a transmitter equipped with $\log_2 q + \log_2 N_c$ RF mirrors. Then, for every $\rv{u} \in \Delta$, the projected sequence of points at the receiver are of the form 
We take the cardinality of MBM constellation set, equal to the Galois field size, i.e., $M = q$. This requires a transmitter equipped with $\log_2 q$ RF mirrors. Then, for every $\rv{u} \in \Delta$, the projected sequence of points at the receiver are of the form 
\begin{IEEEeqnarray}{c}
(\rv{h}_{u[1]}, ... , \rv{h}_{u[{N_c}]}), 
\end{IEEEeqnarray}
where for all $0 \leq i \leq N_c-1$, $0 \leq u[i] \leq q-1, \ \rv{h}_{u[i]} \in \mathcal{H}$. 
%where $\rv{h}_{u[1]} \in \mathcal{S}^1, \rv{h}_{u[2]} \in \mathcal{S}^2$, and so on.
%\quad \text{for some} \> \rv{u} \in \Delta,
%where the mapping $g$ is according to
%\begin{IEEEeqnarray}{c}
%g(\rv{u}) := (g_1(u_1), ... , g_N(u_N)).
%\end{IEEEeqnarray}
In a nutshell, encoder operations include: 1) mapping a message from the set ${1, 2, .., M^{K_c}}$ into an MDS codeword $\rv{u} \in \Delta$; and 2) mapping the MDS codeword $\rv{u}$ into a sequence of MBM constellation points, $(\rv{h}_{u[1]}, ... , \rv{h}_{u[{N_c}]})$.
%Given the set of channel realizations $\rv{h}^M$, and an MDS code  $\Delta$ over finite field of order $M$, the symbols of a codeword in $\Delta$ are mapped into sequences of MBM constellation points. Mapping is carried out separately for each coordinate. Consequently, for every $\rv{u} = (u_1, ..., u_{N_c}) \in \Delta$ the projected codewords at the receiver (ignoring AWGN) are of the form 
%\begin{IEEEeqnarray}{c}
%\rv{h} = (\rv{h}_{u_1}, ... ,\rv{h}_{u_{N_c}}).
%\quad \text{for some} \> \rv{u} \in \Delta,
%\end{IEEEeqnarray}
%where the mapping $g$ is according to
%\begin{IEEEeqnarray}{c}
%g(\rv{u}) := (g_1(u_1), ... , g_N(u_N)).
%\end{IEEEeqnarray}
%In a nutshell, the encoder operations are: (1) Map a message from the set ${1, 2, .., M^{K_c}}$ into an MDS codeword $\rv{u} \in \Delta$, (2) Map codeword $\rv{u}$ into a sequence of MBM constellation points, $\rv{h} = (\rv{h}_{u_1}, ... ,\rv{h}_{u_{N_c}})$.

\subsection{Error Probability Analysis: Hard-decision Decoder }
This section studies the word error probability of MDS-coded MBM over the sample space of the ensemble of media-based constellation sets and AWGN. We use the method developed in \cite{limited_independence} for bounding the tail probability of weakly dependent random variables. The decoder is assumed to use hard-decision error correction. A code with minimum distance $D$ is capable of correcting $t$ errors up to $t = \lfloor (D-1)/2\rfloor$.

%For each fixed $\rv{u} \in \Delta$, the sequence of the constellation points $(\rv{h}_{u_1}, ... ,\rv{h}_{u_{N_c}})$ and the error probability ${\phi}(\rv{u})$ given $\rv{u}$ is sent, are random variables defined on ensemble of channel realizations (notation bar is used to indicate ensemble averages).
%For the i.i.d. Rayleigh fading and additive white Gaussian noise $\rv{z}$, 
For the codeword $\rv{u} \in \Delta$, the noisy projected received sequence at the receiver is of the form
% Ehsan: Please check the above sentence -> looks good.
\begin{IEEEeqnarray}{L}
(\rv{y}[1], ..., \rv{y}[N_c]) = (\rv{h}_{u[1]}, ... , \rv{h}_{u[{N_c}]})  + (\rv{z}[1], ..., \rv{z}[N_c]).
%\\
%\rv{y} = \rv{y}[1], ..., \rv{y}[{N_c}]. 
\IEEEeqnarraynumspace
%\\
%\rv{u} = (u_1, ..., u_N), \quad u_i \in GF(q) \>, \> 1 \leq i \leq N.
\end{IEEEeqnarray}

%$\begin{IEEEeqnarray}{L}
%rv{h} = (\rv{h}_{u[1]}, ..., \rv{h}_{u[N_c]}), \ \rv{h}_i = (h_{i, 1}, ..., h_{i, K}) \>, \> 1 \leq i \leq N, \IEEEeqnarraynumspace \\
%\rv{y} = (\rv{y}_1, ..., \rv{y}_{N_c}), \ \rv{y}_i = (y_{i, 1}, ..., y_{i, K}) \>, \> 1 \leq i \leq N_c. 
%\IEEEeqnarraynumspace
%\\
%\rv{u} = (u_1, ..., u_N), \quad u_i \in GF(q) \>, \> 1 \leq i \leq N.
%\end{IEEEeqnarray}

Initially, for each coordinate $i, 1 \leq i \leq N_c$ of the received vector, the decoder finds the closest point in the constellation set to the received signal $\rv{y}[i]$. This results in the estimate sequence $(\widehat{\rv{h}}[1], ..., \widehat{\rv{h}}[{N_c}])$ for channel gains and an associated message sequence $\widehat{\rv{u}} := (\widehat{u}[1], ..., \widehat{u}[{N_c}])$ which is not necessarily a valid codeword in the set $\Delta$. Subsequently, a hard-decision MDS decoder (e.g. Reed-Solomon) corrects for the errors in $\widehat{\rv{u}}$. 

Let $(e[1], ..., e[N_c])$ be the error indicator vector, where
\begin{equation}
e[i] := \left\{ \,
\begin{IEEEeqnarraybox}[][c]{l?s}
\IEEEstrut
0 & if $\widehat{u}[i] = {u}[i]$, \\
1 & if $\widehat{u}[i] \neq {u}[i]$.
\IEEEstrut
\end{IEEEeqnarraybox}
\right.
\label{eq:example_left_right1}
\end{equation}
A decoding error occurs if 
\begin{IEEEeqnarray}{c}
e := \sum_{i=1}^{N_c} e[i] \geq t+1.
\end{IEEEeqnarray}
Define the symmetric multi-linear polynomial
\begin{IEEEeqnarray}{c}
S_j(z_1, ..., z_{N_c}) := \sum_{1 \leq i_1 \leq i_2 ... \leq i_j \leq N_c} z_{i_1} z_{i_2} ... z_{i_{j}}.
\end{IEEEeqnarray}
Note that the events $\{ e \geq t+1\}$ and $\{S_{t+1}(e[1], ..., e[{N_c}]) \geq 1 \}$ are identical. The reason is, if the number of coordinates in error is less than or equal to $t$, the value of the polynomial $S_{t+1}$ is zero; Conversely, if the number of errors is greater than $t$, $S_{t+1}$ will take a value equal or greater than one.  
By Markov's inequality
\begin{IEEEeqnarray}{LL}
\nonumber
& Pr(e \geq t+1)  \\
&= Pr(S_{t+1}(e[1], ..., e[{N_c}]) \geq 1 ) \\
& \leq \mathbb{E} \big[ S_{t+1}(e[1], ..., e[{N_c}]) \big] \\
& = \sum_{1 \leq i_1 \leq i_2 ... \leq i_{t+1} \leq N_c} \mathbb{E} \big[e[{i_1}] e[{i_2}] ... e[{i_{t+1}]} \big].
\end{IEEEeqnarray}
The above expectation is over AWGN noise and the ensemble of MBM constellation sets.

Next, we use the property of MDS codes where symbols in any set of coordinates of size less than code dimension, $K_c$, are independent of each other. Taking $t+1 \leq K_c$, we can write

\begin{IEEEeqnarray}{LL}
\nonumber
& Pr(e \geq t+1)  \\
\nonumber
& \leq \sum_{1 \leq i_1 \leq i_2 ... \leq i_{t+1} \leq N_c} \mathbb{E} \big[e[{i_1}] e[{i_2}] ... e[{i_{t+1}}] \big] \\
\label{eq:independent _mds}
&= \sum_{1 \leq i_1 \leq i_2 ... \leq i_{t+1} \leq N_c} \mathbb{E} \big[e[i_1]\big] \mathbb{E} \big[e[{i_2}]\big] ... \mathbb{E} \big[e[{i_{t+1}]} \big] \\
\label{eq:identical_mds}
&= \binom{N_c}{t+1} \ \big(\mathbb{E} \big[e[1]\big]\big)^{t+1} \\
&= \binom{N_c}{t+1} \ {P_e}^{t+1} \\ 
%& \leq \binom{N_c}{t+1} \ \bigg(\frac{ ({q}-1)\Gamma(\frac{n+1}{2})}{ 2 \sqrt{\pi}\,\Gamma(\frac{n+2}{2})} \Big(1+\frac{\mathsf{SNR}}{2}\Big)^{-\frac{n}{2}}\bigg)^{t+1} \\
& \leq \binom{N_c}{t+1} \ \bigg(\frac{ (q-1)\Gamma(\frac{n+1}{2})}{ 2 \sqrt{\pi}\,\Gamma(\frac{n+2}{2})} \bigg)^{t+1} \Big(1+\frac{\mathsf{SNR}}{2}\Big)^{-(t+1)K}.
\label{eq:diversity_increase}
\IEEEeqnarraynumspace
\end{IEEEeqnarray}
Equation \eqref{eq:independent _mds} follows because: 1) MDS code property that symbols in any $t+1 \leq K_c$ coordinates are independent of each other; and 2) AWGN noise is independent across coordinates.  
Equation \eqref{eq:identical_mds} follows, noting that AWGN, message symbols, and fading gains have identical distribution across all the coordinates of a codeword.
% Ehsan, the above sentence is awkward when you say: "both with respect to AWGN and symbol probabilities." -> revised. 

The exponent in  \eqref{eq:diversity_increase} shows that applying an MDS code with an error correction capability of $t$ to MBM achieves a diversity order equal to $(t+1) \times K$ (using a single transmit unit). 
%2) the constellation points across different coordinates comes from a disjoint subset;
\end{appendices}

%% file: main.bbl
\begin{thebibliography}{10}
\providecommand{\url}[1]{#1}
\csname url@samestyle\endcsname
\providecommand{\newblock}{\relax}
\providecommand{\bibinfo}[2]{#2}
\providecommand{\BIBentrySTDinterwordspacing}{\spaceskip=0pt\relax}
\providecommand{\BIBentryALTinterwordstretchfactor}{4}
\providecommand{\BIBentryALTinterwordspacing}{\spaceskip=\fontdimen2\font plus
\BIBentryALTinterwordstretchfactor\fontdimen3\font minus
  \fontdimen4\font\relax}
\providecommand{\BIBforeignlanguage}[2]{{%
\expandafter\ifx\csname l@#1\endcsname\relax
\typeout{** WARNING: IEEEtran.bst: No hyphenation pattern has been}%
\typeout{** loaded for the language `#1'. Using the pattern for}%
\typeout{** the default language instead.}%
\else
\language=\csname l@#1\endcsname
\fi
#2}}
\providecommand{\BIBdecl}{\relax}
\BIBdecl

\bibitem{c0}
A.~K. Khandani, ``Media-based modulation: A new approach to wireless
  transmission,'' in \emph{2013 IEEE International Symposium on Information
  Theory}, 2013, pp. 3050--3054.

\bibitem{isit2014}
------, ``Media-based modulation: Converting static rayleigh fading to awgn,''
  in \emph{2014 IEEE International Symposium on Information Theory}, 2014, pp.
  1549--1553.

\bibitem{NN1}
G.~J. Foschini and M.~Gans, ``On limits of wireless communications in a fading
  environment when using multiple antennas,'' \emph{Wireless Personal
  Communications}, vol.~6, pp. 311--335, 1998.

\bibitem{NN2}
G.~J. Foschini, ``Layered space-time architecture for wireless communication in
  a fading environment when using multi-element antennas,'' \emph{Bell Labs
  Technical Journal}, vol.~1, no.~2, pp. 41--59, 1996.

\bibitem{NN3}
{N. Chiurtu, B. Rimoldi, and E. Telatar}, ``On the capacity of multi-antenna
  gaussian channels,'' in \emph{Proceedings. 2001 IEEE International Symposium
  on Information Theory (IEEE Cat. No.01CH37252)}, 2001, pp. 53--.

\bibitem{NN4}
{V. Tarokh, N. Seshadri, and A. R. Calderbank}, ``Space-time codes for high
  data rate wireless communication: performance criterion and code
  construction,'' \emph{IEEE Transactions on Information Theory}, vol.~44,
  no.~2, pp. 744--765, 1998.

\bibitem{NN5}
{L. Zheng and D. N.C. Tse}, ``Diversity and multiplexing: a fundamental
  tradeoff in multiple-antenna channels,'' \emph{IEEE Transactions on
  Information Theory}, vol.~49, no.~5, pp. 1073--1096, 2003.

\bibitem{8758978}
E.~Basar, ``Media-based modulation for future wireless systems: A tutorial,''
  \emph{IEEE Wireless Communications}, vol.~26, no.~5, pp. 160--166, 2019.

\bibitem{9834835}
{E. Seifi, A. K. Khandani}, ``Converting a 1×k static rayleigh channel to k
  parallel awgn using media-based modulation,'' in \emph{2022 IEEE
  International Symposium on Information Theory (ISIT)}, 2022, pp. 3109--3113.

\bibitem{lmimo_mbm}
\BIBentryALTinterwordspacing
{E. Seifi, M. Atamanesh, and A. K. Khandani}, ``Media-based mimo: A new
  frontier in wireless communications,'' 2015. [Online]. Available:
  \url{https://arxiv.org/abs/1507.07516}
\BIBentrySTDinterwordspacing

\bibitem{7511273}
E.~Seifi, M.~Atamanesh, and A.~K. Khandani, ``Media-based mimo: Outperforming
  known limits in wireless,'' in \emph{2016 IEEE International Conference on
  Communications (ICC)}, 2016, pp. 1--7.

\bibitem{new-ak1}
{A. K. Khandani and E. Bateni}, ``A practical, provably unbreakable approach to
  physical layer security,'' in \emph{16th Canadian Workshop on Information
  Theory (CWIT), 2019}, 2019, pp. 1--6.

\bibitem{new-ak2}
\BIBentryALTinterwordspacing
{S. Mohajer Hamidi, A. K. Khandani, and E. Bateni}, ``A secure key sharing
  algorithm exploiting phase reciprocity in wireless channels,'' \emph{CoRR},
  vol. abs/2111.15046, 2021. [Online]. Available:
  \url{https://arxiv.org/abs/2111.15046}
\BIBentrySTDinterwordspacing

\bibitem{c1}
{O. N. Alrabadi, A. Kalis, C. B. Papadias, and R. Prasad}, ``Aerial modulation
  for high order psk transmission schemes,'' in \emph{2009 1st International
  Conference on Wireless Communication, Vehicular Technology, Information
  Theory and Aerospace Electronic Systems Technology}, 2009, pp. 823--826.

\bibitem{c2}
{O. N. Alrabadi, C. B. Papadias, A. Kalis, and R. Prasad}, ``A universal
  encoding scheme for mimo transmission using a single active element for psk
  modulation schemes,'' \emph{IEEE Transactions on Wireless Communications},
  vol.~8, no.~10, pp. 5133--5142, 2009.

\bibitem{c3}
R.~Bains, ``On the usage of parasitic antenna elements in wireless
  communication systems,'' 2008.

\bibitem{Index}
{N. Ishikawa, S. Sugiura, and L. Hanzo}, ``50 years of permutation, spatial and
  index modulation: From classic rf to visible light communications and data
  storage,'' \emph{IEEE Communications Surveys \& Tutorials}, vol.~20, pp.
  1905--1938, 2018.

\bibitem{N4}
{R. Mesleh, H. Haas, C. W. Ahn, and S. Yun}, ``Spatial modulation - a new low
  complexity spectral efficiency enhancing technique,'' in \emph{2006 First
  International Conference on Communications and Networking in China}, 2006,
  pp. 1--5.

\bibitem{N5}
J.~Jeganathan, A.~Ghrayeb, and L.~Szczecinski, ``Spatial modulation: optimal
  detection and performance analysis,'' \emph{IEEE Communications Letters},
  vol.~12, no.~8, pp. 545--547, 2008.

\bibitem{N6}
M.~D. Renzo, H.~Haas, and P.~M. Grant, ``Spatial modulation for
  multiple-antenna wireless systems: a survey,'' \emph{IEEE Communications
  Magazine}, vol.~49, no.~12, pp. 182--191, 2011.

\bibitem{N7}
J.~Jeganathan, A.~Ghrayeb, L.~Szczecinski, and A.~Ceron, ``Space shift keying
  modulation for mimo channels,'' \emph{IEEE Transactions on Wireless
  Communications}, vol.~8, no.~7, pp. 3692--3703, 2009.

\bibitem{N8}
\BIBentryALTinterwordspacing
J.~Jeganathan, ``Space shift keying modulation for mimo channels,'' Master's
  thesis, Concordia University, 2008, unpublished. [Online]. Available:
  \url{https://spectrum.library.concordia.ca/id/eprint/975947/}
\BIBentrySTDinterwordspacing

\bibitem{N9}
Z.~Bouida, H.~El-Sallabi, A.~Ghrayeb, and K.~A. Qaraqe, ``Reconfigurable
  antenna-based space-shift keying (ssk) for mimo rician channels,'' \emph{IEEE
  Transactions on Wireless Communications}, vol.~15, no.~1, pp. 446--457, 2016.

\bibitem{c4-new}
{R. Y. Mesleh, H. Haas, S. Sinanovic, C. W. Ahn, and S. Yun}, ``Spatial
  modulation,'' \emph{IEEE Transactions on Vehicular Technology}, vol.~57,
  no.~4, pp. 2228--2241, 2008.

\bibitem{N3}
{M. Di Renzo, H. Haas, A. Ghrayeb, S. Sugiura, and L. Hanzo}, ``Spatial
  modulation for generalized mimo: Challenges, opportunities, and
  implementation,'' \emph{Proceedings of the IEEE}, vol. 102, no.~1, pp.
  56--103, 2014.

\bibitem{N10}
J.~Jeganathan, A.~Ghrayeb, and L.~Szczecinski, ``Generalized space shift keying
  modulation for mimo channels,'' in \emph{2008 IEEE 19th International
  Symposium on Personal, Indoor and Mobile Radio Communications}, 2008, pp.
  1--5.

\bibitem{N6n}
A.~Younis, N.~Serafimovski, R.~Mesleh, and H.~Haas, ``Generalised spatial
  modulation,'' in \emph{2010 Conference Record of the Forty Fourth Asilomar
  Conference on Signals, Systems and Computers}, 2010, pp. 1498--1502.

\bibitem{N6nn}
J.~Wang, S.~Jia, and J.~Song, ``Generalised spatial modulation system with
  multiple active transmit antennas and low complexity detection scheme,''
  \emph{IEEE Transactions on Wireless Communications}, vol.~11, no.~4, pp.
  1605--1615, 2012.

\bibitem{N6nnn}
J.~Fu, C.~Hou, W.~Xiang, L.~Yan, and Y.~Hou, ``Generalised spatial modulation
  with multiple active transmit antennas,'' in \emph{2010 IEEE Globecom
  Workshops}, 2010, pp. 839--844.

\bibitem{qsm}
R.~Mesleh, S.~S. Ikki, and H.~M. Aggoune, ``Quadrature spatial modulation,''
  \emph{IEEE Transactions on Vehicular Technology}, vol.~64, no.~6, pp.
  2738--2742, 2015.

\bibitem{new-ak0}
Q.~Wu and R.~Zhang, ``Intelligent reflecting surface enhanced wireless network:
  Joint active and passive beamforming design,'' in \emph{2018 IEEE Global
  Communications Conference (GLOBECOM)}, 2018, pp. 1--6.

\bibitem{new-ak3}
{Q. Wu, S. Zhang, B. Zheng, C. You and R. Zhang}, ``Intelligent reflecting
  surface-aided wireless communications: A tutorial,'' \emph{IEEE Transactions
  on Communications}, vol.~69, no.~5, pp. 3313--3351, 2021.

\bibitem{new-ak4}
{S. Nandan and M. A. Rahiman}, ``Intelligent reflecting surface (irs) assisted
  mmwave wireless communication systems: A survey,'' in \emph{Journal of
  Communications}, 2022.

\bibitem{forney1}
G.~Forney, ``On the hamming distance properties of group codes,'' \emph{IEEE
  Transactions on Information Theory}, vol.~38, no.~6, pp. 1797--1801, 1992.

\bibitem{9817680}
E.~Seifi, M.~Atamanesh, and A.~K. Khandani, ``Performance evaluation of
  media-based modulation in comparison with spatial modulations and legacy
  siso/mimo,'' in \emph{2022 17th Canadian Workshop on Information Theory
  (CWIT)}, 2022, pp. 80--85.

\bibitem{mbm_dmt}
\BIBentryALTinterwordspacing
E.~Seifi and A.~K. Khandani, ``Diversity multiplexing trade-off and selection
  gain in media-based modulation,'' \emph{CoRR}, vol. abs/2202.02882, 2022.
  [Online]. Available: \url{https://arxiv.org/abs/2202.02882}
\BIBentrySTDinterwordspacing

\bibitem{chalk}
J.~H.~H. Chalk, ``The optimum pulse-shape for pulse communication,''
  \emph{Journal of the Institution of Electrical Engineers}, vol.~97, pp.
  88--92, 1950.

\bibitem{Tech-rep1}
\BIBentryALTinterwordspacing
A.~K. Khandani, ``Media-based modulation,'' Univ. Waterloo, Tech. Rep.
  [Online]. Available: \url{https://cst.uwaterloo.ca/reports/media-report.pdf}
\BIBentrySTDinterwordspacing

\bibitem{789668}
{M.-S. Alouini and A. J. Goldsmith}, ``A unified approach for calculating error
  rates of linearly modulated signals over generalized fading channels,''
  \emph{IEEE Transactions on Communications}, vol.~47, no.~9, pp. 1324--1334,
  1999.

\bibitem{gauss_ref}
{R. W. Barnard, K. C. Richards, and H. C. Tiedeman}, ``A survey of some bounds
  for gauss' hypergeometric function and related bivariate means,''
  \emph{Journal of Mathematical Inequalities Volume Number}, vol.~4, pp.
  45--52, 01 2010.

\bibitem{limited_independence}
\BIBentryALTinterwordspacing
J.~P. Schmidt, A.~Siegel, and A.~Srinivasan, ``Chernoff–hoeffding bounds for
  applications with limited independence,'' \emph{SIAM Journal on Discrete
  Mathematics}, vol.~8, no.~2, pp. 223--250, 1995. [Online]. Available:
  \url{https://doi.org/10.1137/S089548019223872X}
\BIBentrySTDinterwordspacing

\end{thebibliography}
